\documentclass[useAMS,usenatbib]{mn2e}
\usepackage[dvips]{graphicx}
\usepackage{hyperref}

\newcommand{\bz}{$\langle B_z \rangle$}

\newcommand{\msun}{$M_{\odot}$}

\newcommand{\kms}{km\,s$^{-1}$}

\newcommand{\vsini}{$v \sin i$}
\newcommand{\teff}{$T_{\rm eff}$}

\newcommand{\mnras}{$MNRAS$}
\newcommand{\apj}{$ApJ$}
\newcommand{\apjl}{$ApJ$}
\newcommand{\aj}{$AJ$}
\newcommand{\aap}{$A\&A$}

\newcommand{\aaps}{$A\&AS$}

\newcommand{\apjs}{$ApJS$}

\newcommand{\physscr}{Phys.\ Scr.}

\title[Atmospheric parameters of the Magnetic B Stars]{The Magnetic Early B-type Stars II: stellar atmospheric parameters in the era of Gaia}
\author[M.\ Shultz et al.]{M.\ E.\ Shultz$^{1}$\thanks{E-mail:
mshultz@udel.edu},
G.\ A.\ Wade$^2$,
Th.\ Rivinius$^3$,
E.\ Alecian$^{5}$,
C.\ Neiner$^6$,
\newauthor{V.\ Petit$^7$, J.\ P.\ Wisniewski$^{8}$, and the MiMeS and BinaMIcS Collaborations}
\footnotemark[1]\thanks{Based on observations obtained at the Canada-France-Hawaii Telescope (CFHT) which is operated by the National Research Council of Canada, the Institut National des Sciences de l'Univers of the Centre National de la Recherche Scientifique of France, and the University of Hawaii; at the La Silla Observatory, ESO Chile with the MPA 2.2 m telescope; and at the Bernard Lyot Telescope.}\\
$^1$Annie Jump Cannon Fellow, Department of Physics and Astronomy, University of Delaware, 217 Sharp Lab, Newark, Delaware, 19716, USA\\
$^2$Department of Physics and Space Science, Royal Military College of Canada, Kingston, Ontario K7K 7B4, Canada\\
$^3$ESO - European Organisation for Astronomical Research in the Southern Hemisphere, Casilla 19001, Santiago 19, Chile\\
$^5$Universit\'e Grenoble Alpes, IPAG, F-38000 Grenoble, France\\
$^6$LESIA, Observatoire de Paris, PSL Research University, CNRS, Sorbonne Universit\'es, UPMC Univ. Paris 06, Univ. Paris Diderot,\\
Sorbonne Paris Cit\'e, 5 place Jules Janssen, F-92195 Meudon, France\\
$^7$Department of Physics and Astronomy, University of Delaware, 217 Sharp Lab, Newark, Delaware, 19716, USA\\
$^{8}$Homer L. Dodge Department of Physics \& Astronomy, The University of Oklahoma, 440 W. Brooks Street, Norman, OK 73019, USA\\
}
\begin{document}

\date{}

\pagerange{\pageref{firstpage}--\pageref{lastpage}} \pubyear{2002}

\maketitle

\label{firstpage}

\begin{abstract}
Atmospheric parameters determined via spectral modelling are unavailable for many of the known magnetic early B-type stars. We utilized high-resolution spectra together with NLTE models to measure effective temperatures \teff~and surface gravities $\log{g}$ of stars for which these measurements are not yet available. We find good agreement between our \teff~measurements and previous results obtained both photometrically and spectroscopically. For $\log{g}$, our results are compatible with previous spectroscopic measurements; however, surface gravities of stars previously determined photometrically have been substantially revised. We furthermore find that $\log{g}$ measurements obtained with HARPSpol are typically about 0.1 dex lower than those from comparable instruments. Luminosities were determined using Gaia Data Release 2 parallaxes. We find Gaia parallaxes to be unreliable for bright stars ($V<6$ mag) and for binaries; in these cases we reverted to Hipparcos parallaxes. In general we find luminosities systematically lower than those previously reported. Comparison of $\log{g}$ and $\log{L}$ to available rotational and magnetic measurements shows no correlation between either parameter with magnetic data, but a clear slow-down in rotation with both decreasing $\log{g}$ and increasing $\log{L}$, a result compatible with the expectation that magnetic braking should lead to rapid magnetic spindown that accelerates with increasing mass-loss. 
\end{abstract}


\section{Introduction}

\begin{table*}
\caption[]{Spectropolarimetric data summary and references for magnetic detections for the 5 stars added to the sample. The first row gives the names of the stars. The second row contains remarks as to special properties. The third row gives the spectral type. The fourth through seventh rows give the number of spectropolarimetric observations available for each instrument. The eighth row gives the reference for the original magnetic detection. Rows 9 to 15 give, respectively: the projected rotational velocity \vsini; the rotational period $P_{\rm rot}$; the epoch used to determine the zero-point of the phase curve (typically the time of maximum $|\langle B_z \rangle|$); the peak observed value of \bz; the mean value $B_0$ of the sinusoidal fit to \bz; the semi-amplitude $B_1$ of the first harmonic of the sinusoidal fit to \bz; and the semi-amplitude $B_2$ of the second harmonic. Reference key: \protect\cite{2012A&A...542A..55P}$^a$; \protect\cite{2017A&A...605A.104B}$^b$; \protect\cite{2014A&A...562A.143F}$^c$; This work$^d$; \protect\cite{2015ApJ...811L..26W}$^e$; \protect\cite{2017MNRAS.467L..81H}$^f$; \protect\cite{2016A&A...587A...7P}$^g$; \protect\cite{2017MNRAS.471.1543H}$^h$; \protect\cite{2017A&A...597L...6C}$^i$; \protect\cite{2017MNRAS.472..400H}$^j$.}
\label{newstars}
\begin{tabular}{l r r r r r}
\hline
\hline
Star & HD\,43317 & HD\,47777 & HD\,345439 & CPD $-57^\circ$ 3509 & CPD $-62^\circ$ 2124 \\
\hline
Remarks & SPB & He Be & -- & -- & -- \\
Spec. Type & B3\,IV & B3\,V & B2\,IV & B2\,IV & B2\,IV \\
ESPaDOnS & -- & 13 & -- & -- & -- \\
Narval & 34 & -- & -- & -- & -- \\
HARPSpol & -- & -- & -- & 1 & 1 \\
FORS2 & -- & -- & 18 & 20 & 17 \\
Detection & {\cite{2013A&A...557L..16B}} & {\cite{2014A&A...562A.143F}} & {\cite{2015A&A...578L...3H}} & {\cite{2016A&A...587A...7P}} & {\cite{2017A&A...597L...6C}} \\
\vsini~(\kms) & $115\pm9$$^a$ & $60\pm5$$^{c}$ & $270\pm20^e$ & $35\pm2^g$ &  $35\pm5^i$ \\
$P_{\rm rot}$ (d) & 0.897673(4)$^b$ &  2.6415(6)$^{c,d}$ & $0.77018(2)^{e,f}$ & 6.3626(3)$^h$ & 2.62809(5)$^j$ \\
JD0 - 2400000 (d) & 56185.8380$^b$ & 54461.8(2)$^d$ & 56926.0425$^f$ & 56984.04(6)$^h$ & 57444.146(8)$^j$ \\
\bz$_{\rm max}$ (kG) & $0.30\pm0.01$$^b$ & $0.68\pm0.09^d$ & $2.5\pm0.1^f$ & $1.07\pm0.07^h$ & $6.8\pm0.5^j$ \\
$B_0$ (kG) & $0.045\pm0.016$$^b$ & $-0.02\pm0.04^d$ & $0.9\pm0.1^f$ & $0.18\pm0.05^h$ & $5.5\pm0.1^{d,j}$ \\
$B_1$ (kG) & $0.221\pm0.022$$^b$ & $-0.62\pm0.03^d$ & $1.6\pm0.1^f$ & $0.89\pm0.06^h$ & $-1.2\pm0.2^{d,j}$ \\
$B_2$ (kG) & -- & -- & -- & -- & $1.0\pm0.2^{d,j}$ \\
\hline\hline
\end{tabular}
\end{table*}

About 1 in 15 early-type stars possesses a detectable magnetic field \citep{grun2012c,2017MNRAS.465.2432G}. Their magnetic fields are typically strong ($10^2 - 10^4$ G), topologically simple (mostly tilted dipoles), stable over at least thousands of rotational cycles \citep[e.g.][]{2018MNRAS.475.5144S}, and their strength show no correlation with rotation (unlike what would be expected for magnetic fields maintained by contemporaneous dynamos). These properties lead to their characterization as fossil magnetic fields \citep[e.g.][]{2015IAUS..305...61N}. Magnetic OB stars are particularly interesting due to their magnetically confined winds, which often lead to magnetospheres that can be detected via X-ray, ultraviolet, optical, and infrared emission lines \citep[e.g.][hereafter P13]{petit2013}. Magnetic wind confinement also leads to rapid spindown \citep[e.g.][]{ud2009}, an effect which should intensify with increasing mass-loss rate and increasing magnetic field strength. 

Magnetospheres can be divided into those in which rotation plays a negligible role (Dynamical Magnetospheres or DMs), and those in which centrifugal support due to rapid rotation is decisive in sculpting the circumstellar plasma distribution (Centrifugal Magnetospheres or CMs). P13 introduced a rotation-magnetic confinement diagram, and showed that the position of a star on the diagram is broadly predictive of its magnetospheric H$\alpha$ emission status. O-type stars (with high mass-loss rates and typically very slow rotation) are almost invariably predicted to have DMs, and always possess detectable H$\alpha$ emission. Conversely, B-type stars only possess detectable magnetospheric emission when they are both very rapidly rotating and very strongly magnetized, i.e.\ when they are predicted to have very large CMs. 

For most of the stars studied by P13, rotational periods and surface magnetic field strengths were unknown, meaning that only limiting values of their magnetic and rotational properties were available, and their positions on the rotation-magnetic confinement diagram were only limiting values. In consequence, the conditions under which CMs become detectable were not observationally constrained. 

\citet[][hereafter Paper I]{2018MNRAS.475.5144S} presented magnetic field measurements, rotational periods, and projected surface rotational velocities \vsini~for all known main-sequence early B-type stars in the P13 sample for which sufficient magnetic data had been obtained for accurate characterization of their surface magnetic properties. Before these results can be used to obtain surface rotational properties and magnetic oblique rotator models, fundamental stellar parameters (masses and radii) must also be determined, which in turn require stellar atmospheric parameters \teff, $\log{g}$, and $\log{L}$. For many of the stars, only photometric measurements of the first two parameters are available. The newly available Gaia Data Release 2 \citep[DR2;][]{2018A&A...616A...1G} parallaxes mean that distances and luminosities can be obtained with higher precision for many of the more distant stars in the sample. The purpose of this paper is to combine the available high-resolution spectroscopic data with Gaia DR2 parallaxes to determine high-precision atmospheric parameters. 

An overview of the sample and observations is provided in \S~\ref{sec:sample}. \S~\ref{sec:surf_pars} describes the measurements and the resulting properties of the sample, together with an examination of some of the systematics arising from different measurement methods. Results are discussed, and conclusions drawn, in \S~\ref{sec:disc}. Previously unreported magnetic measurements are provided for HD\,47777 in Appendix \ref{hd47777}. Surface gravity measurements of individual single stars are detailed in Appendix \ref{sec:logg_single}, and those of binary systems in Appendix \ref{sec:logg_bin}.

\section{Sample and observations}\label{sec:sample}

The sample consists of essentially all known magnetic main sequence stars with spectral types between B5 and B0. The selection criteria and properties of the sample were described in Paper I. One star, HD 35912, has been removed as we demonstrated in Paper I the absence of a detectable surface magnetic field. The study is based primarily upon an extensive database of high-resolution ESPaDOnS, Narval, and HARPSpol spectropolarimetry, in some cases supplemented with FEROS spectroscopy; these data were also described in Paper I. The majority of these data were acquired by the MiMeS and BinaMIcS Large Programs (LPs). The basic observational techniques and strategy of the MiMeS LPs, as well as the reduction and analysis of ESPaDOnS, Narval, and HARPSpol data, were described by \cite{2016MNRAS.456....2W}. The BinaMIcS LPs used the same instruments as MiMeS. Additional observations were acquired by the BRIght Target Explorer Constellation polarimetric survey \citep[BRITEpol;][]{neiner2017ppas}, the B-fields in OB stars \citep[BOB;][]{2015A&A...582A..45F,2017A&A...599A..66S} LP at the European Southern Observatory, and by various independent ESPaDOnS observing programs (listed in Paper I) at the Canada France Hawaii Telescope. 


Since Paper I five stars satisfying the selection criteria have been added to this sample. The magnetic and rotational properties of these stars, along with the available high- and low-resolution spectropolarimetric measurements, are given together with the relevant references in Table 1. In the case of HD\,47777 the magnetic data are published here for the first time, and are described in Appendix \ref{hd47777}. The final sample consists of 56 stars, listed in Table \ref{lumteffloggtab}.


\section{Atmospheric parameters}\label{sec:surf_pars}

\begin{table*}
\caption[Physical parameters.]{Stellar surface parameters. A superscript $p$ after the name indicates a star for which P13 utilized photometric calibrations to determine the surface parameters; a superscript $c$ indicates a star for which P13 determined luminosities using {\sc chorizos}. The 3$^{\rm rd}$ column indicates, for the spectroscopic binaries, which component the parameters relate to; the magnetic component is indicated with a superscript $m$. The 4$^{\rm th}$ column indicates whether the star is chemically peculiar, and if this is He-w(eak) or He-s(trong). References for spectroscopic analyses are provided in the final column, where superscript $l$ indicates $\log{L}$, $t$ indicates \teff, and $g$ indicates $\log{g}$.}
\label{lumteffloggtab}
\begin{tabular}{l l l | l r r r l}
\hline
\hline
Star Name & Alt. Name & Comp. & CP? & $\log{(L/L\odot)}$    & \teff (kK) & $\log{g}$ & References \\
\hline
HD\,3360 & $\zeta$\,Cas &  -- & -- & 3.82$\pm$0.06 & 20.8$\pm$0.2 & 3.80$\pm$0.05 & {\cite{2014AA...566A...7N}}$^{ltg}$  \\
HD\,23478 & ALS\,14589 &  -- & He-s & 3.2$\pm$0.2 & 20.0$\pm$2.0 & 4.20$\pm$0.20 & {\cite{2015MNRAS.451.1928S}}$^{tg}$ This work$^{l}$  \\
HD\,25558 & 40\,Tau & A & -- & 2.8$\pm$0.3 & 16.9$\pm$0.8 & 4.20$\pm$0.20 & {\cite{2014MNRAS.438.3535S}}$^{ltg}$  \\
 & & B$^{m}$ & -- & 2.6$\pm$0.4 & 16.3$\pm$0.8 & 4.25$\pm$0.25 & {\cite{2014MNRAS.438.3535S}}$^{ltg}$  \\
HD\,35298$^c$ & V\,1156\,Ori &  -- & He-w & 2.4$\pm$0.1 & 15.8$\pm$0.8 & 4.25$\pm$0.12 & This work$^{ltg}$  \\
HD\,35502$^c$ & -- & A$^{m}$ & He-s & 3.0$\pm$0.1 & 18.4$\pm$0.6 & 4.30$\pm$0.20 & {\cite{2016MNRAS.460.1811S}}$^{ltg}$  \\
 & & Ba & -- & 1.4$\pm$0.3 &  8.9$\pm$0.3 & 4.30$\pm$0.30 & {\cite{2016MNRAS.460.1811S}}$^{ltg}$  \\
 & & Bb & -- & 1.4$\pm$0.3 &  8.9$\pm$0.3 & 4.30$\pm$0.30 & {\cite{2016MNRAS.460.1811S}}$^{ltg}$  \\
HD\,36485$^c$ & $\delta$\,Ori\,C & A$^{m}$ & He-s & 3.1$\pm$0.2 & 20.0$\pm$2.0 & 4.20$\pm$0.20 & {\cite{leone2010}}$^{t}$ This work$^{lg}$  \\
 & & B & -- & 1.6$\pm$0.2 & 10.0$\pm$2.0 & 4.30$\pm$0.20 & {\cite{leone2010}}$^{t}$ This work$^{lg}$  \\
HD\,36526$^p$ & V\,1099\,Ori &  -- & He-w & 2.3$\pm$0.2 & 15.0$\pm$2.0 & 4.10$\pm$0.14 & This work$^{ltg}$  \\
HD\,36982$^p$ & LP\,Ori &  -- & He-s & 3.0$\pm$0.2 & 22.0$\pm$2.0 & 4.40$\pm$0.20 & {\cite{petit2012a}}$^{t}$ This work$^{lg}$  \\
HD\,37017$^p$ & V\,1046\,Ori & A$^{m}$ & He-s & 3.4$\pm$0.2 & 21.0$\pm$2.0 & 4.10$\pm$0.20 & {\cite{1998AA...337..183B}}$^{t}$ This work$^{lg}$  \\
 & & B & -- & 2.1$\pm$0.3 & 14.0$\pm$1.5 & 4.25$\pm$0.25 & {\cite{1998AA...337..183B}}$^{t}$ This work$^{lg}$  \\
HD\,37058$^c$ & V\,359\,Ori &  -- & He-w & 2.9$\pm$0.1 & 18.6$\pm$0.6 & 4.17$\pm$0.07 & This work$^{ltg}$  \\
HD\,37061 & NU\,Ori & Aa & -- & 4.35$\pm$0.09 & 30.0$\pm$0.5 & 4.20$\pm$0.10 & {\cite{2011AA...530A..57S}}$^{ltg}$  \\
 & & Ab & -- & 2.7$\pm$0.3 & 17.0$\pm$2.0 & 4.30$\pm$0.10 & {\cite{2019MNRAS.482.3950S}}$^{ltg}$  \\
 & & C$^{m}$ & He-s & 3.3$\pm$0.3 & 22.0$\pm$1.0 & 4.30$\pm$0.10 & {\cite{2019MNRAS.482.3950S}}$^{ltg}$  \\
HD\,37479$^p$ & $\sigma$\,Ori\,E &  -- & He-s & 3.5$\pm$0.2 & 23.0$\pm$2.0 & 4.20$\pm$0.20 & {\cite{1989AA...224...57H}}$^{t}$ This work$^{lg}$  \\
HD\,37776$^p$ & V\,901\,Ori &  -- & He-s & 3.3$\pm$0.2 & 22.0$\pm$1.0 & 4.25$\pm$0.20 & {\cite{2007AA...468..263C}}$^g$ This work$^{tl}$  \\
HD\,43317 & HR\,2232 &  -- & -- & 2.95$\pm$0.08 & 17.4$\pm$1.0 & 4.07$\pm$0.10 & {\cite{2012A&A...542A..55P}}$^{g}$ This work$^{lg}$  \\
HD\,44743 & $\beta$\,CMa &  -- & -- & 4.41$\pm$0.06 & 24.7$\pm$0.3 & 3.78$\pm$0.08 & {\cite{2015AA...574A..20F}}$^{ltg}$  \\
HD\,47777 &              & --  & He-s & 3.42$\pm$0.15 & $22.0 \pm 1.0$ & $4.20 \pm 0.10$ & {\cite{2014A&A...562A.143F}}$^{ltg}$  \\
HD\,46328 & $\xi^1$\,CMa &  -- & -- & 4.5$\pm$0.1 & 27.0$\pm$1.0 & 3.78$\pm$0.07 & {\cite{2017MNRAS.471.2286S}}$^{ltg}$  \\
HD\,52089 & $\epsilon$\,CMa &  -- & -- & 4.35$\pm$0.05 & 22.5$\pm$0.3 & 3.40$\pm$0.08 & {\cite{2015AA...574A..20F}}$^{ltg}$  \\
HD\,55522 & HR\,2718 &  -- & He-s & 3.0$\pm$0.2 & 17.4$\pm$0.4 & 3.95$\pm$0.06 & {\cite{2004AA...413..273B}}$^{t}$ This work$^{lg}$  \\
HD\,58260$^c$ & ALS\,14015 &  -- & He-s & 3.2$\pm$0.3 & 19.3$\pm$1.3 & 4.2$\pm$0.2 & This work$^{lt}${\cite{2007AA...468..263C}}$^{g}$  \\
HD\,61556 & HR\,2949 &  -- & He-w & 3.1$\pm$0.2 & 18.5$\pm$0.8 & 4.10$\pm$0.15 & {\cite{2015MNRAS.449.3945S}}$^{ltg}$  \\
HD\,63425 & -- &  -- & -- & 4.49$\pm$0.07 & 29.5$\pm$1.0 & 4.00$\pm$0.10 & {\cite{2011MNRAS.412L..45P}}$^{ltg}$  \\
HD\,64740 & HR\,3089 &  -- & He-s & 3.8$\pm$0.2 & 24.5$\pm$1.0 & 4.01$\pm$0.09 & {\cite{1990ApJ...358..274B}}$^{t}$ This work$^{lg}$  \\
HD\,66522$^p$ & ALS\,16280 &  -- & He-s & 3.5$\pm$0.2 & 20.8$\pm$2.1 & 3.88$\pm$0.23 & This work$^{ltg}$  \\
HD\,66665 & -- &  -- & -- & 4.7$\pm$0.2 & 28.5$\pm$1.0 & 3.90$\pm$0.10 & {\cite{2011MNRAS.412L..45P}}$^{ltg}$  \\
HD\,66765$^c$ & ALS\,14050 &  -- & He-s & 3.4$\pm$0.2 & 20.0$\pm$2.0 & 4.13$\pm$0.20 & {\cite{alecian2014}}$^{t}$ This work$^{lg}$  \\
HD\,67621$^p$ & ALS\,14055 &  -- & He-w & 3.3$\pm$0.1 & 21.0$\pm$0.6 & 4.18$\pm$0.10 & {\cite{alecian2014}}$^{t}$ This work$^{lg}$  \\
HD\,96446 & V\,430\,Car &  -- & He-s & 3.8$\pm$0.2 & 23.0$\pm$1.0 & 3.74$\pm$0.10 & This work$^{ltg}$  \\
HD\,105382$^p$ & HR\,4618 &  -- & He-s & 3.0$\pm$0.2 & 18.0$\pm$0.5 & 4.13$\pm$0.07 & {\cite{2001AA...366..121B}}$^{t}$ This work$^{lg}$  \\
HD\,121743$^p$ & $\phi$\,Cen &  -- & He-w & 3.6$\pm$0.2 & 21.0$\pm$1.3 & 4.02$\pm$0.12 & {\cite{alecian2014}}$^{t}$ This work$^{lg}$  \\
HD\,122451$^p$ & $\beta$\,Cen & Aa & -- & 4.5$\pm$0.2 & 25.0$\pm$2.0 & 3.55$\pm$0.11 & {\cite{2016A&A...588A..55P}}$^{tl}$ This work$^{g}$  \\
 & & Ab$^{m}$ & -- & 4.4$\pm$0.2 & 23.0$\pm$2.0 & 3.55$\pm$0.11 & {\cite{2016A&A...588A..55P}}$^{tl}$ This work$^{g}$  \\
HD\,125823$^p$ & a\,Cen &  -- & He-w & 3.2$\pm$0.2 & 19.0$\pm$2.0 & 4.14$\pm$0.12 & {\cite{2010AA...520A..44B}}$^{t}$ This work$^{lg}$  \\
HD\,127381 & $\sigma$\,Lup &  -- & He-s & 3.76$\pm$0.06 & 23.0$\pm$1.0 & 4.02$\pm$0.10 & {\cite{henrichs2012}}$^{ltg}$  \\
HD\,130807$^c$ & $o$\,Lup & A$^{m}$ & He-w & 2.7$\pm$0.2 & 17.0$\pm$1.0 & 4.25$\pm$0.10 & {\cite{2018arXiv180805503B}}$^{t}$ This work$^{lg}$  \\
 & & B & -- & 2.5$\pm$0.2 & 14.0$\pm$1.0 & 4.25$\pm$0.10 & {\cite{2018arXiv180805503B}}$^{t}$ This work$^{lg}$  \\
HD\,136504$^c$ & $\epsilon$\,Lup & A$^{m}$ & -- & 3.7$\pm$0.2 & 20.5$\pm$0.5 & 3.97$\pm$0.15 & This work$^{ltg}$  \\
 & & B$^{m}$ & -- & 3.3$\pm$0.2 & 18.5$\pm$0.5 & 4.13$\pm$0.15 & This work$^{ltg}$  \\
HD\,142184 & HR\,5907 &  -- & He-s & 2.8$\pm$0.1 & 18.5$\pm$0.5 & 4.31$\pm$0.05 & {\cite{grun2012}}$^{ltg}$  \\
HD\,142990$^p$ & V\,913\,Sco &  -- & He-w & 2.9$\pm$0.1 & 18.0$\pm$0.5 & 4.15$\pm$0.11 & This work$^{ltg}$  \\
HD\,149277$^p$ & ALS\,14369 & A$^{m}$ & He-s & 3.5$\pm$0.2 & 20.0$\pm$2.0 & 3.75$\pm$0.15 & This work$^{ltg}$  \\
 & & B & -- & 3.2$\pm$0.2 & 19.0$\pm$2.0 & 3.85$\pm$0.15 & This work$^{ltg}$  \\
HD\,149438 & $\tau$\,Sco &  -- & -- & 4.5$\pm$0.1 & 32.0$\pm$1.0 & 4.00$\pm$0.10 & {\cite{2006AA...448..351S}}$^{ltg}$  \\
HD\,156324$^p$ & ALS\,4060 & Aa$^{m}$ & He-s & 3.8$\pm$0.6 & 22.0$\pm$3.0 & 4.00$\pm$0.30 & {\cite{alecian2014}}$^{tg}$ {\cite{2018MNRAS.475..839S}}$^{l}$  \\
 & & Ab & -- & 2.0$\pm$0.9 & 15.0$\pm$1.5 & 4.30$\pm$0.30 & {\cite{alecian2014}}$^{tg}$ {\cite{2018MNRAS.475..839S}}$^{l}$  \\
 & & B & -- & 2.0$\pm$0.9 & 14.0$\pm$1.5 & 4.30$\pm$0.30 & {\cite{alecian2014}}$^{tg}$ {\cite{2018MNRAS.475..839S}}$^{l}$  \\
HD\,156424$^p$ & ALS\,17405 &  -- & He-s & 3.5$\pm$0.4 & 20.0$\pm$3.0 & 3.99$\pm$0.10 & {\cite{alecian2014}}$^{t}$ This work$^{lg}$  \\
HD\,163472 & V\,2052\,Oph &  -- & -- & 3.8$\pm$0.1 & 25.2$\pm$1.1 & 4.20$\pm$0.11 & {\cite{neiner2003b}}$^{ltg}$  \\
HD\,164492C & EM*\,LkHA\,123 & A$^{m}$ & He-s & 4.1$\pm$0.3 & 26.0$\pm$2.0 & 4.25$\pm$0.25 & {\cite{2017MNRAS.465.2517W}}$^{ltg}$  \\
 & & Ba & -- & 4.1$\pm$0.3 & 24.0$\pm$2.0 & 4.00$\pm$0.40 & {\cite{2017MNRAS.465.2517W}}$^{ltg}$  \\
 & & Bb & -- & 2.7$\pm$0.3 & 15.0$\pm$2.0 & 4.00$\pm$0.40 & {\cite{2017MNRAS.465.2517W}}$^{ltg}$  \\
\hline\hline
\end{tabular}
\end{table*}
 
\begin{table*}
\contcaption{}
\label{lumteffloggtab:continued}
\begin{tabular}{l l l | l r r r l}
\hline
\hline
Star Name & Alt. Name & Comp. & CP? & $\log{(L/L\odot)}$    & \teff (kK) & $\log{g}$ & References \\
\hline
HD\,175362$^c$ & Wolff's\,Star &  -- & He-w & 2.6$\pm$0.1 & 17.6$\pm$0.4 & 4.24$\pm$0.10 & This work$^{ltg}$  \\
HD\,176582 & HR\,7185 &  -- & He-w & 2.9$\pm$0.1 & 17.0$\pm$1.0 & 4.00$\pm$0.10 & {\cite{bohl2011}}$^{g}$ This work$^{lt}$  \\
HD\,182180 & HR\,7355 &  -- & He-s & 3.1$\pm$0.2 & 19.8$\pm$1.4 & 4.25$\pm$0.05 & {\cite{rivi2013}}$^{ltg}$  \\
HD\,184927 & V\,1671\,Cyg &  -- & He-s & 3.6$\pm$0.2 & 22.0$\pm$1.0 & 3.90$\pm$0.23 & {\cite{2015MNRAS.447.1418Y}}$^{ltg}$  \\
HD\,186205$^c$ & ALS\,10427 &  -- & He-s & 3.8$\pm$0.2 & 19.6$\pm$0.8 & 3.84$\pm$0.17 & This work$^{ltg}$  \\
HD\,189775$^c$ & HR\,7651 &  -- & He-w & 2.9$\pm$0.1 & 17.5$\pm$0.6 & 4.12$\pm$0.08 & This work$^{ltg}$  \\
HD\,205021 & $\beta$\,Cep &  -- & -- & 4.3$\pm$0.1 & 25.0$\pm$1.0 & 3.80$\pm$0.15 & {\cite{2010AA...515A..74L}}$^{ltg}$  \\
HD\,208057$^c$ & 16\,Peg &  -- & -- & 3.0$\pm$0.1 & 16.5$\pm$1.2 & 4.00$\pm$0.16 & This work$^{ltg}$  \\
HD\,345439 & ALS\,10681 &  -- & He-s & 4.0$\pm$0.3 & 23.0$\pm$2.0 & 4.29$\pm$0.19 & This work$^{ltg}$  \\
ALS\,3694$^p$ & CPD\,$-48^\circ8684$ &  -- & He-s & 3.8$\pm$0.2 & 25.0$\pm$1.0 & 4.00$\pm$0.10 & This work$^{ltg}$  \\
CPD\,$-57^\circ 3509$ & -- &  -- & He-s & 3.8$\pm$0.2 & 23.6$\pm$0.2 & 4.05$\pm$0.10 & {\cite{2017A&A...597L...6C}}$^{tg}$ This work$^{l}$  \\
CPD\,$-57^\circ 3509$ & -- &  -- & He-s & 3.9$\pm$0.3 & 23.8$\pm$0.2 & 4.15$\pm$0.10 & {\cite{2016A&A...587A...7P}}$^{tg}$ This work$^{l}$  \\
\hline\hline
\end{tabular}
\end{table*}



This section collects prior determinations of atmospheric parameters from the literature, and describes the steps taken to obtain more precise constraints when the high-resolution spectroscopic data described in Paper I provide the opportunity to improve on previous measurements. 

We began with the surface parameters provided by P13, who collected spectroscopic modelling measurements from the literature determined (for hotter stars) with Non-Local Thermodynamic Equilibrium (NLTE) {\sc fastwind} or {\sc tlusty} model atmospheres \citep{2003ApJS..146..417L}, or (for cooler stars) LTE models such as {\sc ATLAS} \citep{1979ApJS...40....1K}. Where spectral modelling is already available, literature values are adopted without modification. 



In those cases for which spectral modelling was unavailable, P13 used photometry to derive \teff~and $\log{g}$, with appropriate spectral type calibrations. In several cases spectral modelling has since been performed. For the remaining stars, we present new measurements based on spectroscopic modelling below (\S~\ref{subsec:teff}-\ref{subsec:lum}). Table \ref{lumteffloggtab} collects the stellar surface parameters, together with the references when values were adopted from the literature. In the end, literature values were adopted without modification for 20 stars; entirely new values are presented for 15 stars; and for the remaining 21 stars, one parameter or more has been modified. 


\subsection{Effective temperatures}\label{subsec:teff}

   \begin{figure}
   \centering
   \includegraphics[width=8.5cm]{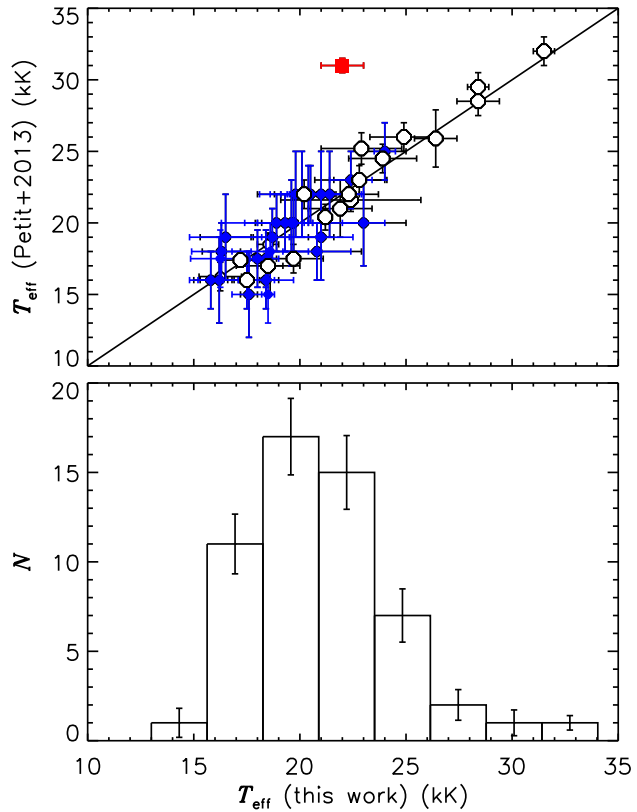}
      \caption[Comparison of \teff~measurements from EW ratios to photometric and spectroscopic modelling determinations.]{{\em Top}: comparison of measurements of \teff~obtained via EW ratios with those obtained from the literature. Filled blue circles indicate stars for which P13 determined \teff~using photometric data; open black circles indicate stars for which spectral modelling has already been performed. The filled red square indicates HD 37061/NU Ori (see text). {\em Bottom}: the distribution of adopted effective temperatures.}
         \label{teff_compare}
   \end{figure}


Photometric \teff~calibrations are often inaccurate for chemically peculiar stars, since their non-standard, variable surface chemical abundance patterns (in particular He, Si, and Fe) redistribute flux across the spectrum in a fashion that is unique to each star and thus impossible to properly account for without detailed modelling. As a sanity check on photometric \teff~determinations we used equivalent width (EW) ratios of \teff-sensitive spectral lines of different ionizations but the same atomic species \citep{1992ApJ...400..681G}, and compared these to EW ratios measured from a grid of model spectra. These measurements were performed for all stars for which high-resolution spectra are available. This is less precise than detailed comparison of observed to synthetic spectra, but it is based on the same physics, yields similar results \citep[e.g.][]{2015MNRAS.449.3945S,2017MNRAS.471.2286S}, is computationally cheaper, and is to first order independent of abundance peculiarities.

EWs were measured using mean (i.e.\ rotationally averaged) spectra created from all available ESPaDOnS, Narval, and HARPSpol observations for each star, thus maximizing the S/N and minimizing the potential effects of stellar variability due to e.g.\ chemical spots and/or pulsations. For the hotter stars (\teff$\ge 25$ kK), EWs of He {\sc i} 587.6 nm and He {\sc i} 667.8 nm vs.\ He {\sc ii} 468.6 nm, and Si {\sc iii} 455.3 nm and 456.8 nm vs.\ Si~{\sc iv} 411.6 nm were compared. For cooler stars the ratios used were Si {\sc ii} 413.1 nm, 505.6 nm, 634.7 nm, and 637.1 nm vs.\ Si {\sc iii} 455.3 nm and 456.8 nm; P {\sc ii} 604.3 nm vs.\ P {\sc iii} 422.2 nm; S {\sc ii} 564.0 nm vs S {\sc iii} 425.4 nm; and Fe {\sc ii} 516.9 nm vs.\ Fe {\sc iii} 507.4 nm and 512.7 nm. These lines were selected by searching Vienna Atomic Line Database (VALD3: \citealt{piskunov1995, ryabchikova1997, kupka1999, kupka2000,2015PhyS...90e4005R}) line lists with the criteria that the lines be both isolated and strong within the \teff~range of interest. Many of the sample stars are chemically peculiar He-weak or He-strong stars (see Table \ref{lumteffloggtab}), and may therefore possess numerous spectral lines that would not be expected in a star with standard solar abundances. The broad spectral lines of rapid rotators may also be strongly blended. These considerations required the line lists to be individually tailored for each star by excluding obvious blends. When one of the ionizations does not appear at all in the spectrum, the chemical species in question was discarded from consideration (although this did provide an additional upper or lower bound on \teff). 

We compared the EW ratios measured from the mean spectra to a grid of EW ratios determined from the non-LTE solar metallicity grid of BSTAR2006 synthetic spectra \citep{lanzhubeny2007}, i.e.\ essentially the method described by \cite{2015MNRAS.449.3945S}. The grid was limited to the range of the star's approximate $\log{g}$. The \teff~was calculated as the mean value across the grid, with the uncertainty obtained from the standard deviation of these values; since only those regions of the grid corresponding to $\log{g}$ were included, the uncertainty also includes the uncertainty in $\log{g}$. 


Special care was required for spectroscopic binaries. Where possible, we measured EW ratios from individual (rather than mean) spectra, using only those observations in which the stellar components are clearly separated. The final \teff~was determined from the mean across all such observations and all chemical ionizations examined. This was possible for HD\,136504 and HD\,149277. It was not practical for the remaining systems, but in these cases detailed spectral modelling is generally already available in the literature and these values were adopted without modification. 



Fig.\ \ref{teff_compare} compares our \teff~measurements to the values adopted by P13. Photometric \teff~values are indicated with blue solid circles, and spectroscopic \teff~measurements by black open circles. Our measurements are consistent with those from spectral modelling, suggesting they are fairly reliable. They are also consistent with those from photometry, albeit more precise. The only significant outlier is NU Ori (filled red square in Fig.\ \ref{teff_compare}). In this case it was determined that the magnetic field detection is associated with a previously undetected companion, rather than with the B0V primary as originally assumed \citep{2008MNRAS.387L..23P,2019MNRAS.482.3950S}. The magnetic star's \teff~was inferred from orbital and evolutionary models. After NU Ori, the next most significant change in \teff~is in ALS 3694, which shows no Si~{\sc ii} or Si~{\sc iv} lines in its spectrum, but does possess fairly prominent Si~{\sc iii} lines. Despite the low S/N, a weak He~{\sc ii} 468.6 nm line can also be discerned. These indicate \teff$=23\pm2$ kK, 3 kK hotter than (although formally consistent with) the photometric determination of 20$\pm$3 kK \citep{land2007}. 

As the error bars from EW ratios are smaller than measurements obtained from photometry (with a median uncertainty ratio of 60\%), and as they are additionally independent of reddening, we adopted these values in preference to the photometric measurements used by P13. 

The final \teff~distribution is shown in the bottom panel of Fig.\ \ref{teff_compare}. The histogram uncertainties were determined by a Monte Carlo process, in which $10^4$ synthetic datasets were created with the values of individual datapoints varying randomly within Gaussian distrbutions normalized to the (presumed 1$\sigma$) error bars. The error bars in each bin represent the standard deviation in bin number across all synthetic datasets. The distribution peaks at about 19 kK, and is approximately Gaussian, with an upper range of 32 kK and a lower cutoff of 15 kK. The MiMeS survey, from which the majority of the sample was drawn, focused primarily upon the hottest stars, and declines in completeness from 30\% at B0 to about 10\% at B5 \citep{2016MNRAS.456....2W}. This is the most likely explanation for the declining number of stars in the sample at the cooler end. 

\subsection{Surface gravities}\label{subsec:logg}

   \begin{figure}
   \centering
   \includegraphics[width=8.5cm]{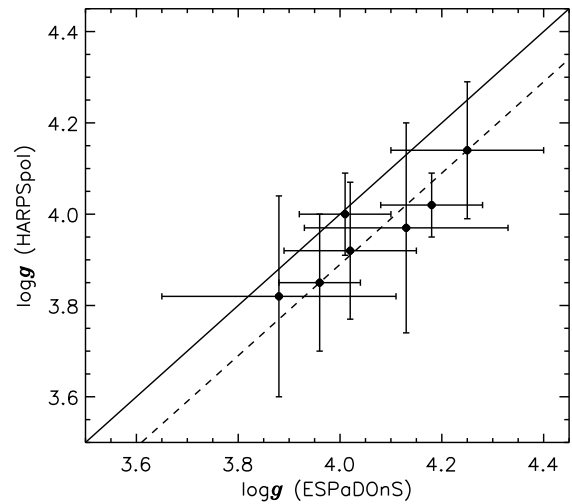}
      \caption[]{Comparison of $\log{g}$ measured with HARPSpol to values obtained from ESPaDOnS.}
         \label{esp_harps_logg}
   \end{figure}

   \begin{figure}
   \centering
   \includegraphics[trim = 0 0 0 0, width=8.5cm]{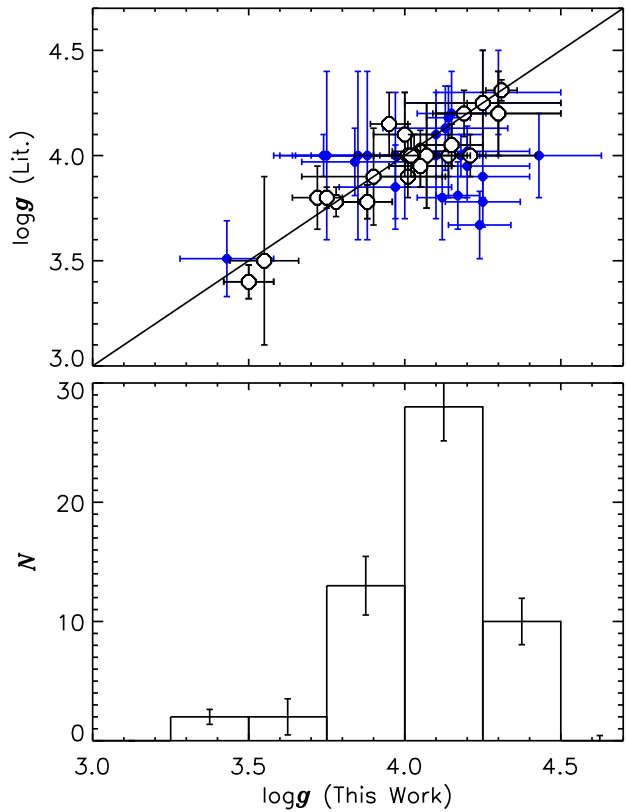}
      \caption[]{{\em Top}: Surface gravities from the literature as a function of those measured in this work. Stars for which P13 used a photometric calibration are indicated by filled blue circles; open circles represent stars for which spectroscopic measurements are available. {\em Bottom}: histogram of $\log{g}$. The median value is about 4.05.}
         \label{logg_compare}
   \end{figure}

   \begin{figure}
   \centering
   \includegraphics[width=8.5cm]{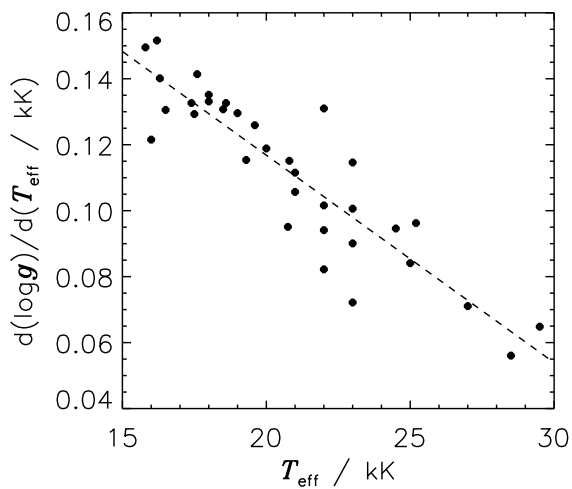}
      \caption[]{Correlation of the slope of the rate of change of $\log{g}$ with \teff~as a function of \teff. The dashed line indicates the best linear fit.}
         \label{dlogg_dteff}
   \end{figure}

Spectroscopic surface gravities are not available in the literature for many of the sample stars. While surface gravity can be determined photometrically, this is a less sensitive diagnostic than the pressure-broadened wings of H Balmer lines. 

We used H$\beta$ in the majority of cases. In comparison to higher-numbered Balmer lines, H$\beta$ has a high S/N in ESPaDOnS and Narval spectra, and is generally free of blending with strong metallic or He lines. Some stars display strong magnetospheric H Balmer line emission; this is most prominent in H$\alpha$ but also affects H$\beta$, therefore for these stars H$\gamma$ was often used instead as the next-best available line. 

Both H$\beta$ and H$\gamma$ are close to the edges of their respective spectral orders in ESPaDOnS/Narval spectra. To avoid warping of the line wings, the two overlapping orders of the unnormalized spectra were first merged, with the merging wavelength chosen as the point at which the flux uncertainties intersect, and the merged spectra were then normalized using a linear fit between continuum regions. In order to maximize the S/N, initially unnormalized spectra were co-added, with merging and normalization performed after co-addition. 

We determined $\log{g}$ with a goodness-of-fit test. We convolved synthetic BSTAR2006 {\sc tlusty} spectra \citep{lanzhubeny2007} with rotational profiles corresponding to the \vsini~values found in Paper I, and then calculated the reduced $\chi^2$ for each synthetic spectrum, with an integration range extending from 483 nm to 489 nm for H$\beta$ and 431 to 437 nm for H$\gamma$. In most cases we used a range of 3.5 to 4.5 in $\log{g}$. Balmer lines are weakly sensitive to \teff, therefore we tested fits at the minimum, mean, and maximum \teff~(using the values and uncertainties adopted in \S~\ref{subsec:teff}). The rotationally broadened cores of H Balmer lines may be subject to effects that may not have been accounted for by {\sc tlusty} \citep[e.g.\ the core-wing anomaly;][]{2002ApJ...578L..75K}, therefore we excluded the region inside $\pm$\vsini. Where relevant we also excluded the range of velocities containing the majority of the emission (as evaluated by eye from H$\alpha$, which is much more sensitive to emission). For each \teff~we then fit a low-order polynomial to the reduced $\chi^2$ as a function of $\log{g}$ in order to locate the $\chi^2$ minimum for that \teff; the final value of $\log{g}$ and its uncertainty were respectively the mean of the values at the $\chi^2$ minima for minimum, mean, and maximum \teff, and one-half of the range of these values. Due to the very high S/N, systematic uncertainty from the models dominates over the contribution from photon noise. The resulting best-fit line profiles for single stars are shown in Appendix \ref{sec:logg_single}. The special considerations involved in modelling the H$\beta$ lines of spectroscopic binary stars, and the results of those analyses, are given in Appendix \ref{sec:logg_bin}.

The spectroscopic dataset is somewhat heterogeneous. While the majority of stars were observed with ESPaDOnS and/or Narval, which are identical instruments yielding indistinguishable results, in some cases FEROS and HARPSpol measurements are also available, while in some other cases only HARPSpol measurements are available (see Paper I, Table 1 and Table \ref{newstars} in the present work). For stars with data from multiple spectrographs, $\log{g}$ was measured using mean spectra from each available instrument. ESPaDOnS, Narval, and FEROS all yield compatible results. However, $\log{g}$ measurements performed with HARPSpol are systematically about 0.1 dex lower than measurements performed using ESPaDOnS, as illustrated in Fig.\ \ref{esp_harps_logg}. This is likely due to the narrower wavelength range of the spectral orders of the HARPS spectrograph, combined with the reduction pipeline, which does not provide un-normalized spectra; since the orders are shorter than the line widths of H lines, it is very likely that these are overnormalized, leading to lower apparent surface gravities. Where ESPaDOnS, Narval, or FEROS data are available, the values found from these instruments were adopted. Where only HARPSpol data are available (HD 96446, HD 105382, HD 122451, CPD\,$-57^\circ 3509$, and CPD\,$-62^\circ 2124$) we increased $\log{g}$ by 0.1 dex.

The surface gravities measured here are compared to those adopted by P13 in the top panel of Fig.\ \ref{logg_compare}. As in Fig.\ \ref{teff_compare}, open circles denote values determined via spectral modelling, while filled blue circles indicate stars for which P13 used photometric calibrations. There is good agreement between our spectral modelling and results from the literature. The photometric measurements cluster around $\log{g}=4.0$; our spectroscopic measurements are more dispersed. 

The lower panel of Fig.\ \ref{logg_compare} shows the distribution of adopted $\log{g}$ measurements. Histogram errors were determined using the same Monte Carlo process decribed in \S~\ref{subsec:teff}. $\log{g}$ values span the main sequence, from about 3.5 to 4.5, but peak at around 4.1. Thus, while the sample in principle probes the entire main sequence, it is somewhat biased towards stars in the first half of the main sequence (between about 4.0 and 4.3).

The uncertainties in \teff~and $\log{g}$ are correlated, as a higher \teff~requires a higher $\log{g}$ to obtain an equally good fit. This introduces a tilt in the error ellipse on the \teff-$\log{g}$ diagram, thus affecting the uncertainties in stellar parameters derived from evolutionary models. To quantify this, for each star the slope ${\rm d}\log{g}/{\rm d}T_{\rm eff}$ was determined from the $\chi^2$ in the \teff-$\log{g}$ plane. These are shown in Fig.\ \ref{dlogg_dteff}, where we find the calibration

\begin{equation}\label{dlogg_dteff_calibration}
\frac{{\rm d}\log{g}}{{\rm d}T_{\rm eff}} = (0.24 - 0.006~T_{\rm eff})~{\rm kK}^{-1},
\end{equation}

\noindent for \teff~in kK. This relationship should be used when constraining the radii, masses, and ages of stars from evolutionary models on the \teff-$\log{g}$ diagram. 

\subsection{Luminosities}\label{subsec:lum}

\begin{table*}
\caption[Photometric data for luminosity determination.]{Photometric data for luminosity determination: parallax $\pi$, $V$ magnitude, Distance Modulus $DM$, extinction $A_{\rm V}$, absolute magnitude $M_V$, Bolometric Correction $BC$, and bolometric magnitude $M_{\rm bol}$. The fourth column indicates the origin of the parallax measurement: G(aia), H(ipparcos), or the mean Gaia parallax of open (cl)uster members.}
\label{phottab}
\begin{tabular}{l | r r c r r r r r}
Star & $V$   & $\pi$ & Origin  & $DM$  & $A_{\rm V}$ & $M_V$ & $BC$  & $M_{\rm bol}$ \\
Name & (mag) & (mas) &   & (mag) & (mag)       & (mag) & (mag) & (mag)         \\
\hline
HD\,3360 & 3.66 & $ 5.5\pm 0.2$  & H & $ 6.29\pm0.06$ & $<0.01$ & $-2.64\pm0.07$ & $-2.00\pm0.02$ & $-4.64\pm0.09$ \\ 
HD\,23478 & 6.69 & $3.47\pm0.06$ & G & $ 7.30\pm0.04$ & $0.6\pm0.1$ & $-1.2\pm0.2$ & $-2.0\pm0.5$ & $-3.2\pm0.6$ \\ 
HD\,25558 & 5.33 & $ 5.1\pm 0.3$ & H & $ 6.5\pm0.1$ & $ 0.06\pm0.05$ & $-1.2\pm0.2$ & $-1.4\pm0.1$ & $-2.6\pm0.3$ \\ 
HD\,35298 & 7.91 & $2.69\pm0.06$ & G & $ 7.85\pm0.05$ & $  0.01\pm0.05$ & $  0.06\pm0.07$ & $-1.3\pm0.3$ & $-1.3\pm0.4$ \\ 
HD\,35502 & 7.34 & $2.61\pm0.06$ & G & $ 7.92\pm0.05$ & $0.30\pm0.04$ & $-0.88\pm0.09$ & $-1.8\pm0.2$ & $-2.6\pm0.3$ \\ 
HD\,36485 & 6.83 & $2.33\pm0.09$ & cl & $ 8.16\pm0.15$ & $ 0.0\pm0.1$ & $-1.4\pm0.2$ & $-2.0\pm0.5$ & $-3.4\pm0.7$ \\ 
HD\,36526 & 8.29 & $2.44\pm0.08$ & G & $ 8.06\pm0.07$ & $0.1\pm0.1$ & $ 0.1\pm0.2$ & $-1.2\pm0.5$ & $-1.0\pm0.7$ \\ 
HD\,36982 & 8.46 & $2.45\pm0.06$ & G & $ 8.05\pm0.05$ & $1.0\pm0.1$ & $-0.6\pm0.2$ & $-2.2\pm0.4$ & $-2.9\pm0.6$ \\ 
HD\,37017 & 6.56 & $2.4\pm0.2$   & cl & $ 8.1\pm0.2$ & $0.2\pm0.1$ & $-1.7\pm0.3$ & $-2.1\pm0.5$ & $-3.9\pm0.7$ \\ 
HD\,37058 & 7.30 & $2.60\pm0.05$ & G & $ 7.92\pm0.04$ & $ 0.08\pm0.04$ & $-0.70\pm0.08$ & $-1.8\pm0.2$ & $-2.5\pm0.3$ \\ 
HD\,37061 & 6.83 & $2.7\pm0.3$   & cl & $ 8.59\pm0.05$ & $2.080\pm0.010$ & $-3.84\pm0.05$ & $-2.4\pm0.2$ & $ -6.2\pm0.2$ \\ 
HD\,37479 & 6.61 & $2.28\pm0.09$ & G & $ 8.21\pm0.09$ & $0.2\pm0.1$ & $-1.8\pm0.2$ & $-2.4\pm0.4$ & $-4.2\pm0.6$ \\ 
HD\,37776 & 6.96 & $2.28\pm0.06$ & G & $ 8.21\pm0.06$ & $0.20\pm0.05$ & $-1.5\pm0.1$ & $-2.2\pm0.3$ & $-3.7\pm0.4$ \\ 
HD\,43317 & 6.61 & $2.92\pm0.06$ & G & $ 7.68\pm0.04$ & $<0.06$ & $-1.06\pm0.07$ & $-1.6\pm0.2$ & $-2.6\pm0.2$ \\ 
HD\,44743 & 1.97 & $ 6.6\pm 0.2$ & H & $ 5.89\pm0.06$ & $ 0.030\pm0.010$ & $-3.96\pm0.09$ & $-2.41\pm0.03$ & $ -6.4\pm0.1$ \\ 
HD\,46328 & 4.33 & $2.4\pm0.2$   & H & $ 8.1\pm0.2$ & $0.11\pm0.05$ & $-2.93\pm0.09$ & $-2.61\pm0.09$ & $ -5.6\pm0.2$ \\ 
HD\,47777 & 7.93 & $1.4\pm0.1$   & G & $9.25\pm0.15$ & $0.19[m0.05$ & $-1.5\pm0.2$ & $-2.3\pm0.3$ & $-3.8\pm0.5$ \\
HD\,52089 & 1.50 & $ 8.1\pm 0.1$ & H & $ 5.47\pm0.04$ & $ 0.040\pm0.010$ & $-4.01\pm0.05$ & $-2.17\pm0.04$ & $ -6.18\pm0.09$ \\ 
HD\,55522 & 5.89 & $3.9\pm0.4$   & H & $ 7.0\pm0.2$ & $<0.02$ & $-1.2\pm0.2$ & $-1.6\pm0.2$ & $-2.8\pm0.5$ \\ 
HD\,58260 & 6.73 & $2.4\pm0.3$   & H & $ 8.1\pm0.3$ & $0.150\pm0.010$ & $-1.5\pm0.5$ & $-1.8\pm0.2$ & $-3.3\pm0.6$ \\
HD\,61556 & 4.43 & $ 7.2\pm 1.1$ & H & $ 5.7\pm0.3$ & $<0.05$ & $-1.3\pm0.3$ & $-1.8\pm0.3$ & $-3.1\pm0.6$ \\ 
HD\,63425 & 6.94 & $0.87\pm0.04$ & G & $10.30\pm0.10$ & $0.36\pm0.02$ & $-3.7\pm0.1$ & $-2.82\pm0.08$ & $ -6.6\pm0.2$ \\ 
HD\,64740 & 4.63 & $4.3\pm0.2$   & H & $ 6.83\pm0.07$ & $ 0.05\pm0.05$ & $-2.2\pm0.1$ & $-2.5\pm0.3$ & $-4.8\pm0.4$ \\ 
HD\,66522 & 7.19 & $2.01\pm0.03$ & G & $ 8.49\pm0.03$ & $0.8\pm0.1$ & $-2.1\pm0.1$ & $-2.1\pm0.5$ & $-4.2\pm0.6$ \\ 
HD\,66665 & 7.81 & $0.40\pm0.09$ & G & $12.0\pm0.5$ & $0.21\pm0.05$ & $-4.4\pm0.5$ & $-2.74\pm0.08$ & $ -7.1\pm0.6$ \\ 
HD\,66765 & 6.62 & $2.1\pm0.1$   & G & $ 8.4\pm0.1$ & $ 0.1\pm0.1$ & $-1.9\pm0.2$ & $-2.0\pm0.5$ & $-3.8\pm0.7$ \\ 
HD\,67621 & 6.32 & $2.81\pm0.08$ & G & $ 7.76\pm0.06$ & $ 0.05\pm0.03$ & $-1.49\pm0.09$ & $-2.1\pm0.3$ & $-3.6\pm0.4$ \\ 
HD\,96446 & 6.69 & $1.85\pm0.07$ & G & $ 8.66\pm0.08$ & $0.31\pm0.05$ & $-2.3\pm0.1$ & $-2.4\pm0.3$ & $-4.7\pm0.4$ \\ 
HD\,105382 & 4.47 & $ 7.4\pm 0.6$ & H & $ 5.6\pm0.2$ & $<0.03$ & $-1.2\pm0.2$ & $-1.7\pm0.2$ & $-2.9\pm0.4$ \\ 
HD\,121743 & 3.81 & $ 6.2\pm 0.2$ & H & $ 6.03\pm0.05$ & $  0.01\pm0.07$ & $-2.23\pm0.09$ & $-2.0\pm0.2$ & $-4.3\pm0.2$ \\ 
HD\,122451 & 0.60 & $ 8.3\pm 0.5$ & H & $ 5.4\pm0.1$ & $0.11\pm0.09$ & $-4.9\pm0.2$ & $-2.4\pm0.2$ & $ -7.3\pm0.4$ \\ 
HD\,125823 & 4.42 & $ 7.1\pm 0.2$ & H & $ 5.73\pm0.05$ & $<0.11$ & $-1.3\pm0.1$ & $-1.9\pm0.4$ & $-3.2\pm0.6$ \\ 
HD\,127381 & 4.42 & $ 5.7\pm 0.2$ & H & $ 6.23\pm0.07$ & $0.14\pm0.05$ & $-2.0\pm0.1$ & $-2.3\pm0.1$ & $-4.2\pm0.2$ \\ 
HD\,130807 & 4.31 & $ 8.1\pm 0.6$ & H & $ 5.4\pm0.1$ & $<0.06$ & $-1.1\pm0.2$ & $-1.4\pm0.3$ & $-2.6\pm0.5$ \\ 
HD\,136504 & 3.37 & $ 6.4\pm 0.7$ & H & $ 6.0\pm0.2$ & $ 0.02\pm0.03$ & $-2.6\pm0.3$ & $-1.98\pm0.06$ & $-4.6\pm0.3$ \\ 
HD\,136504 & 3.37 & $ 6.4\pm 0.7$ & H & $ 6.0\pm0.2$ & $ 0.040\pm0.010$ & $-2.6\pm0.4$ & $-1.73\pm0.07$ & $-4.4\pm0.4$ \\ 
HD\,142184 & 5.40 & $ 7.6\pm 0.4$ & H & $ 5.6\pm0.1$ & $0.28\pm0.03$ & $-0.5\pm0.1$ & $-1.8\pm0.2$ & $-2.2\pm0.4$ \\ 
HD\,142990 & 5.43 & $ 5.9\pm 0.2$ & H & $ 6.15\pm0.08$ & $0.14\pm0.03$ & $-0.9\pm0.1$ & $-1.7\pm0.2$ & $-2.6\pm0.3$ \\ 
HD\,148937 & 2.81 & $ 6.9\pm 0.5$ & H & $ 5.8\pm0.2$ & $0.19\pm0.02$ & $-3.2\pm0.2$ & $-2.97\pm0.07$ & $ -6.2\pm0.3$ \\ 
HD\,149277 & 8.41 & $1.18\pm0.06$ & G & $ 9.6\pm0.1$ & $0.6\pm0.1$ & $-1.8\pm0.2$ & $-2.0\pm0.5$ & $-3.8\pm0.7$ \\ 
HD\,156324 & 8.76 & $0.86\pm0.07$ & G & $10.3\pm0.2$ & $0.9\pm0.2$ & $-2.5\pm0.3$ & $-2.2\pm0.6$ & $-4.8\pm0.9$ \\ 
HD\,156424 & 8.90 & $3.0\pm0.9$   & cl & $ 7.6\pm0.6$ & $0.6\pm0.2$ & $-2.0\pm0.3$ & $-2.0\pm0.6$ & $-4.0\pm1.0$ \\ 
HD\,163472 & 5.83 & $2.4\pm0.4$   & H & $ 8.1\pm0.4$ & $0.90\pm0.05$ & $-3.2\pm0.4$ & $-2.5\pm0.1$ & $ -5.6\pm0.5$ \\ 
HD\,164492 & 6.80 & $0.97\pm0.07$ & cl & $10.4\pm0.7$ & $  0.01\pm0.09$ & $-4.4\pm1.4$ & $-2.7\pm0.4$ & $ -7.1\pm1.8$ \\ 
HD\,175362 & 5.38 & $ 7.6\pm 0.3$ & H & $ 5.59\pm0.07$ & $<0.03$ & $-0.22\pm0.09$ & $-1.6\pm0.2$ & $-1.9\pm0.3$ \\ 
HD\,176582 & 6.40 & $3.32\pm0.04$ & G & $ 7.39\pm0.03$ & $  0.01\pm0.06$ & $-0.99\pm0.06$ & $-1.4\pm0.3$ & $-2.3\pm0.4$ \\ 
HD\,182180 & 6.02 & $4.3\pm0.1$   & G & $ 6.84\pm0.05$ & $0.200\pm0.010$ & $-1.01\pm0.05$ & $-2.0\pm0.4$ & $-3.0\pm0.4$ \\ 
HD\,184927 & 7.44 & $1.53\pm0.05$ & G & $ 9.07\pm0.07$ & $0.18\pm0.05$ & $-1.8\pm0.1$ & $-2.2\pm0.3$ & $-4.1\pm0.4$ \\ 
HD\,186205 & 8.54 & $0.7\pm0.1$   & G & $10.8\pm0.3$ & $0.65\pm0.04$ & $-2.9\pm0.4$ & $-1.9\pm0.3$ & $-4.8\pm0.7$ \\ 
HD\,189775 & 6.11 & $3.92\pm0.06$ & G & $ 7.03\pm0.03$ & $  0.01\pm0.04$ & $-0.92\pm0.05$ & $-1.6\pm0.2$ & $-2.5\pm0.3$ \\ 
HD\,205021 & 3.23 & $4.8\pm0.3$   & H & $ 6.6\pm0.1$ & $0.10\pm0.04$ & $-3.5\pm0.2$ & $-2.4\pm0.1$ & $ -5.9\pm0.3$ \\ 
HD\,208057 & 5.08 & $ 5.2\pm 0.2$ & H & $ 6.43\pm0.10$ & $<0.07$ & $-1.4\pm0.1$ & $-1.4\pm0.2$ & $-2.8\pm0.3$ \\ 
HD\,345439 & 11.11 & $0.44\pm0.04$ & G & $11.8\pm0.2$ & $2.2\pm0.1$ & $-2.9\pm0.3$ & $-2.4\pm0.4$ & $ -5.3\pm0.7$ \\ 
ALS\,3694   & 10.35 & $0.83\pm0.05$ & G & $10.4\pm0.1$ & $1.90\pm0.09$ & $-2.0\pm0.2$ & $-2.4\pm0.4$ & $-4.3\pm0.6$ \\ 
CPD\,$-57^\circ 3509$  & 10.70 & $0.37\pm0.09$ & G & $12.2\pm0.5$ & $1.09\pm0.02$ & $-2.5\pm0.6$ & $-2.5\pm0.2$ & $ -5.0\pm0.8$ \\ 
CPD\,$-62^\circ 2124$  & 10.99 & $0.36\pm0.04$ & G & $12.2\pm0.2$ & $0.960\pm0.010$ & $-2.2\pm0.2$ & $-2.4\pm0.2$ & $-4.6\pm0.5$ \\ 
\hline\hline
\end{tabular}
\end{table*}

   \begin{figure}
   \centering
   \includegraphics[width=8.5cm]{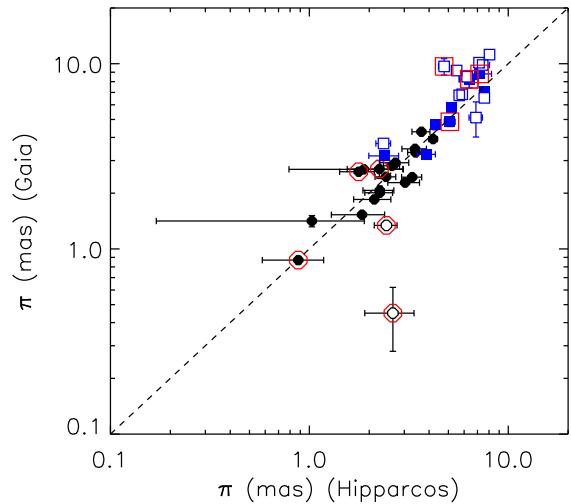}
      \caption[]{Comparison of Hipparcos and Gaia parallaxes. Circles indicate dim stars ($V > 6$), squares bright stars ($V < 6$) for which Gaia parallaxes are expected to be inaccurate. Filled symbols indicate stars for which Gaia and Hipparcos agree within $3\sigma$ of the Hipparcos parallax error. Gaia uncertainties are smaller than the symbol size in most cases. Large red circles indicate binaries. With the exception of 2 binaries, the Hipparcos and Gaia parallaxes are in agreement for all dim stars. The obvious outlier, with a Hipparcos parallax of about 3 mas and a Gaia parallax of about 0.5 mas, is HD\,37017; this star is discussed further in the text.}
         \label{hipparcos_gaia_parallax}
   \end{figure}

   \begin{figure}
   \centering
   \includegraphics[width=8.5cm]{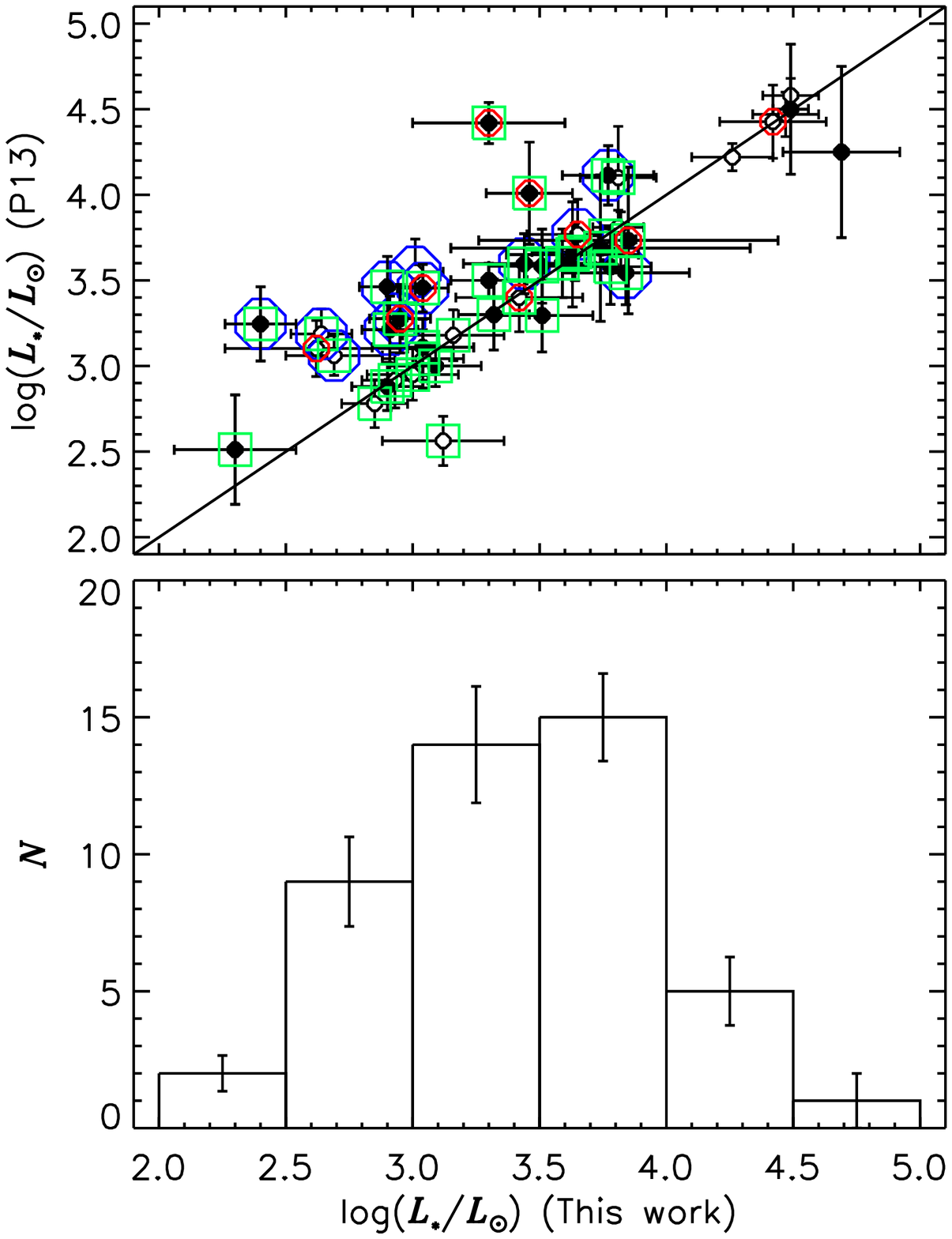}
      \caption[]{{\em Top}: Luminosities adopted by P13 as a function of those determined in this work. Filled circles indicate stars for which Gaia parallaxes were used to determine distances. Small open red circles indicate spectroscopic binaries. Open green squares indicate CP stars, for which a different BC was used than for solar metallicity stars (see text). Large open blue circles indicate stars for which P13 utilized {\sc chorizos} to determine luminosities. {\em Bottom}: Distribution of adopted luminosities.}
         \label{lum_compare}
   \end{figure}

The photometric parameters used to determine $\log{L}$ are provided in Table \ref{phottab}. Visual magnitudes $V$ were obtained from {\sc simbad}. Distance moduli $DM$ were obtained from Hipparcos \citep{perry1997,vanleeuwen2007} or Gaia DR2 \citep[][]{2018A&A...616A...1G}\footnote{Obtained from \url{http://gea.esac.esa.int/archive/}.} parallaxes $\pi$ as $DM = 5\log{(1/\pi)} - 5$, for $\pi$ in arcseconds. 

A comparison of Hipparcos and Gaia parallaxes for our sample is shown in Fig.\ \ref{hipparcos_gaia_parallax}. Most stars agree within 3$\sigma$ of the Hipparcos error bar. Outliers are divided into two classes. The first are binary systems. DR2 treated all sources as single stars; since orbital motion was not accounted for in the astrometric solution, the parallaxes of binary systems may be unreliable. The second class of outliers are relatively bright stars ($V < 6$); since Gaia is optimized for dimmer targets, the parallaxes of bright stars may be unreliable \citep{2018A&A...616A...2L}. Indeed, a bias towards higher Gaia parallaxes is seen in stars with the largest parallaxes, which also tend to be the brightest. When Hipparcos parallaxes were available, we adopted Gaia parallaxes only for systems with $V > 6$ and at least 3$\sigma$ agreement between Gaia and Hipparcos results. In the end, we used DR2 parallaxes for 27 stars, or about half the sample, and retained the Hipparcos parallaxes for 24 stars. 

The most obvious outlier in Fig.\ \ref{hipparcos_gaia_parallax} is HD\,37017, which has a Gaia parallax ($0.5 \pm 0.2$~mas) much smaller than its Hipparcos parallax ($2.6 \pm 0.7$~mas). HD\,37017 is an eccentric spectroscopic binary consisting of two B-type stars, with an orbital period of about 18~d \citep{1998AA...337..183B}. At its Hipparcos distance, its orbital properties predict the components to be separated by about 0.5 mas, which is very close to the Gaia parallax and suggests that this value is affected by the system's orbital motion. HD\,37017 is listed as a possible (although not probable) member of the Ori OB1c cluster by \cite{land2007}; taking the average of the Gaia parallaxes of the stars listed by \citeauthor{land2007} as probable members gives $\pi = 2.4 \pm 0.2$ mas, which is compatible with the Hipparcos parallax but more precise. We adopted the cluster value. 

In 4 additional cases Hipparcos parallaxes are unavailable, or Gaia parallaxes were judged unreliable. These are HD 36485 \citep[an SB2,][]{leone2010}, HD 37061 \citep[an SB3,][]{2019MNRAS.482.3950S}, HD 156424 (no Gaia parallax), and HD 164492C (no Gaia parallax). All of these stars are in clusters. For HD 36485, a member of the Ori OB1b association \citep{land2007}, we utilized the mean distance inferred from the DR2 parallaxes of other association members in the sample \citep[HD 36526, HD 37776, HD 37479;][]{land2007}. The distance to HD 37061 was determined via the distance to nearby Orion Nebula Cluster stars by \cite{2019MNRAS.482.3950S}. HD 156424 is a member of the Sco OB4 association, as is HD 156324 \citep{2005AA...438.1163K}; therefore we adopted the same distance as for HD 156324. The luminosity of HD 164492C was determined by \cite{2017MNRAS.465.2517W} using the distance of cluster members and the orbital properties of the system; \cite{2017MNRAS.465.2517W} found that the orbital properties implied a distance of about 1 kpc, which is confirmed with the DR2 distance to the high-probability cluster member HD 164637 \citep{2000A&AS..146..251B}, $1200^{+400}_{-240}$~pc. 

Extinctions $A_{\rm V}$ were calculated assuming $R_{\rm V} = 3.1$. The intrinsic colours $(B-V)_0$ were determined using BSTAR2006 synthetic spectra \citep{lanzhubeny2007}, with uncertainties derived from the minimum and maximum values obtained within the \teff~and $\log{g}$ error bars. Absolute visual magnitudes $M_{\rm V}$ were then determined using distance moduli $DM$ as determined above. We then calculated bolometric magnitudes and luminosities $M_{\rm bol}$ and $\log{(L_*/L_\odot)}$ by applying Bolometric Corrections $BC$ and assuming $M_{\rm bol,\odot}=4.74$.


For solar metallicity stars, $BC$ was obtained in the same way as by P13, i.e.\ via linear interpolation between the theoretical {\sc tlusty} BSTAR2006 grid \citep{lanzhubeny2007}, but using the values of \teff~and $\log{g}$ found above. For chemically peculiar stars (the majority of the sample, see Table \ref{lumteffloggtab}), the empirical $BC$ developed by \cite{2008AA...491..545N} was used. The extremely high abundances of Fe, Si, and other elements such as Pr, Nd, or Eu lead to flux redistribution from the UV to the optical, necessitating a different $BC$ from that for solar metallicity stars. For chemically normal stars, the $BC$ uncertainty was determined from the uncertainties in \teff~and $\log{g}$. For the CP stars, the $BC$ uncertainty was determined from the uncertainty in \teff, with an additional $\pm0.15$ mag for stars below 19 kK and $\pm0.2$ mag for stars above 19~kK, where the extra uncertainties reflect 1) the intrinsic uncertainty in the empirical $BC$ correction relationship and, 2) that the relationship is only calibrated up to about 19 kK, and is therefore an extrapolation of uncertain reliability at higher \teff. The consequences of utilizing the \cite{2008AA...491..545N} $BC$ are to increase the $BC$ by about $-0.1$ mag and to increase the uncertainty by a factor of about 2. 

Fig.\ \ref{lum_compare} compares the luminosities used by P13 to those adopted here. P13 utilized luminosities derived from three methods: spectral modelling, photometrically via bolometric corrections ($BC$s), and spectrophotometrically via SED fitting using {\sc chorizos}. Only one star, HD 61556, has a significantly higher luminosity than the value assumed by P13; this is due to the higher \teff~determined by \cite{2015MNRAS.449.3945S} via spectroscopic analysis of this star. Generally, the luminosities determined here are systematically lower than those utilized by P13. This is principally for two reasons. The first is that distance can be set as a free parameter in {\sc chorizos}, which seems to have resulted in distance moduli systematically greater (by up to 1.5 dex) than would be inferred from either Hipparcos or Gaia parallaxes. 


The second cause of our systematically lower luminosities is multiplicity. Spectroscopic binaries are highlighted in Fig.\ \ref{lum_compare} with doubled red circles. To determine the luminosities of individual stellar components, we started with the total system luminosity $\log{L_{\rm sys}} = \log{(L_1 + L_2)}$, determined in the usual way from $V$, $DM$, and $BC$. The $BC$ is a function of \teff, thus the same $BC$ shouldn't really be used for both stars. However, the luminosity of the primary is only significantly different from the combined luminosity when all components are close enough in luminosity, mass, and \teff~to contribute similar amounts to the system brightness. We therefore feel justified in using a single $BC$ for all components. Close binaries are believed to be primordial \citep{1994MNRAS.271..999B}, so we assumed the components to be coeval. As a result we could constrain the luminosities of the individual stars using isochrones \citep{ekstrom2012}. We determined $L_1$ and $L_2$ along each isochrone using the mass ratio $M_1/M_2$ (when the orbital parameters are known), or (when they are not) from the luminosity ratio obtained either spectroscopically (i.e.\ EW ratios) or interferometrically. Values for which $\log{L_{\rm sys}}$ fell outside the range determined from photometry were discarded. The components' luminosities were then constrained from the remaining values. This analysis was described in more detail by \cite{2018MNRAS.475..839S} for the case of HD 156324, and by \cite{2019MNRAS.482.3950S} for HD 37061. 

Orbital mass ratios were used for HD 36485 \citep[$M_1/M_2 = 2.6$;][]{leone2010}, HD 37017 \citep[$M_1/M_2=2$;][]{1998AA...337..183B}, and HD 149277 \citep[$M_1/M_2 = 1.1$;][]{2016PhDT.......390S}. Interferometric luminosity ratios are available for HD 130807 \citep{2013MNRAS.436.1694R}, HD 136504 \citep[][]{pablo_epslup} and for HD 122451 \citep{2005MNRAS.356.1362D,2006AA...455..259A,2016A&A...588A..55P}. For HD 25558, HD 35502, and HD 164492C, the luminosities were determined via luminosity ratios and spectrophotometric fitting by \cite{2014MNRAS.438.3535S}, \cite{2016MNRAS.460.1811S}, and \cite{2017MNRAS.465.2517W}, respectively. 

As can be seen in Fig.\ \ref{lum_compare}, the net effect of discarding most {\sc chorizos} luminosities, and correcting for multiplicity, is to reduce the average luminosity. Despite the more precise distances available with Gaia parallaxes, our uncertainties are comparable to those published by P13. This is mainly because the reduced uncertainty in $DM$ is offset by the larger uncertainty in $BC$ due to the utilization of the relationship appropriate for CP stars. 

The lower panel of Fig.\ \ref{lum_compare} shows the distribution of adopted $\log{L}$ values; histogram errors were determined using the Monte Carlo process described in \S~\ref{subsec:teff}. This mirrors the distribution of \teff, in that it is approximately log-normal, peaking around $\log{L} \sim 3.75$. Since lower-luminosity stars are intrinsically more common than stars with a higher luminosity, distribution is almost certainly because the MiMeS survey completeness declines from B0 to B5, having been biased towards more luminous stars. 


\section{Discussion \& Summary}\label{sec:disc}


   \begin{figure}
   \centering
   \includegraphics[width=8.5cm]{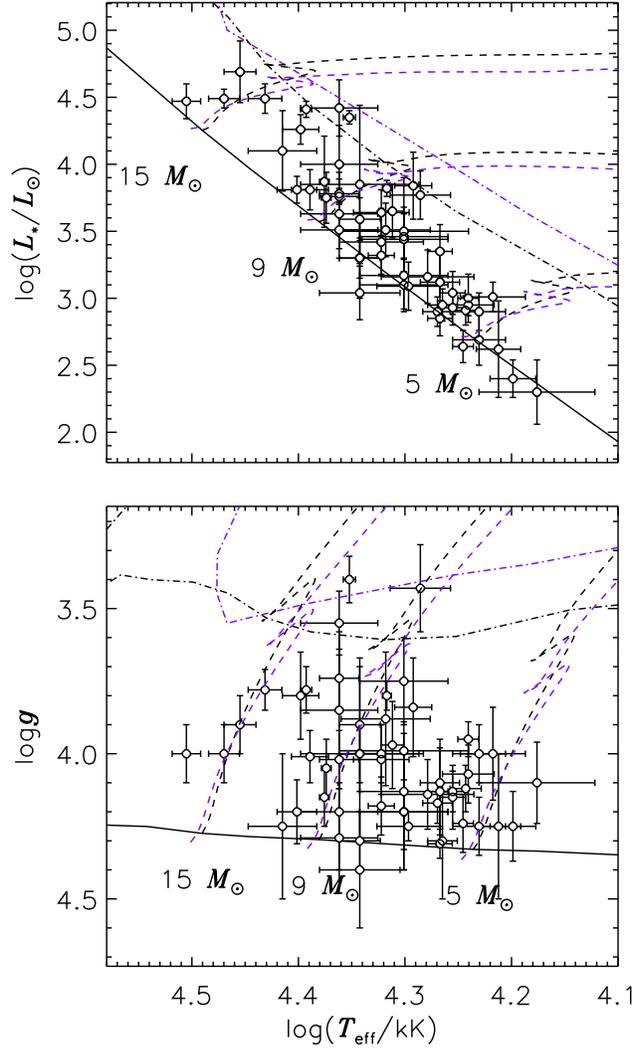}
      \caption[]{HRD (top) and \teff-$\log{g}$ diagram (bottom) for the sample stars. Dashed lines indicate the labelled evolutionary tracks, computed using the Geneva evolutionary models calculated by \protect\cite{ekstrom2012} . The solid and dot-dashed lines indicate the ZAMS and the TAMS. Black lines indicate rotating models; purple lines indicate non-rotating models.}
         \label{hrd_shrd_bstars}
   \end{figure}

Fig.\ \ref{hrd_shrd_bstars} shows the sample stars on the Hertzspung-Russell diagram (HRD) and the \teff-$\log{g}$ diagram, and compares their positions to the rotating ($v_0 / v_{\rm crit} = 0.4$) and non-rotating Geneva evolutionary models calculated by \cite{ekstrom2012}. \teff, $\log{L}$, and $\log{g}$ are approximately consistent, insofar as that stars occupy similar positions on the main sequence in either diagram. The majority of the sample stars have masses between 4 and 15 $M_\odot$, and both their luminosities and surface gravities are consistent with evolutionary statuses between the Zero Age and Terminal Age Main Sequence (ZAMS and TAMS). One star, HD 52089, lies above the rotating TAMS in both diagrams; however, it lies at or below the non-rotating TAMS. While the star's rotation period is not known (Paper I), it has sharp spectral lines and is likely  a slow rotator for which non-rotating models are appropriate. 


   \begin{figure}
   \centering
	\begin{tabular}{cc}
   \includegraphics[trim=75 25 25 75, width=4.cm]{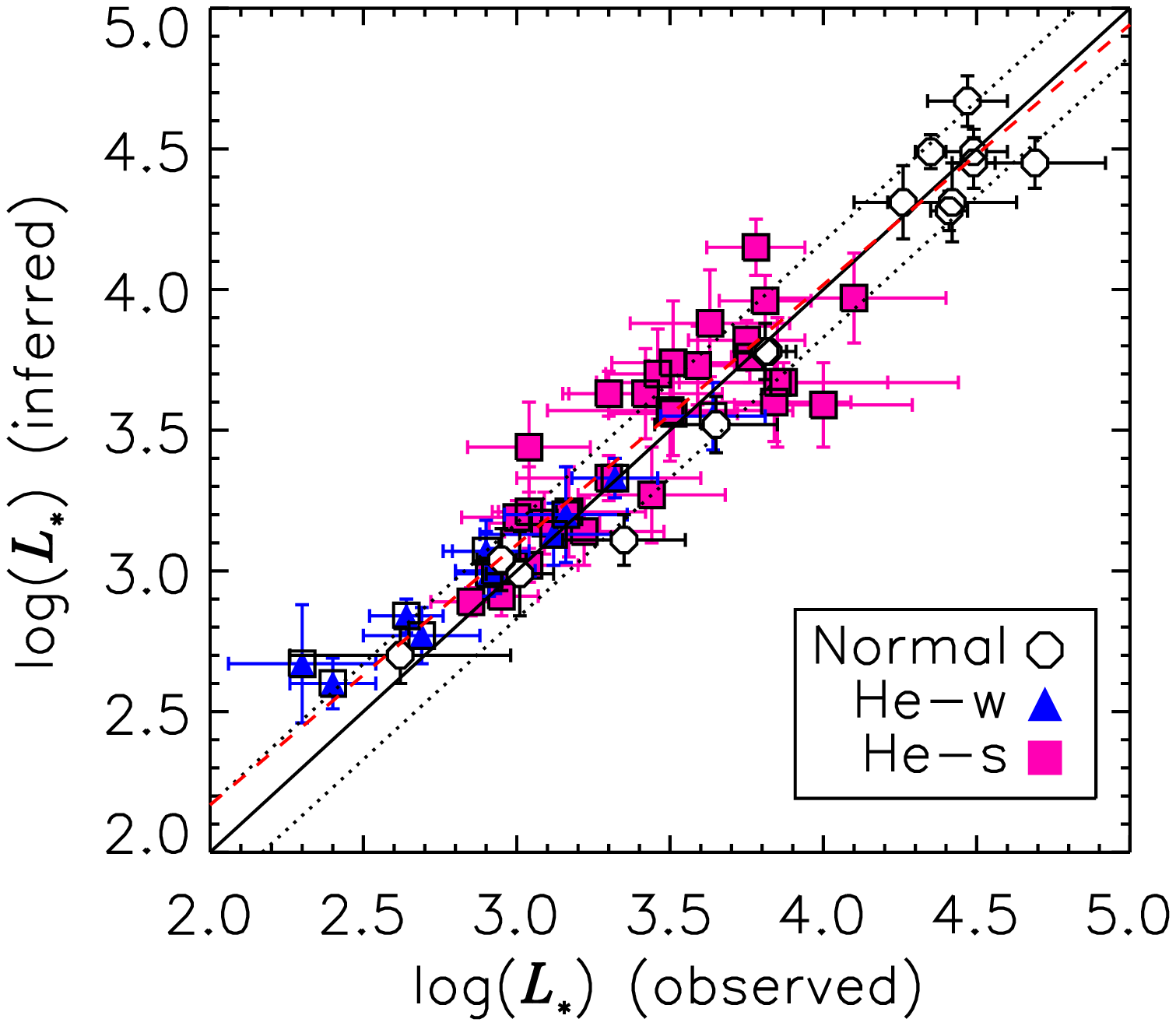} &
   \includegraphics[trim=75 25 25 75, width=4.cm]{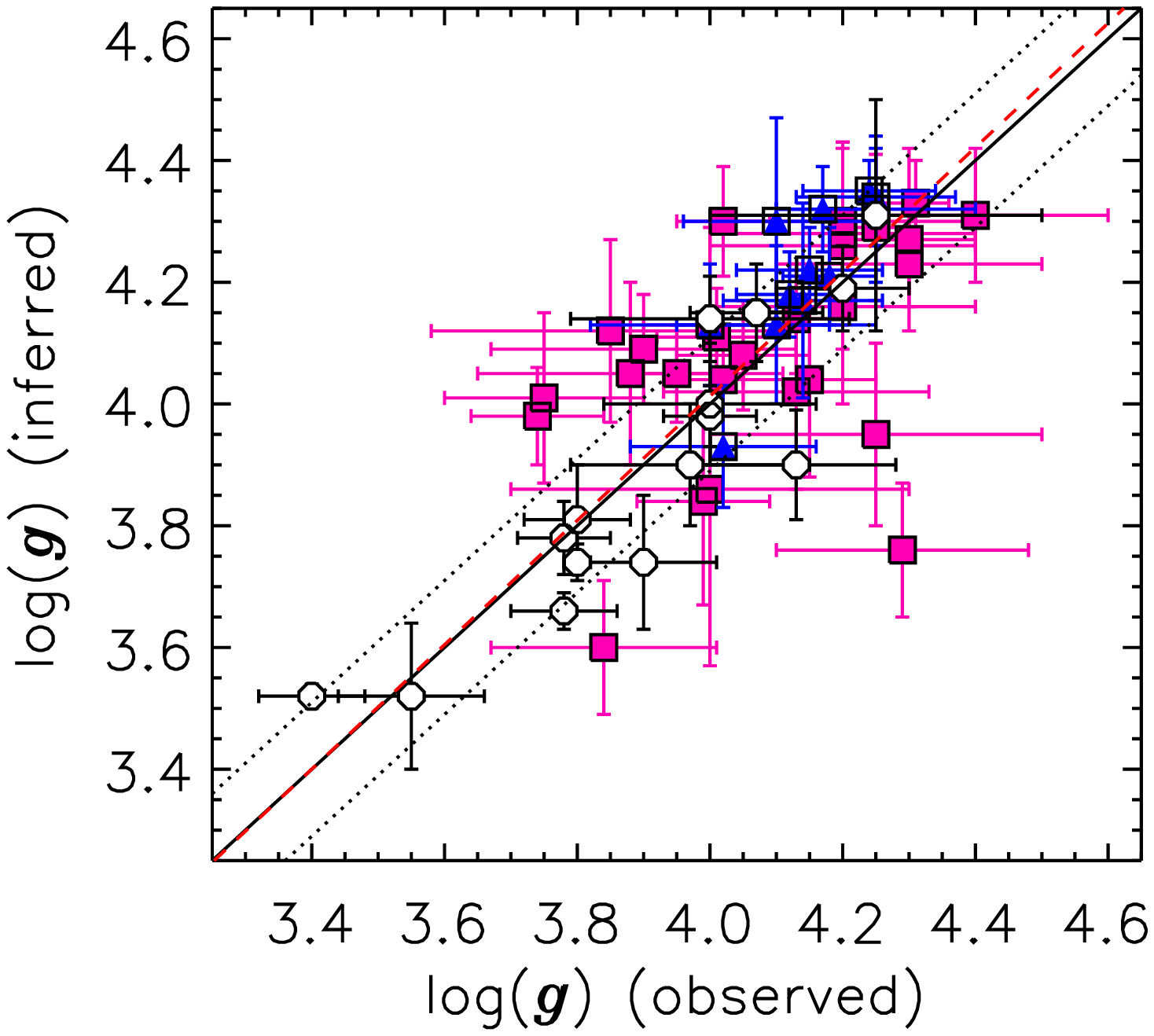} \\
	\end{tabular}
      \caption[]{{\em Left}: Luminosities inferred from $\log{g}$ and \teff~as a function of the measured luminosities. {\em Right}: surface gravities inferred from $\log{L}$ and \teff~as a function of the measured surface gravities. Solid lines indicate $x=y$; dotted lines, the mean measurement errors; red dashed lines show the regressions.}
         \label{in_out}
   \end{figure}

As a check on the consistency of the measured parameters, we used the evolutionary models in Fig.\ \ref{hrd_shrd_bstars} to infer $\log{L}$ from $\log{g}$ and \teff~(Fig.\ \ref{in_out}, left), and to infer $\log{g}$ from $\log{L}$ and \teff~(Fig.\ \ref{in_out}, right). In both cases, linear regression of the measured vs.\ inferred quantities yields a relationship that is compatible with the $x=y$ line within the measured uncertainties. Many of the stars are chemically peculiar, with significant He under- or over-abundances. Since these might affect the surface gravity in systematic ways by reducing or increasing the partial pressure of H, each star's CP type (He-weak, He-strong, or normal i.e.\ no chemical peculiarity) is indicated in Fig.\ \ref{in_out}. Chemically normal and He-w stars both exhibit very good agreement between measured and inferred values. The He-strong sub-sample shows a larger variance, although this is not statistically significant when compared to the typical uncertainties in this sub-sample.

As can be seen in Fig.\ \ref{hrd_shrd_bstars}, there is an apparent deficit of stars with masses less than about 7 \msun~in the second half of the main sequence. This is in contrast to more massive stars, which populate the entirety of the main sequence. The most likely explanation for this is that the sample is incomplete in this mass range. The \teff~and $\log{L}$ distributions in Figs.\ \ref{teff_compare} and \ref{lum_compare} peak at \teff~$\sim 18$~kK and $\log{L} \sim 3.75$. Since the real distributions undoubtedly increase towards cooler lower temperatures and luminosities, the sample cannot be complete below these thresholds. As previously noted the completeness of the MiMeS survey, from which the majority of the sample was drawn, declines monotonically from about 30\% at B0 to about 10\% at B5 \citep[Fig.\ 6, ][]{2016MNRAS.456....2W}. It is also worth pointing out that the Ap stars are evenly distributed across the main sequence \citep[e.g.][]{2006AA...450..763K}. Since this is true of Ap stars and the hotter magnetic B stars, it is highly unlikely that the absence of mid-range B-type stars with low surface gravities reflects an actual deficit of magnetic stars in this mass range and evolutionary stage. Future observations should attempt to address the absence of high-resolution spectropolarimetric data for magnetic stars below about 7 \msun~in the second half of the main sequence, since until such stars are studied in detail the main sequence evolution of stars in this mass range cannot be investigated.


   \begin{figure}
   \centering
   \includegraphics[width=8.5cm]{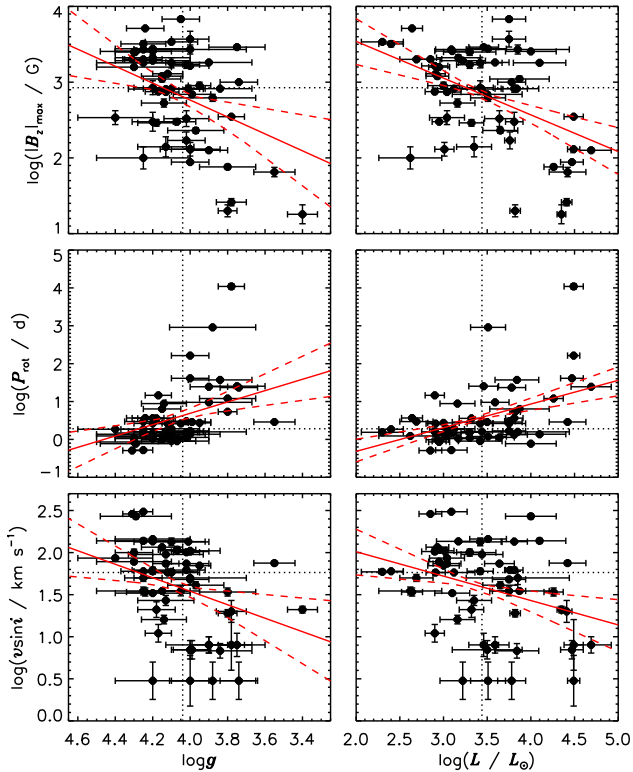}
      \caption[]{{\em Left panels}: from top to bottom, $\log{(|B_{z}|_{\rm max})}$, $\log{P_{\rm rot}}$, and \vsini~as functions of $\log{g}$. Dotted lines show the median value of the x- and y-axes. Solid and dashed red lines show linear regressions and their $1\sigma$ uncertainties. {\em Right panels}: as left, for $\log{L}$. }
         \label{logg_lum_bz_rot}
   \end{figure}

While deriving masses, ages, and oblique rotator models is beyond the scope of this work, the evolution of rotational and magnetic parameters is amenable to a qualitative investigation by means of the surface gravity (which is a proxy for age) and luminosity (which is a proxy for both mass and mass-loss rate) derived here, and the empirical magnetic and rotational properties of the sample presented in Paper I. 

The top panels of Fig.\ \ref{logg_lum_bz_rot} show the maximum observed longitudinal magnetic field $\log{(|B_{z}|_{\rm max})}$ (from Table 2 in Paper I and Table \ref{newstars} in the present work) as a function of $\log{g}$ and $\log{L}$. Linear regressions indicate decreasing magnetic field strength with decreasing surface gravity and increasing luminosity, with respective Pearson's Correlation Coefficients \citep[PCCs;][]{1895RSPS...58..240P} of $0.43 \pm 0.07$ and $-0.46 \pm 0.05$. The former could be consistent with a declining surface magnetic field strength with advancing age due to flux conservation in an expanding stellar atmosphere, while the latter would - surprisingly - suggest a decline in the typical surface magnetic field with increasing mass. 


Rotation is investigated in the middle panels of Fig.\ \ref{logg_lum_bz_rot} ($P_{\rm rot}$), and in the bottom panels (\vsini). There is a clear increase in $P_{\rm rot}$ with decreasing $\log{g}$ (${\rm PCC} = -0.43 \pm 0.08$), along with a decline in \vsini~(${\rm PCC} = 0.36 \pm 0.08$). This is expected given the rapid magnetic braking these stars are predicted to experience. Since magnetic braking should happen more rapidly with higher mass-loss rates (viz., higher luminosities), more luminous stars should be slower rotators \citep[e.g.][]{ud2009,petit2013}. This is apparent for $P_{\rm rot}$, which exhibits an increase with increasing $\log{L}$ (${\rm PCC} = 0.47 \pm 0.04$) and for \vsini~(${\rm PCC} = -0.33 \pm 0.05$). 

Removing the most extreme values from the sample (e.g.\ HD\,46328, with $P_{\rm rot}~\sim 30~{\rm yr}$) has essentially no impact on the relationship between $\log{g}$ and rotation or $\log{(|B_{z}|_{\rm max})}$, nor does it affect the relationship between $\log{L}$ and $\log{(|B_{z}|_{\rm max})}$. This does, however, decrease the PCC between $\log{L}$ and both $P_{\rm rot}$ and \vsini, by about 0.1 in both cases.


It should be emphasized that the analyses of \teff~and $\log{g}$ presented here are not intended as substitutes for detailed spectral modelling. Equivalent width ratios utilizing Si~{\sc ii}, {\sc iii}, and {\sc iv}, and He~{\sc i} and {\sc ii} ionization balances are sensitive effective temperature diagnostics. However, if the abundances of these elements are not only horizontally but also {\em vertically} inhomogeneous, this vertical abundance stratification can lead to systematic errors. Such an effect has been reported for late-type Bp stars \citep{2013AA...551A..30B}. We also implicitly assumed that the only factor affecting the Stark-broadened wings of H lines is atmospheric pressure, which may not be the case. Effects such as He or metallic over- or under-abundances that are not accounted for in solar metallicity models can have a profound impact on H Balmer line wings \citep{1997AA...320..257L}. Another possible contributing factor is magnetic pressure \citep[e.g.][]{shulyak2007,shulyak2010}. At least three stars, HD\,61556, HD\,125823 and HD\,184927, all of which are He-variables, show Balmer line wing variations that are clearly correlated with He variations \citep{2015MNRAS.449.3945S,2015MNRAS.447.1418Y}. Furthermore, some stars display magnetospheric Balmer line emission; while this is much weaker in H$\beta$ and H$\gamma$ than in H$\alpha$, and we have attempted to account for this by ignoring the wavelength regions most affected, its presence may in some cases lead to lower apparent surface gravities. Ultimately surface gravities should be derived together with mean surface abundances at a minimum, and ideally with Doppler imaging in order to include the effects of surface abundance inhomogenities; in the case of stars with detectable magnetospheres $\log{g}$ should also be derived together with a model accounting for emission. 

A key limiting factor in the luminosity uncertainties is the bolometric correction, which is not yet calibrated for CP stars above 19 kK \citep{2008AA...491..545N}. It would be helpful if an improved $BC$, appropriate to He-strong stars, were developed. The availability of precise Gaia parallaxes for a large sample of CP stars should additionally help to calibrate a $BC$ to a higher precision.

The trends in rotation and magnetic field strength explored in Fig.\ \ref{logg_lum_bz_rot} demonstrate that rotation almost certainly decreases with age. However, surface gravity and luminosity are only proxies to age, mass, and mass-loss rate. Paper III will utilize the surface parameters determined here, in conjunction with the magnetic and rotational measurements presented in Paper I, to derive model parameters with which to conduct a more precise investigation of the rotational, magnetic, and magnetospheric properties of the magnetic early B-type stars. 

\section*{Acknowledgements}

Based on observations made with ESO Telescopes at the La Silla Paranal Observatory under programme IDs 187.D-0917(C), 092.A-9018(A), 095.D-0269(A), and 095.A-9007(A). This work has made use of the VALD database, operated at Uppsala University, the Institute of Astronomy RAS in Moscow, and the University of Vienna. This work has made use of data from the European Space Agency (ESA) mission Gaia (\url{https://www.cosmos.esa.int/gaia}), processed by the Gaia Data Processing and Analysis Consortium (DPAC, \url{https://www.cosmos.esa.int/web/gaia/dpac/consortium}). Funding for the DPAC has been provided by national institutions, in particular the institutions participating in the Gaia Multilateral Agreement. This research has made use of the WEBDA database, operated at the Department of Theoretical Physics and Astrophysics of the Masaryk University. MS acknowledges the financial support provided by the European Southern Observatory studentship program in Santiago, Chile; the Natural Sciences and Engineering Research Council (NSERC) Postdoctoral Fellowship program; and the Annie Jump Cannon Fellowship, supported by the University of Delaware and endowed by the Mount Cuba Astronomical Observatory. GAW acknowledges support from an NSERC Discovery Grant. VP acknowledges support from the National Science Foundation under Grant No.\ 1747658. JW acknowledges support from NSF AST-1412110. The MiMeS collaboration acknowledges financial support from the Programme National de Physique Stellaire (PNPS) of INSU/CNRS. We acknowledge the Canadian Astronomy Data Centre (CADC). MES thanks Oleg Kochukhov for advice on the bolometric corrections of CP stars. 

\bibliography{bib_dat.bib}{}


\appendix
\renewcommand{\thetable}{\Alph{section}\arabic{table}}

\section{Magnetic analysis of HD\,47777}\label{hd47777}

   \begin{figure}
   \centering
   \includegraphics[width=8.5cm, trim = 25 25 0 25]{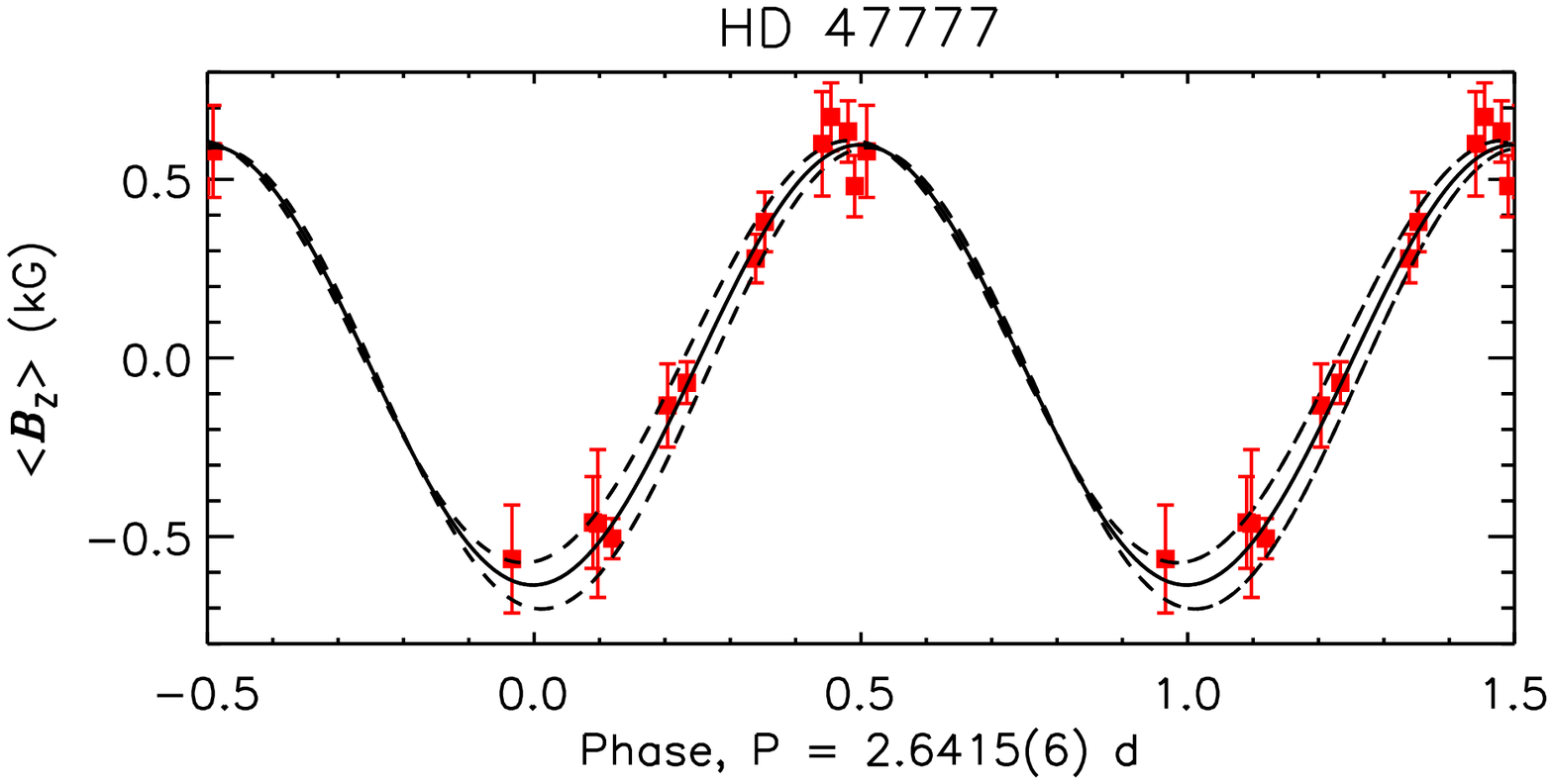} 
      \caption[]{LSD profile \bz~measurements obtained from all He and metallic lines of HD 47777 phased with the rotation period determined here. The solid and dashed curves show the least-squares harmonic fit and the 1$\sigma$ uncertainties.}
         \label{hd47777_bz}
   \end{figure}

HD\,47777 was observed 14 times between 2007 and 2013, with 7 MiMeS observations and 7 observations obtained by 4 CFHT P.I. programs\footnote{Program codes 07BF16, 07BC10B, 12AC03, and 13BC09.} One observation was already reported by \cite{2014A&A...562A.143F}. Sub-exposure times range from 240 s to 1000 s, and the mean peak signal-to-noise (S/N) per spectral pixel of 500 after discarding one observation with a peak S/N of 52. Least-squares deconvolution (LSD) profiles were extracted using a 22 kK Vienna Atomic Line Database \citep[VALD3;][]{piskunov1995,ryabchikova1997,kupka1999,kupka2000,2015PhyS...90e4005R} line mask obtained with an `extract stellar' request, and cleaned and tweaked as described in Paper I. Using the LSD profiles obtained with He and all metallic lines, which have the highest LSD S/N, 5 observations yielded formal definite detections, and 3 gave marginal detections, according to the usual criteria given by \cite{1992AA...265..669D,d1997}.

LSD profiles were also extracted using single-element line masks for He, C, N, O, Al, Si, Mg, and Fe. \bz~was measured from all LSD profiles in the usual way, and H line \bz~measurents were obtained from H$\alpha$, H$\beta$, and H$\gamma$ as described in Paper I. There is some evidence for a systematic discrepancy in results from different lines, with $A_{\rm e} = 0.26 \pm 0.14$ (defined in Paper I), however this is below the 2$\sigma$ significance level and therefore we selected the LSD profiles extracted with all metallic and He lines for modelling (Fig.\ \ref{hd47777_bz}). 

\cite{2014A&A...562A.143F} determined the rotation period to be $2.641(3)$ d using MOST space photometry. This did not quite provide an adequate phasing of the data. Conducting a period search with the \bz~measurements using the same methods described in Paper I yielded $P_{\rm rot} = 2.6415(6)$ d, which is the period used to phase the data in Fig.\ \ref{hd47777_bz}. JD0$=2454461.8(2)$ was determined in the same fashion as in Paper I. As can be seen from the harmonic fit, \bz~is consistent with a first-order sinusoid, indicating a predominantly dipolar surface magnetic field.

\section{Surface gravity measurements of single stars}\label{sec:logg_single}

\begin{figure*}
   \centering
\begin{tabular}{cc}
   \includegraphics[trim = 60 0 60 25, width=8.5cm]{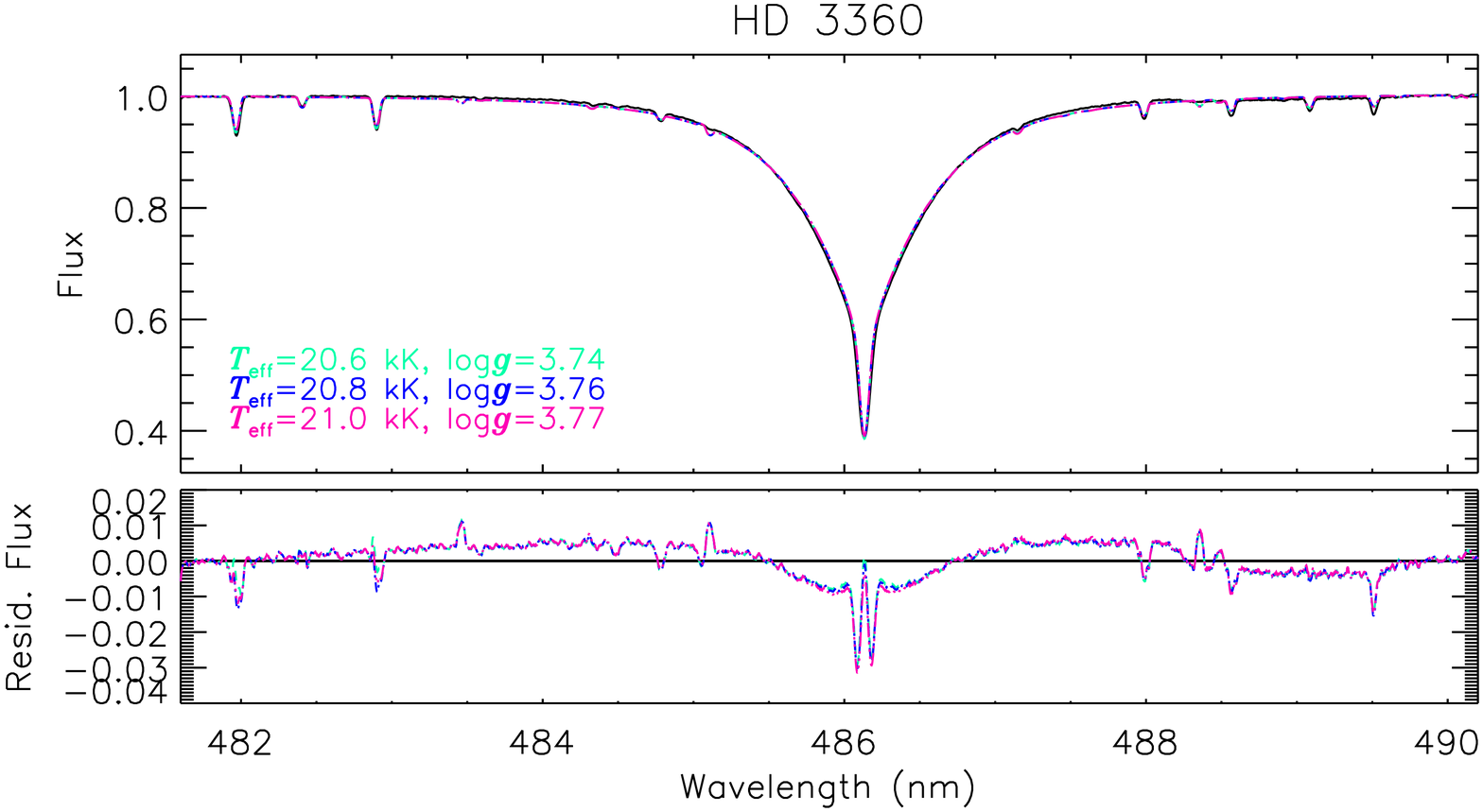} &
   \includegraphics[trim = 60 0 60 25, width=8.5cm]{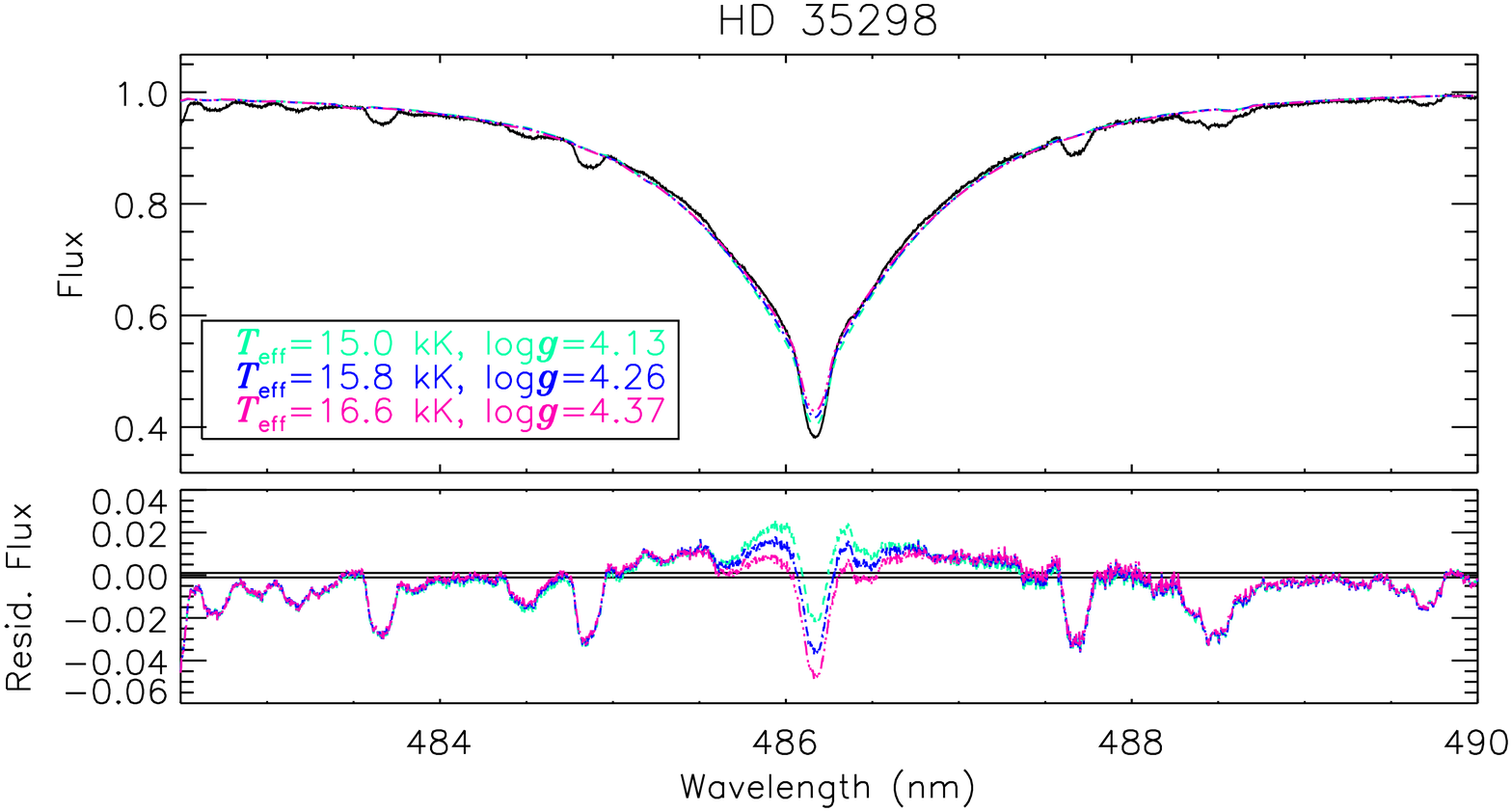} \\ 
   \includegraphics[trim = 60 0 60 25, width=8.5cm]{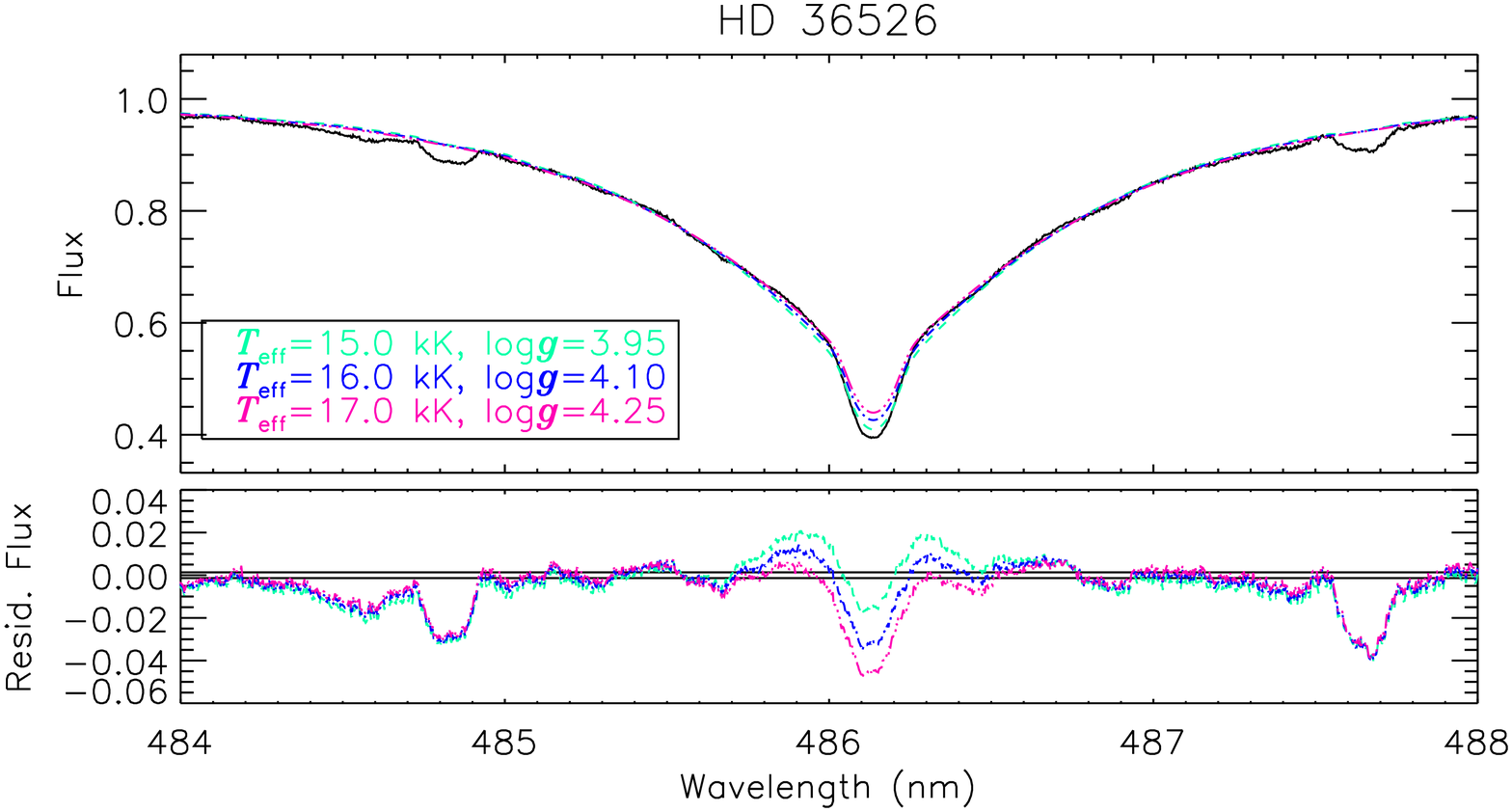} & 
   \includegraphics[trim = 60 0 60 25, width=8.5cm]{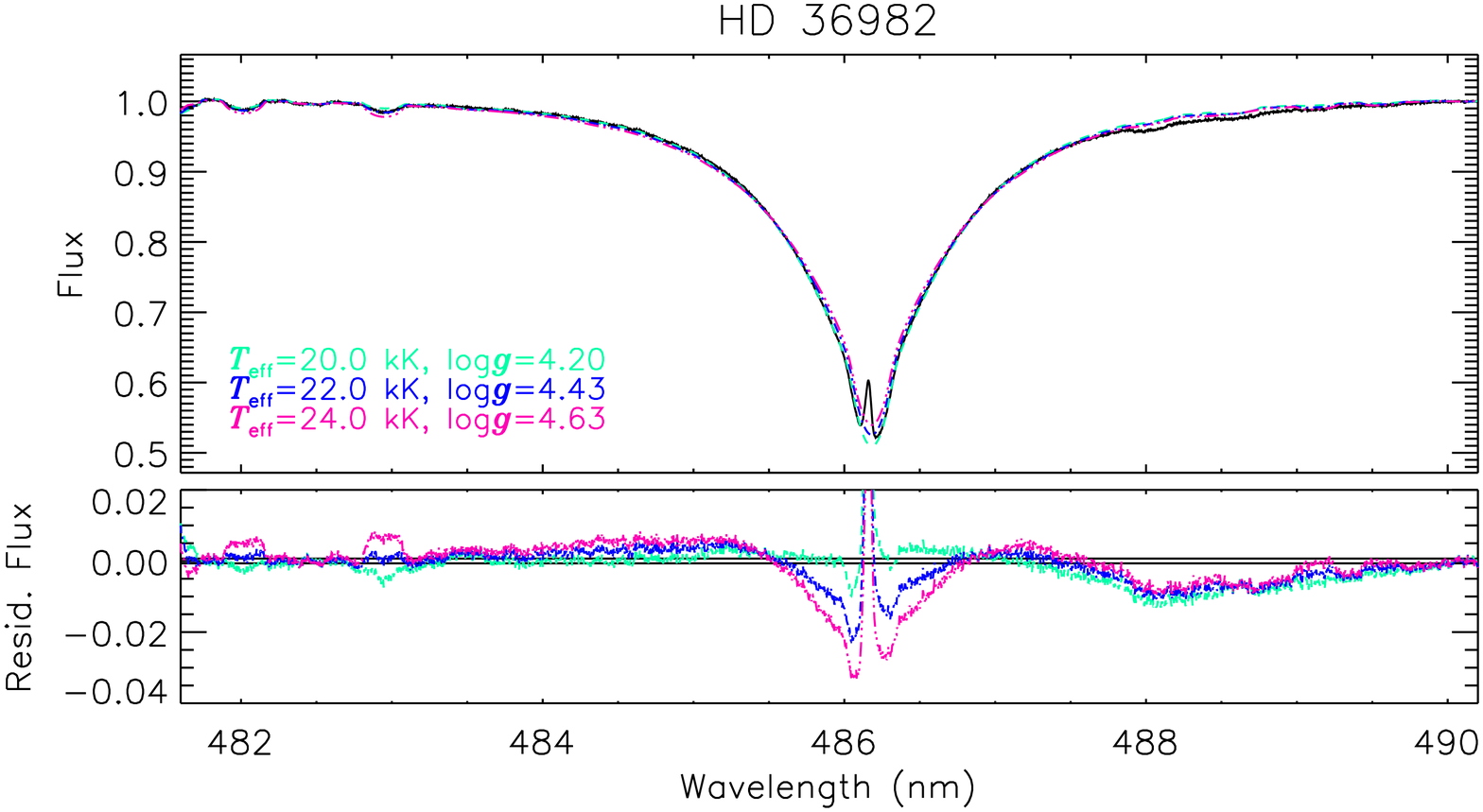} \\ 
   \includegraphics[trim = 60 0 60 25, width=8.5cm]{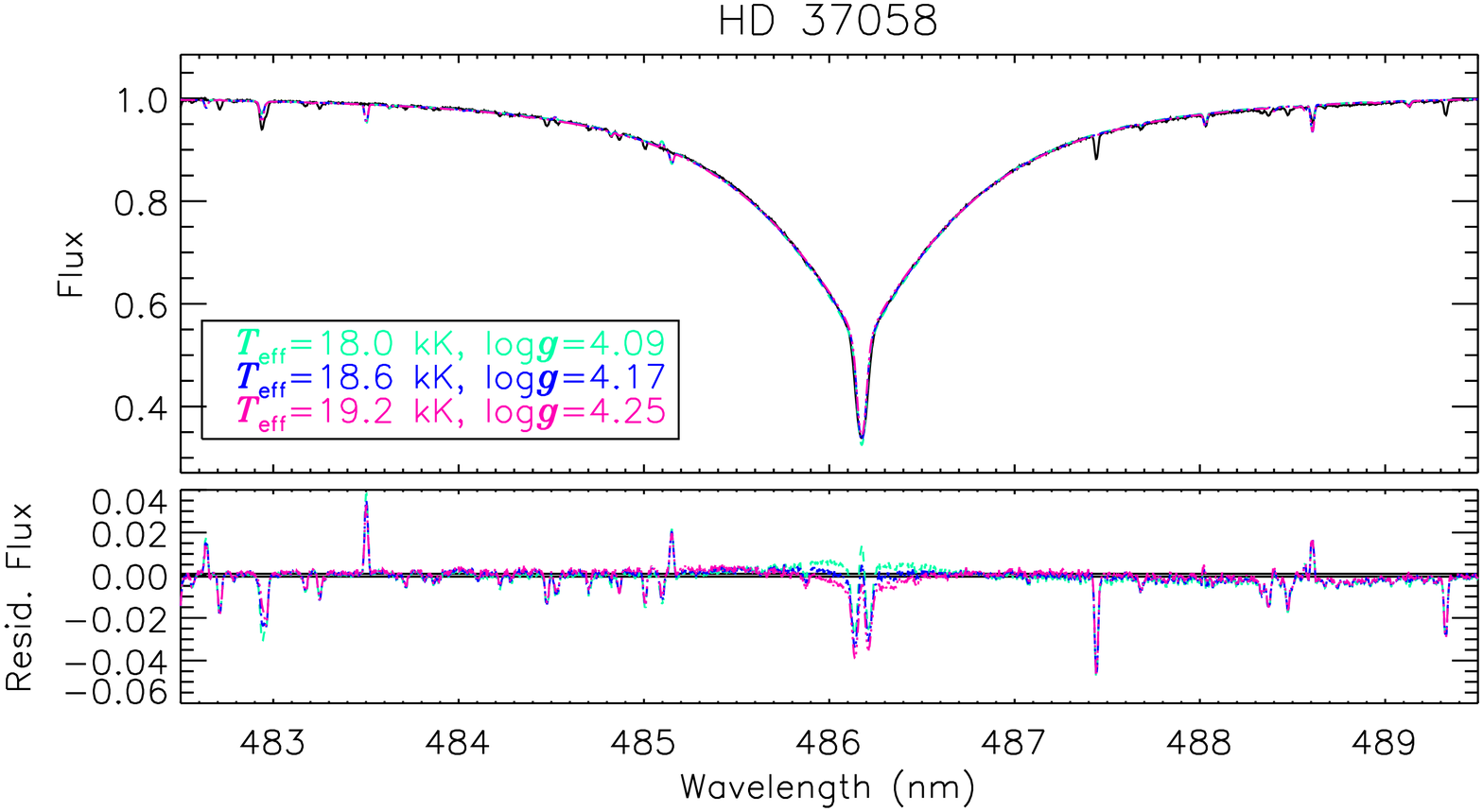} & %
   \includegraphics[trim = 60 0 60 25, width=8.5cm]{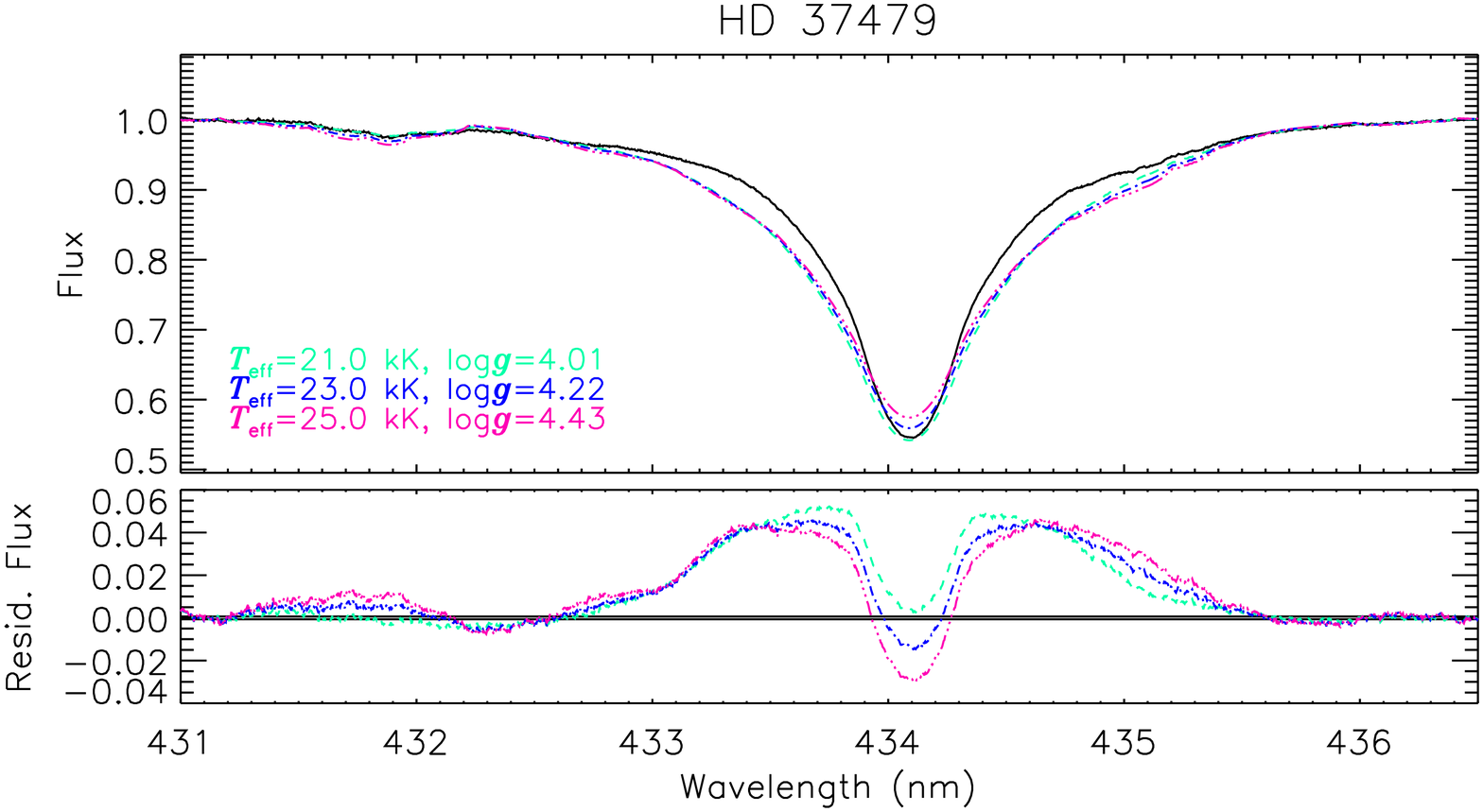} \\ 
   \includegraphics[trim = 60 0 60 25, width=8.5cm]{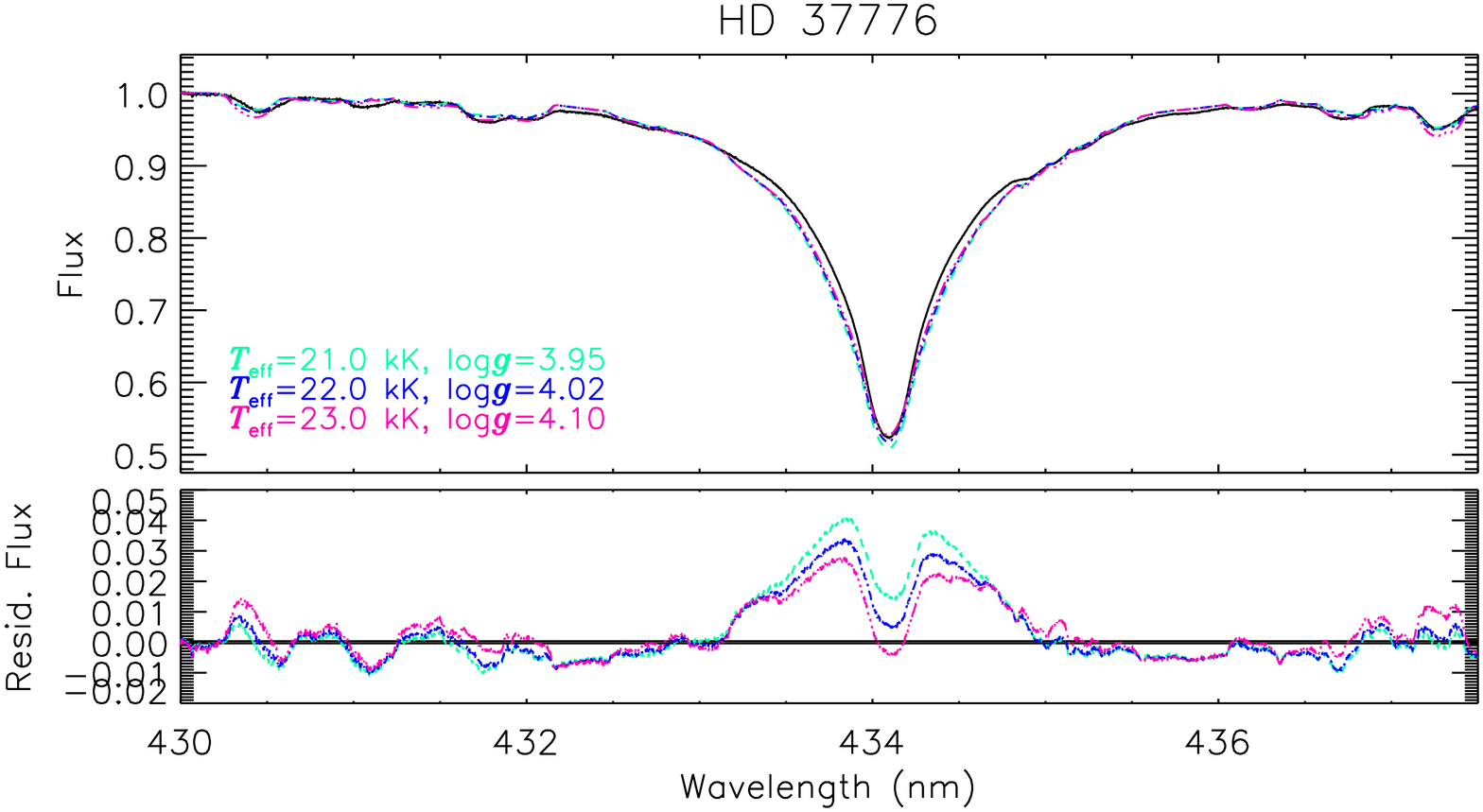} & 
   \includegraphics[trim = 60 0 60 25, width=8.5cm]{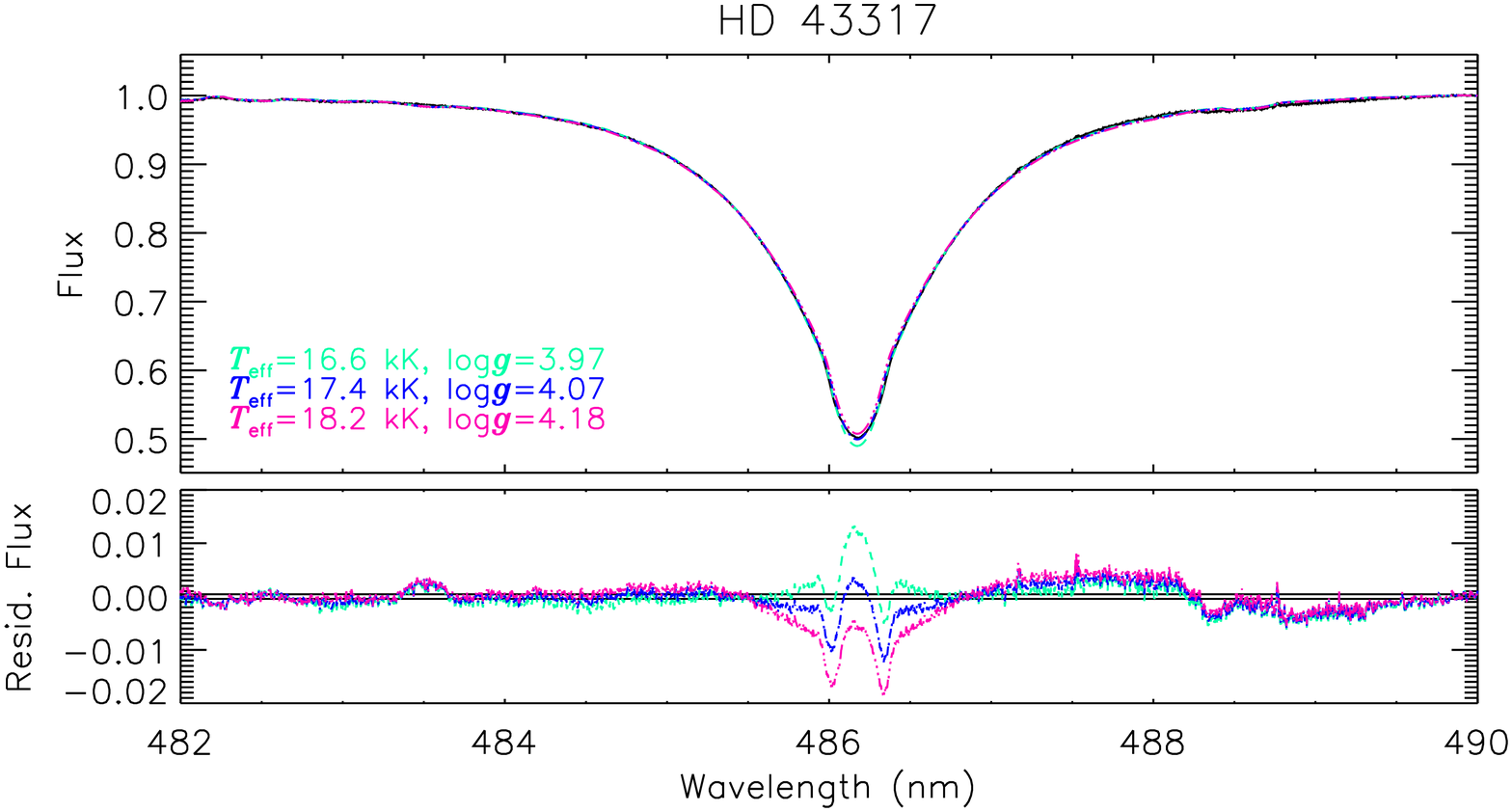} \\ 
\end{tabular}
      \caption[]{Surface gravity determination from H$\beta$ or H$\gamma$. In each pair of sub-panels, the top sub-panel shows the full line and the bottom sub-panel shows the residuals. The mean line profile is shown in black. The dark blue line shows the best-fit model for the \teff; the best-fit model for \teff$-\sigma_{T}$ is shown in light blue; the best-fit model for \teff$+\sigma_{T}$ in purple. Model parameters are indicated in the legend. In the bottom sub-panel the mean flux error bar is indicated by the two horizontal lines above and below 0.}
         \label{balmer1}
   \end{figure*}

\begin{figure*}
   \centering
\begin{tabular}{cc}
   \includegraphics[trim = 60 0 60 25, width=8.5cm]{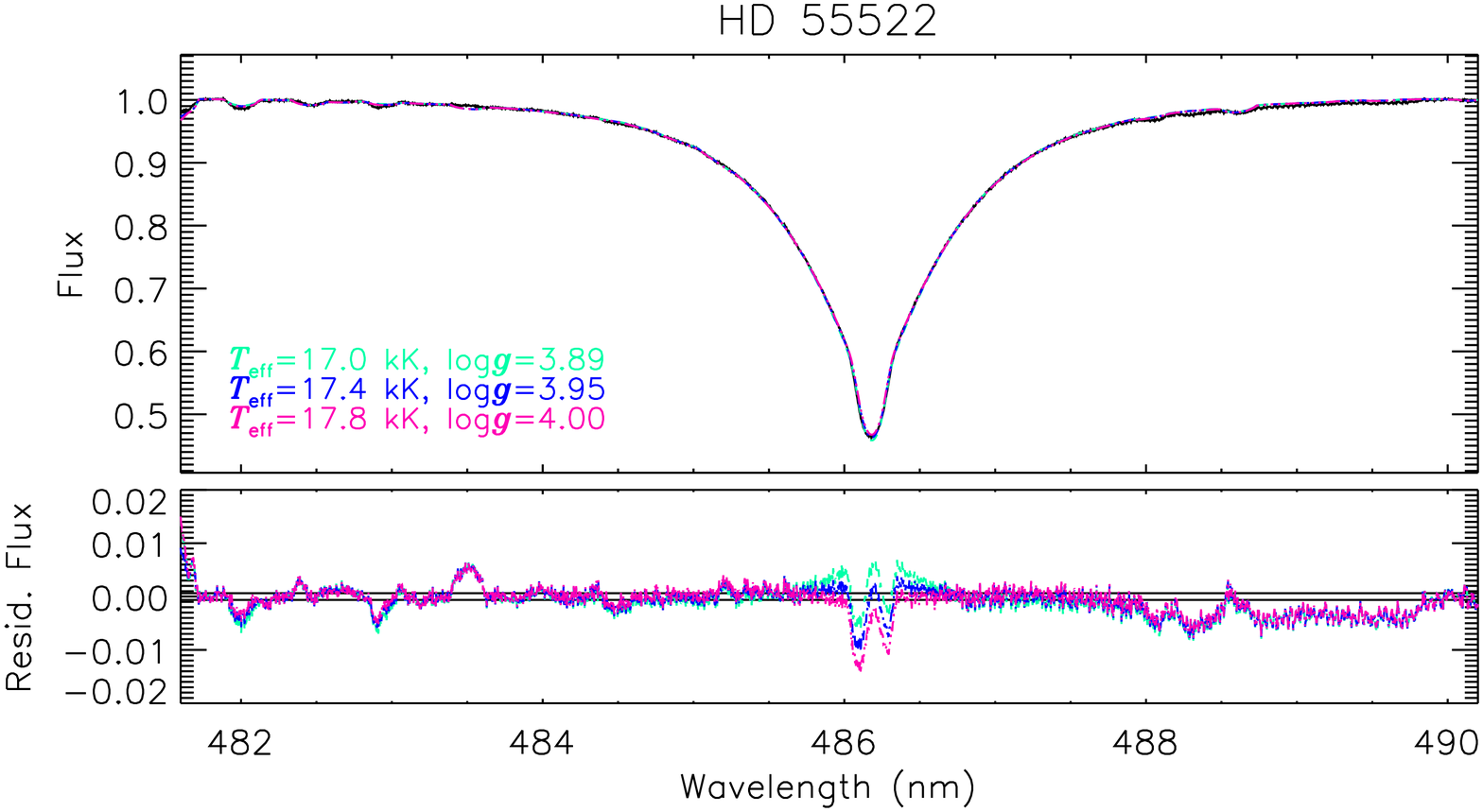} & 
   \includegraphics[trim = 60 0 60 25, width=8.5cm]{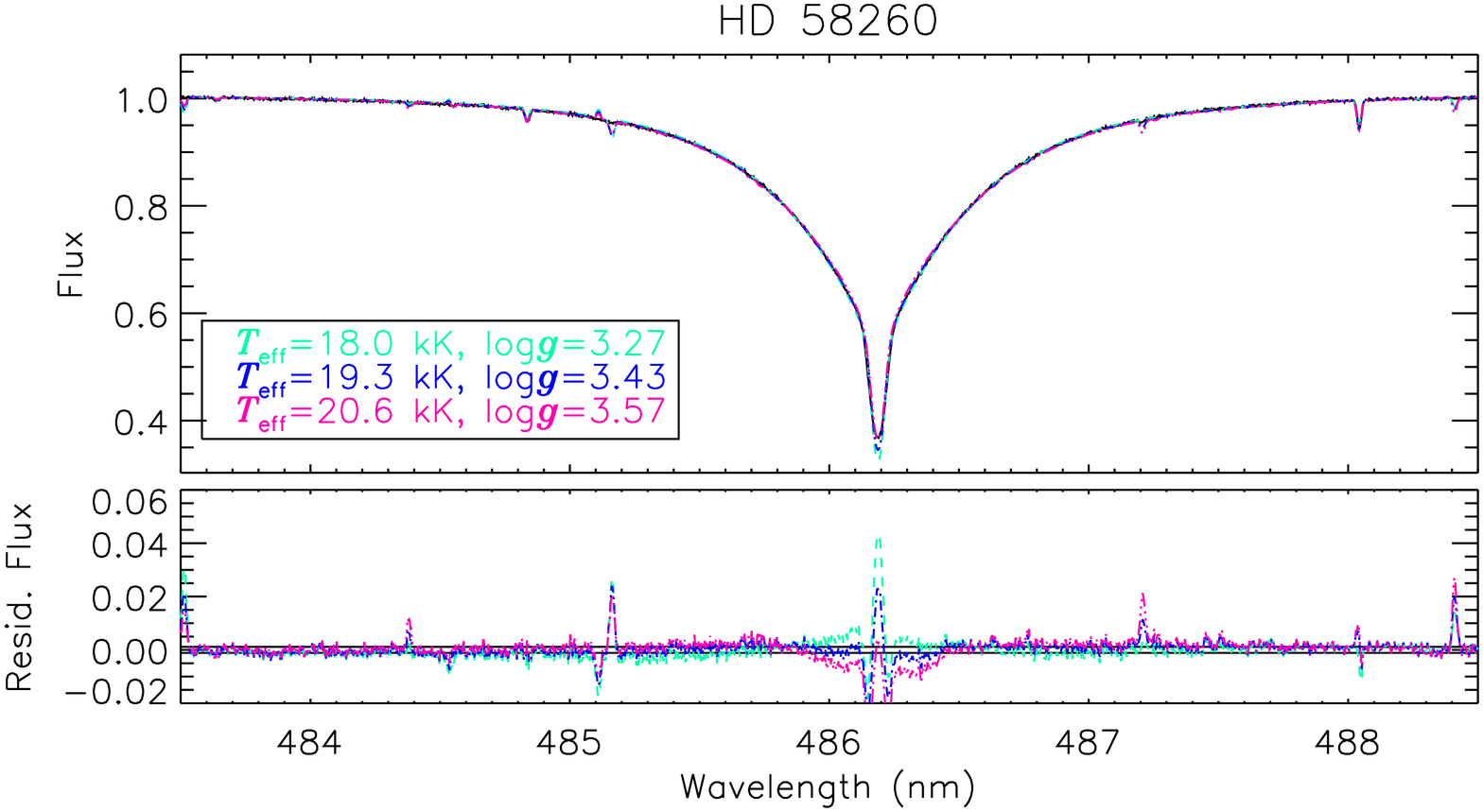} \\ 
   \includegraphics[trim = 60 0 60 25, width=8.5cm]{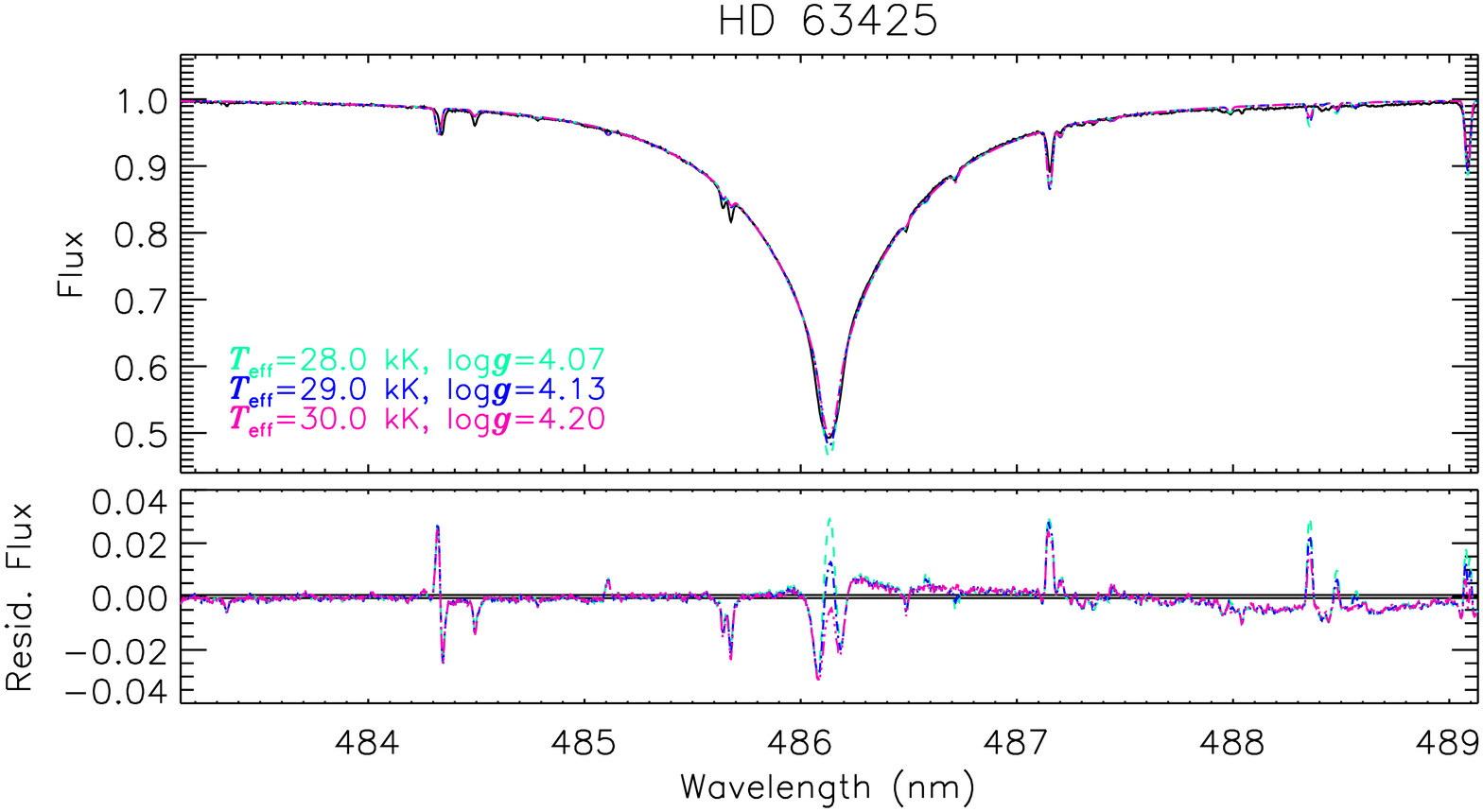} &
   \includegraphics[trim = 60 0 60 25, width=8.5cm]{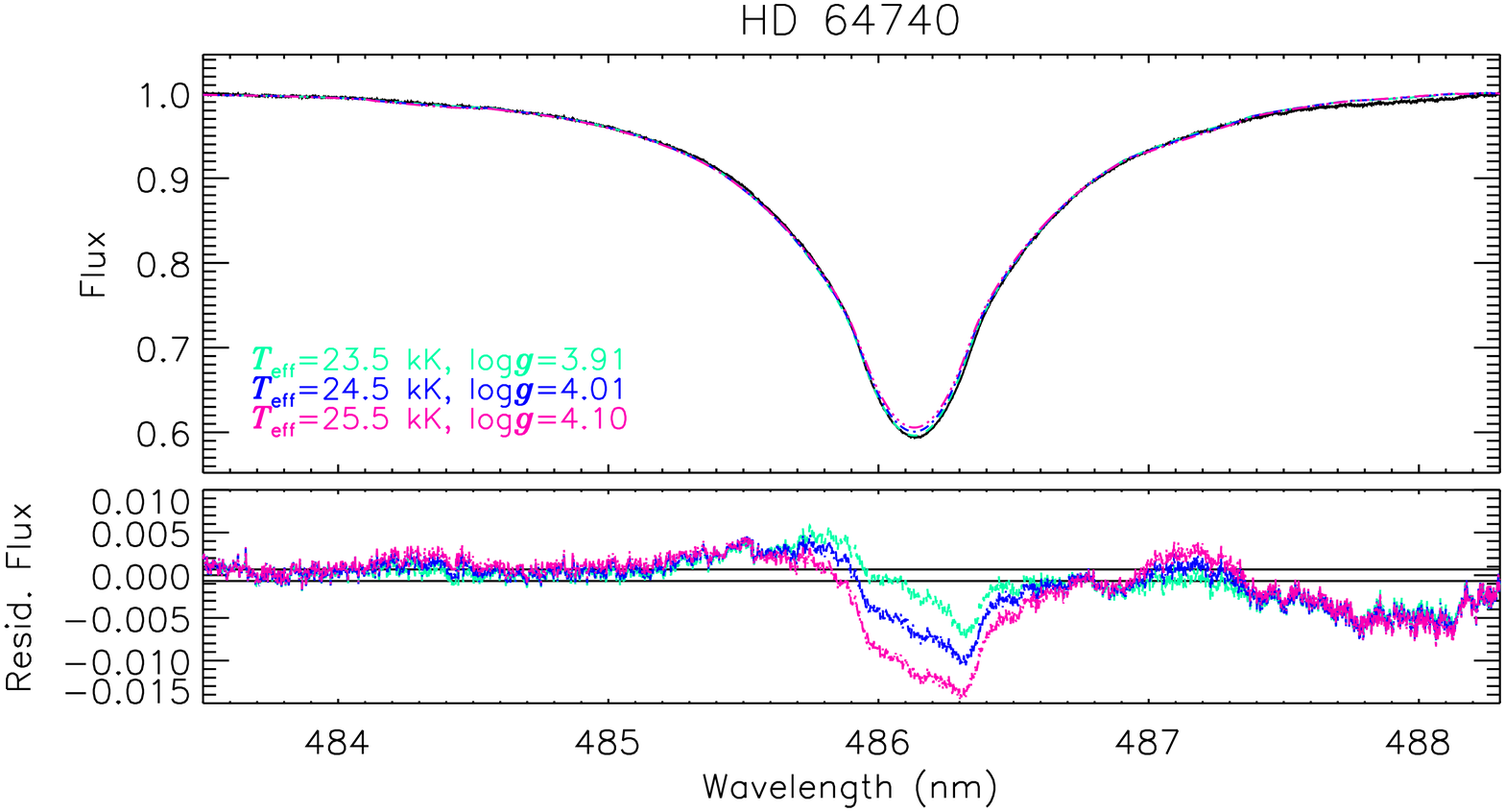} \\ 
   \includegraphics[trim = 60 0 60 25, width=8.5cm]{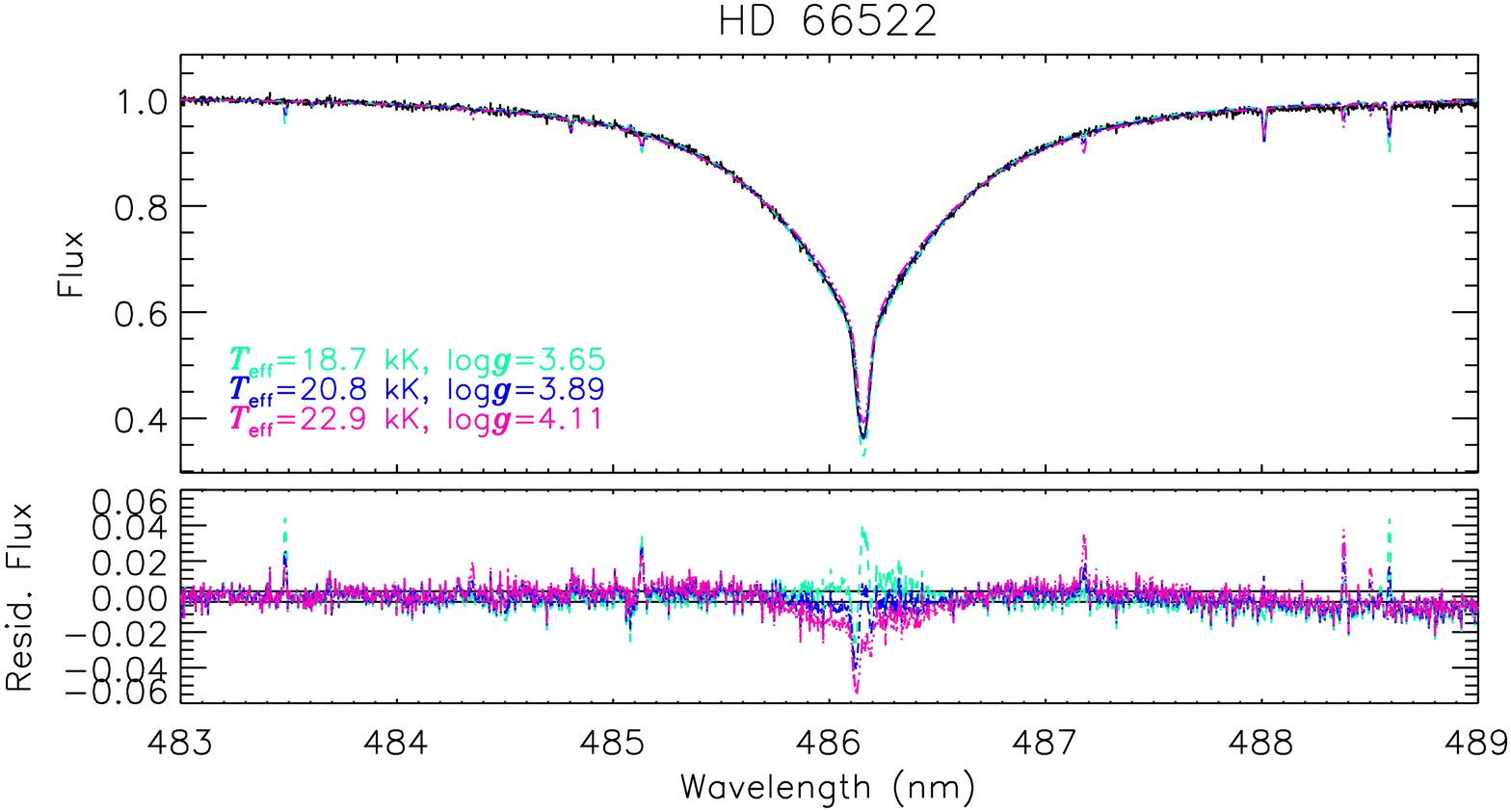} & 
   \includegraphics[trim = 60 0 60 25, width=8.5cm]{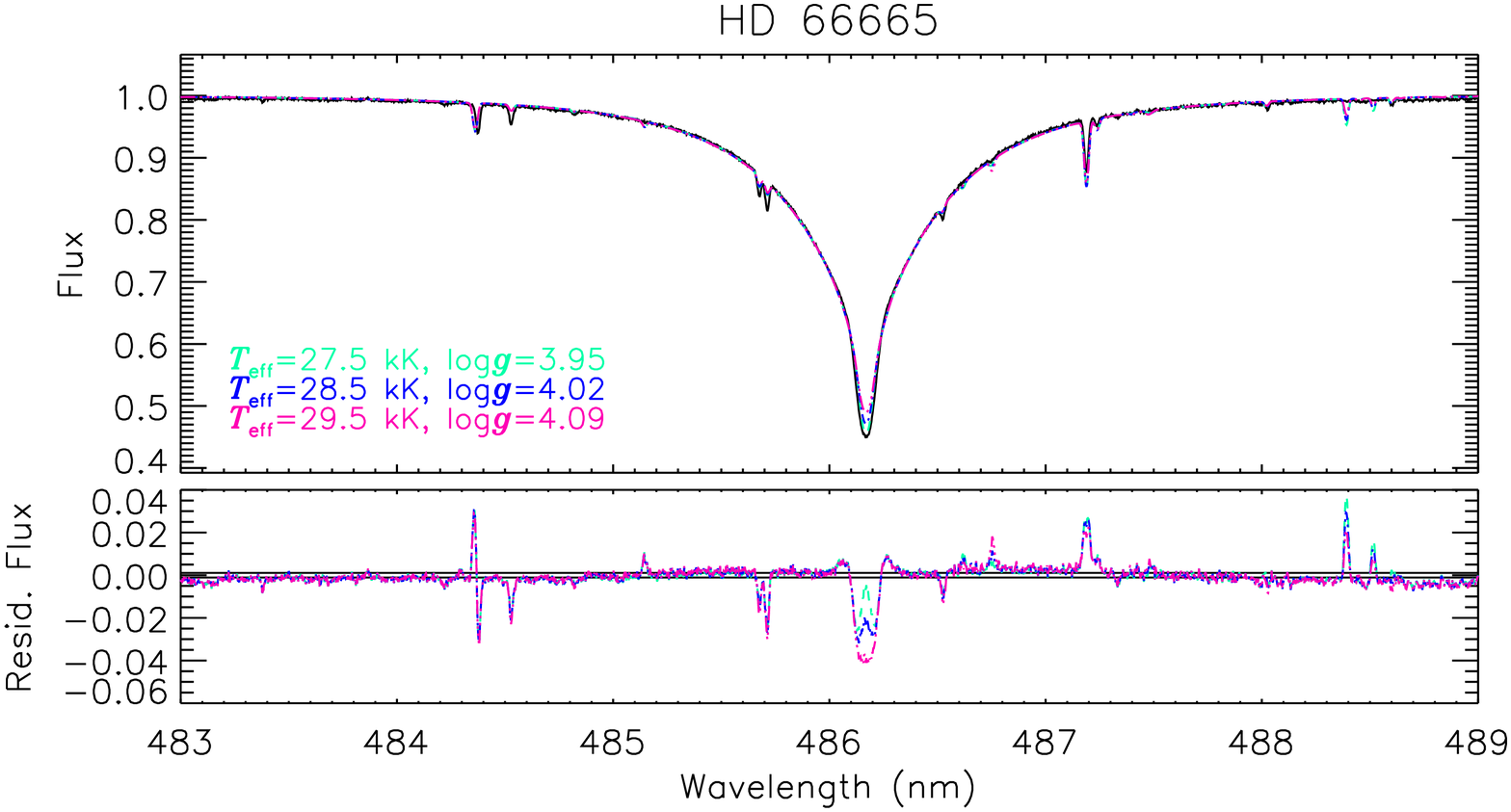} \\ 
   \includegraphics[trim = 60 0 60 25, width=8.5cm]{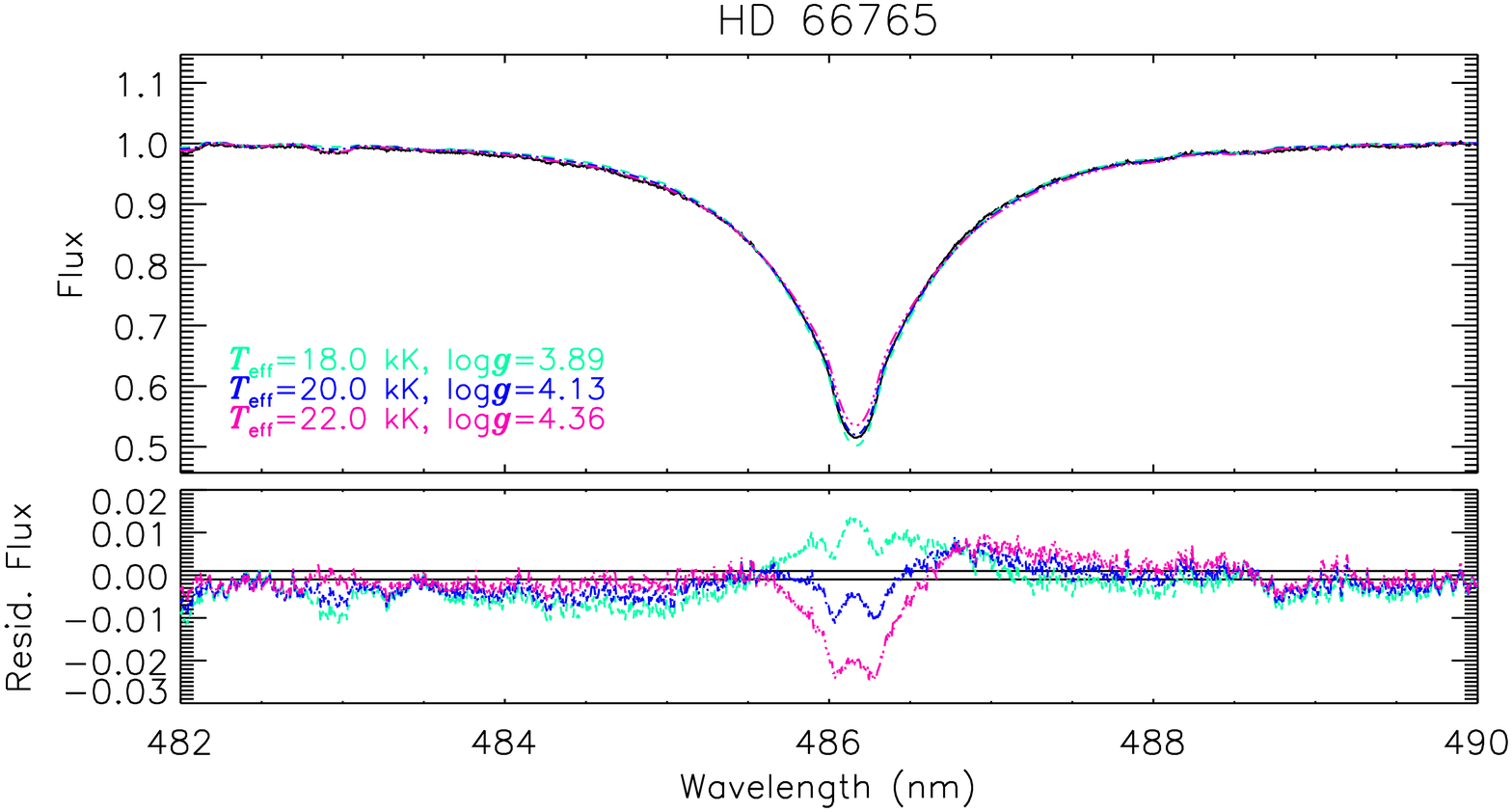} & 
   \includegraphics[trim = 60 0 60 25, width=8.5cm]{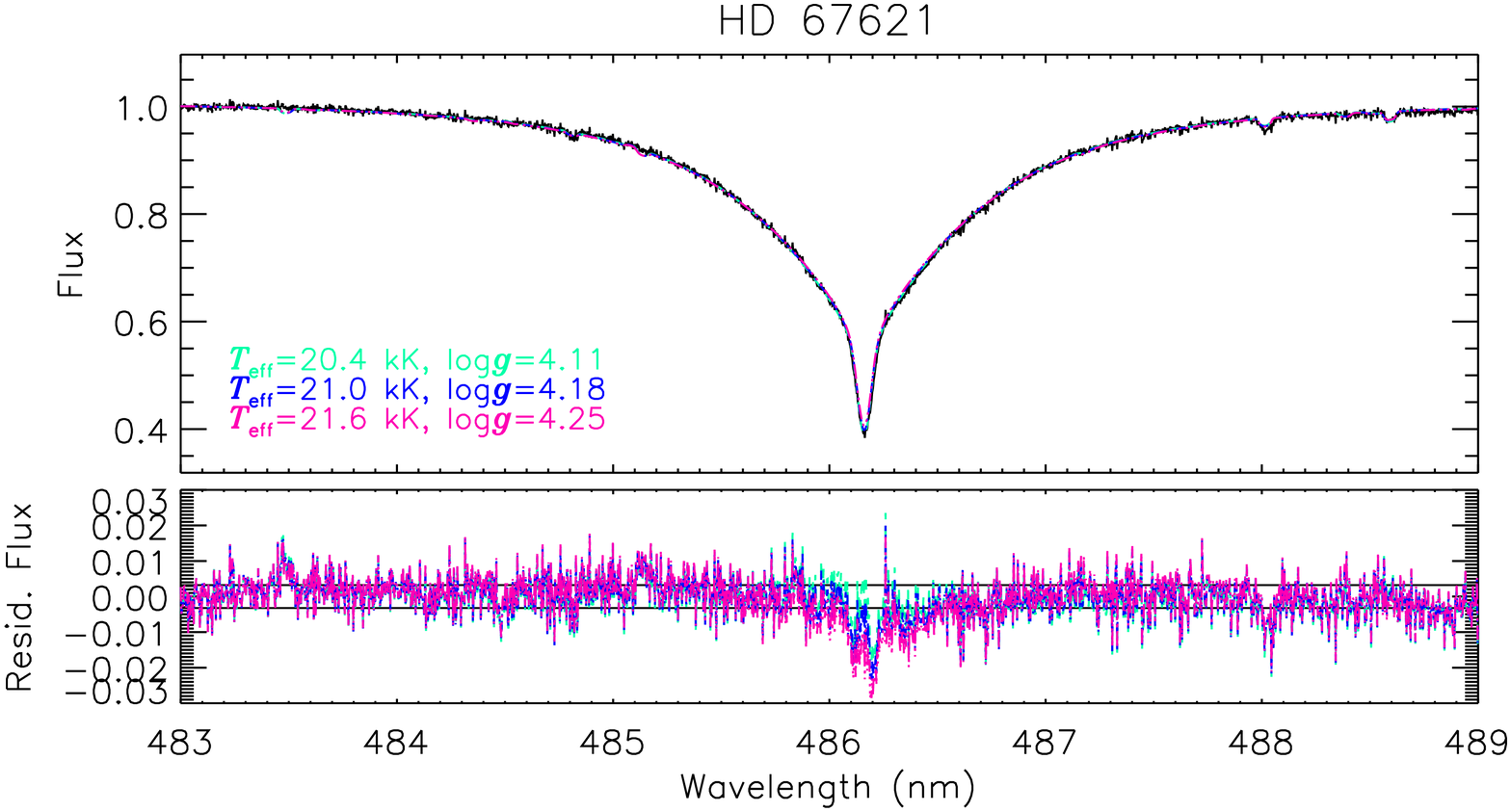} \\ 
   \includegraphics[trim = 60 0 60 25, width=8.5cm]{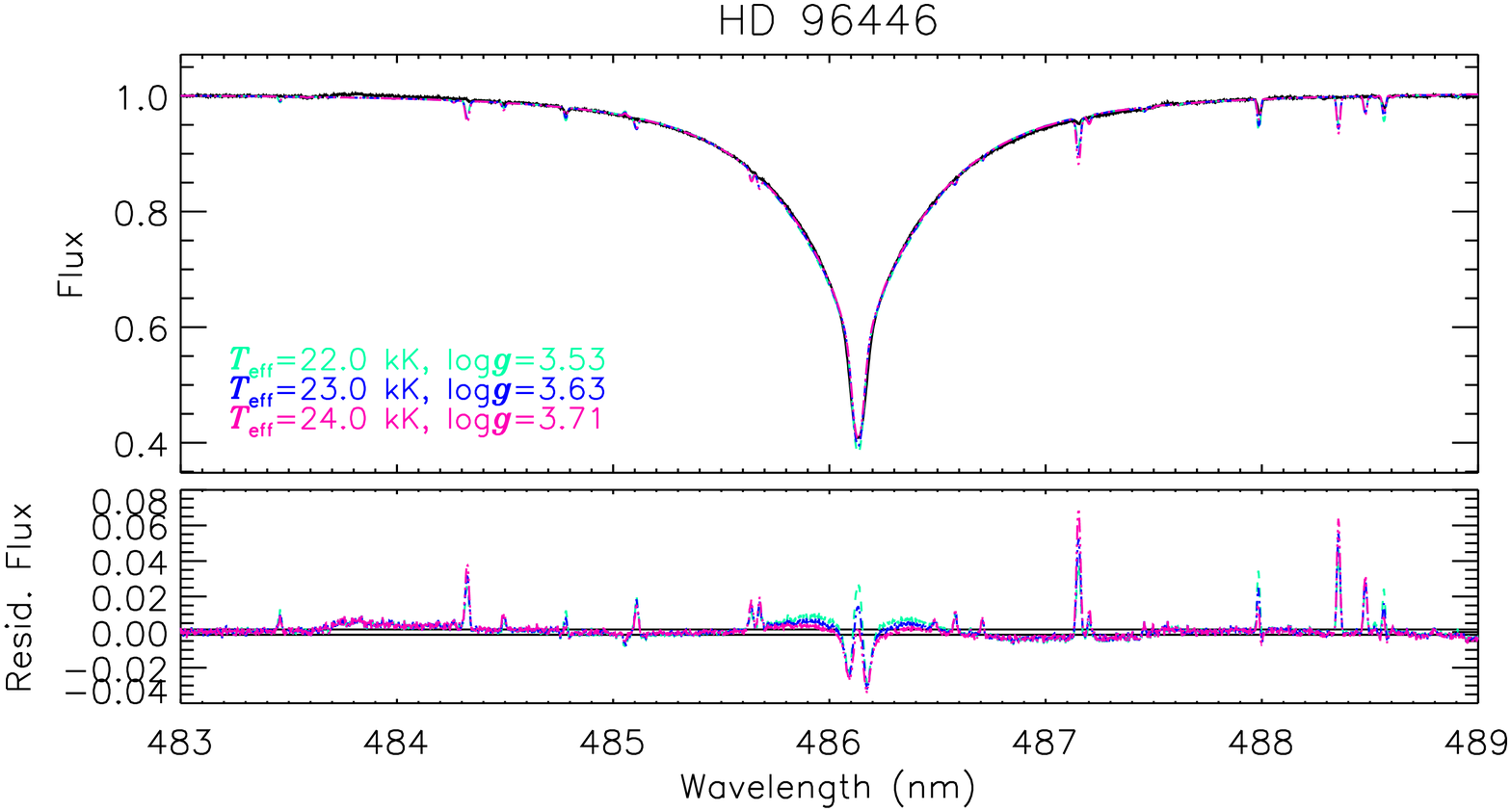} & 
   \includegraphics[trim = 60 0 60 25, width=8.5cm]{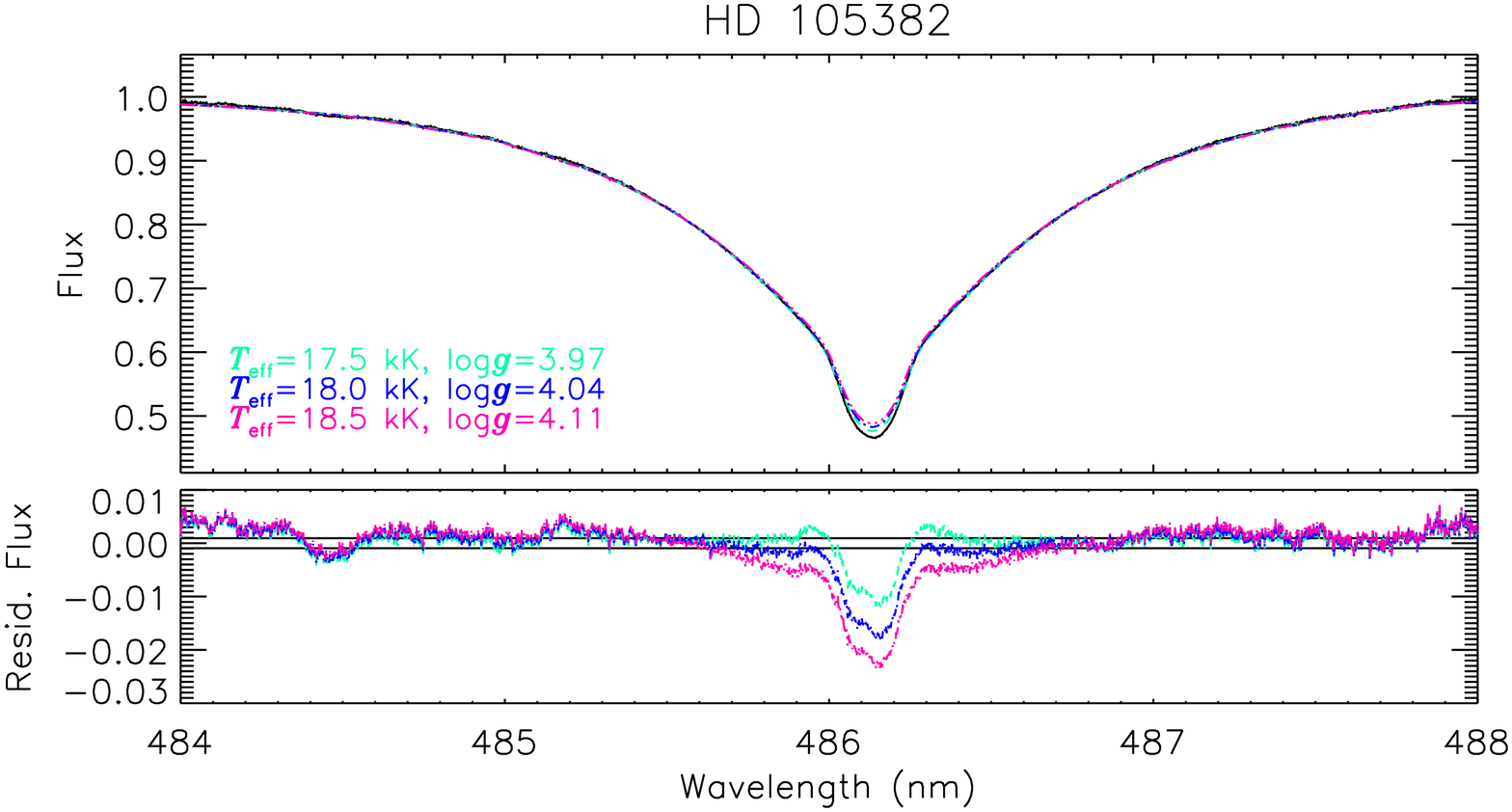} \\ 
\end{tabular}
      \caption[]{As Fig.\ \ref{balmer1}}
         \label{balmer2}
   \end{figure*}

\begin{figure*}
   \centering
\begin{tabular}{cc}
   \includegraphics[trim = 60 0 60 25, width=8.5cm]{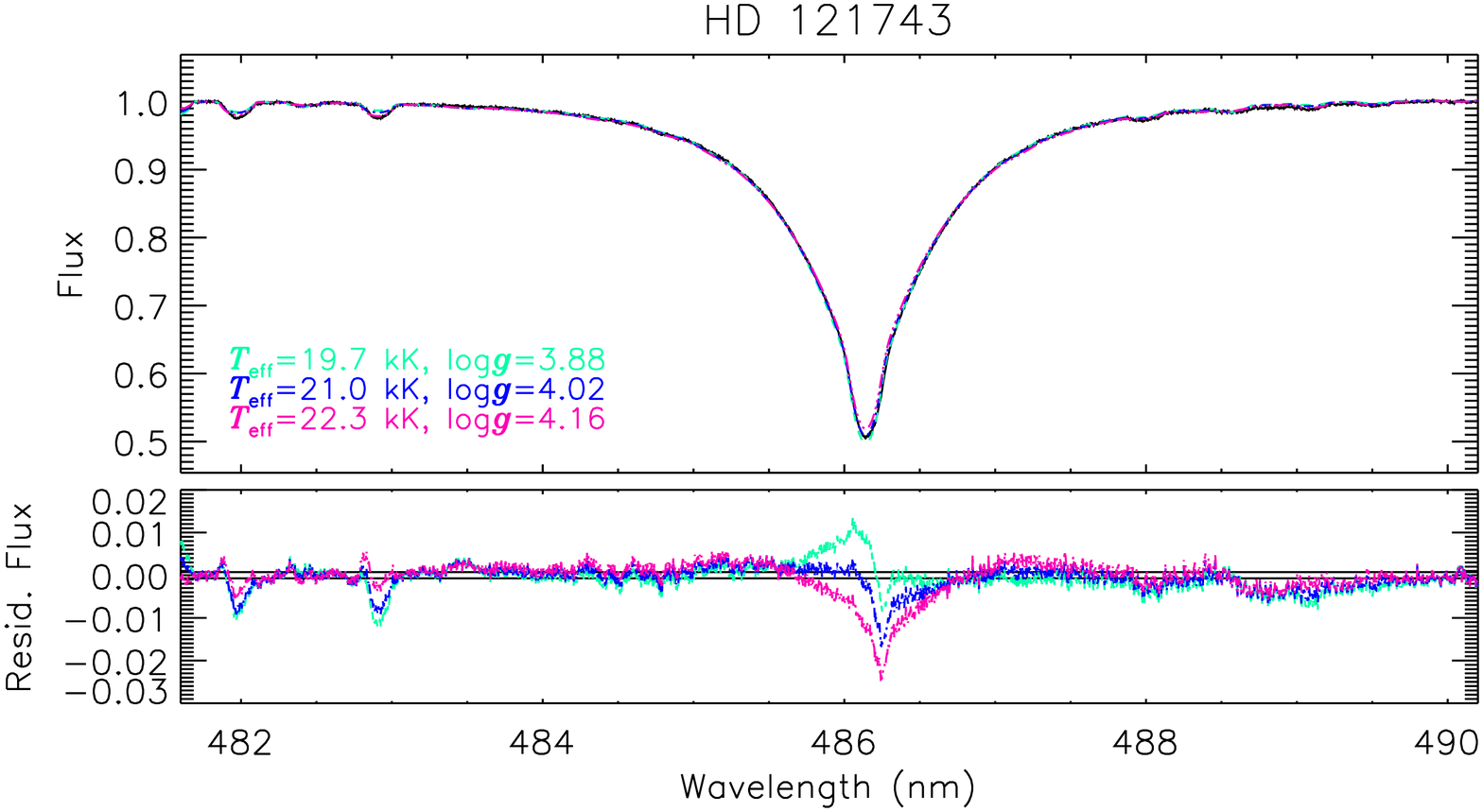} & 
   \includegraphics[trim = 60 0 60 25, width=8.5cm]{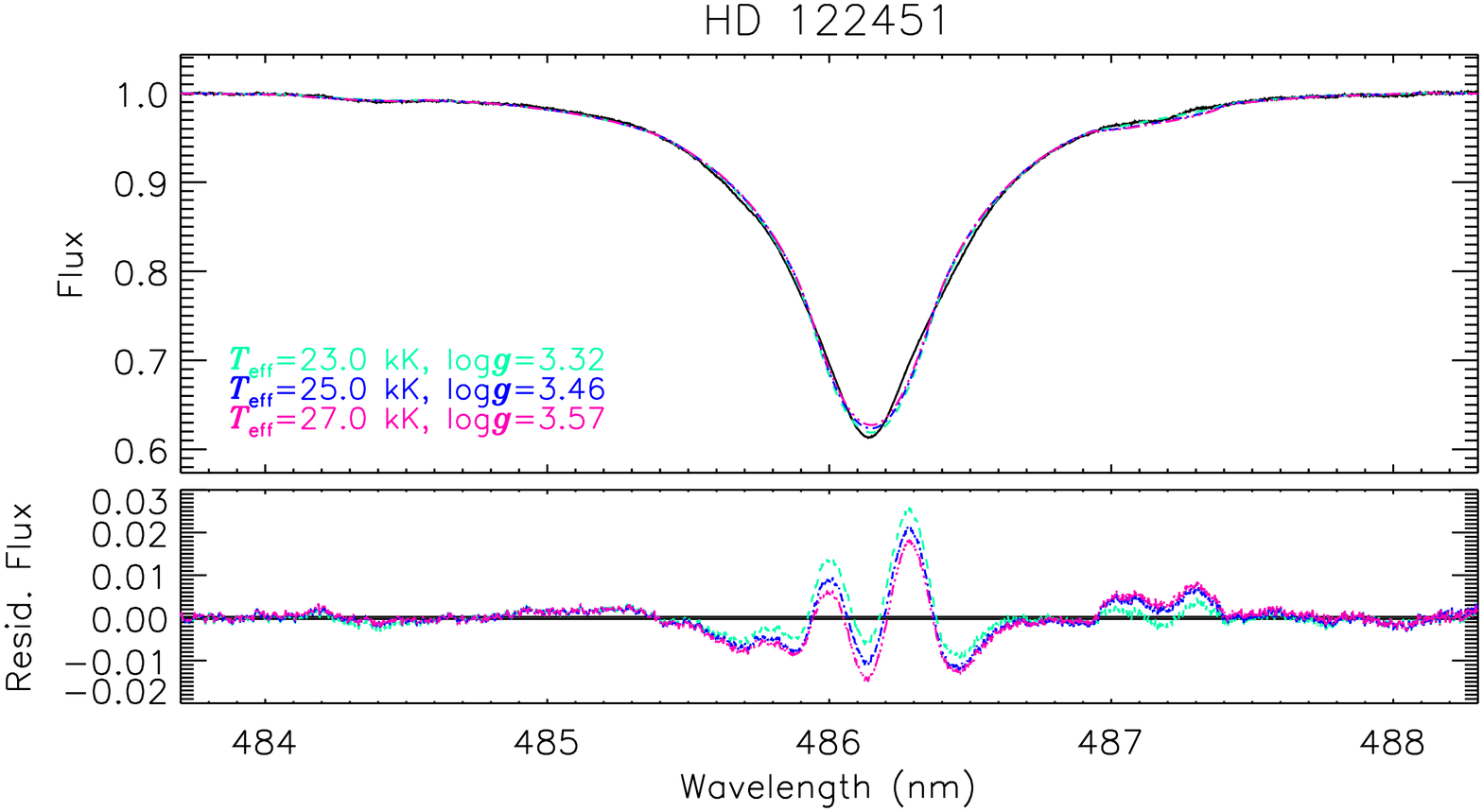} \\
   \includegraphics[trim = 60 0 60 25, width=8.5cm]{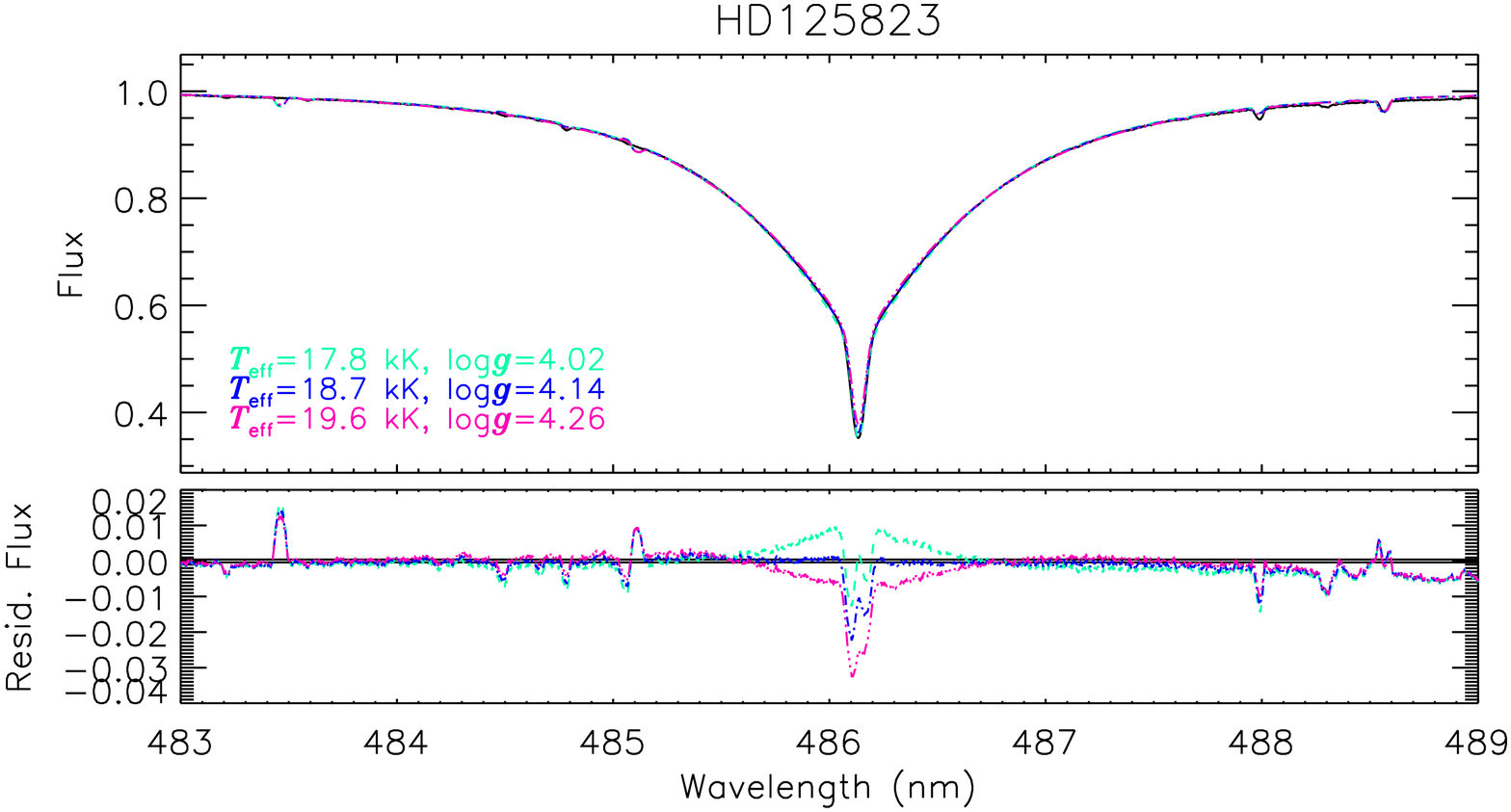} & 
   \includegraphics[trim = 60 0 60 25, width=8.5cm]{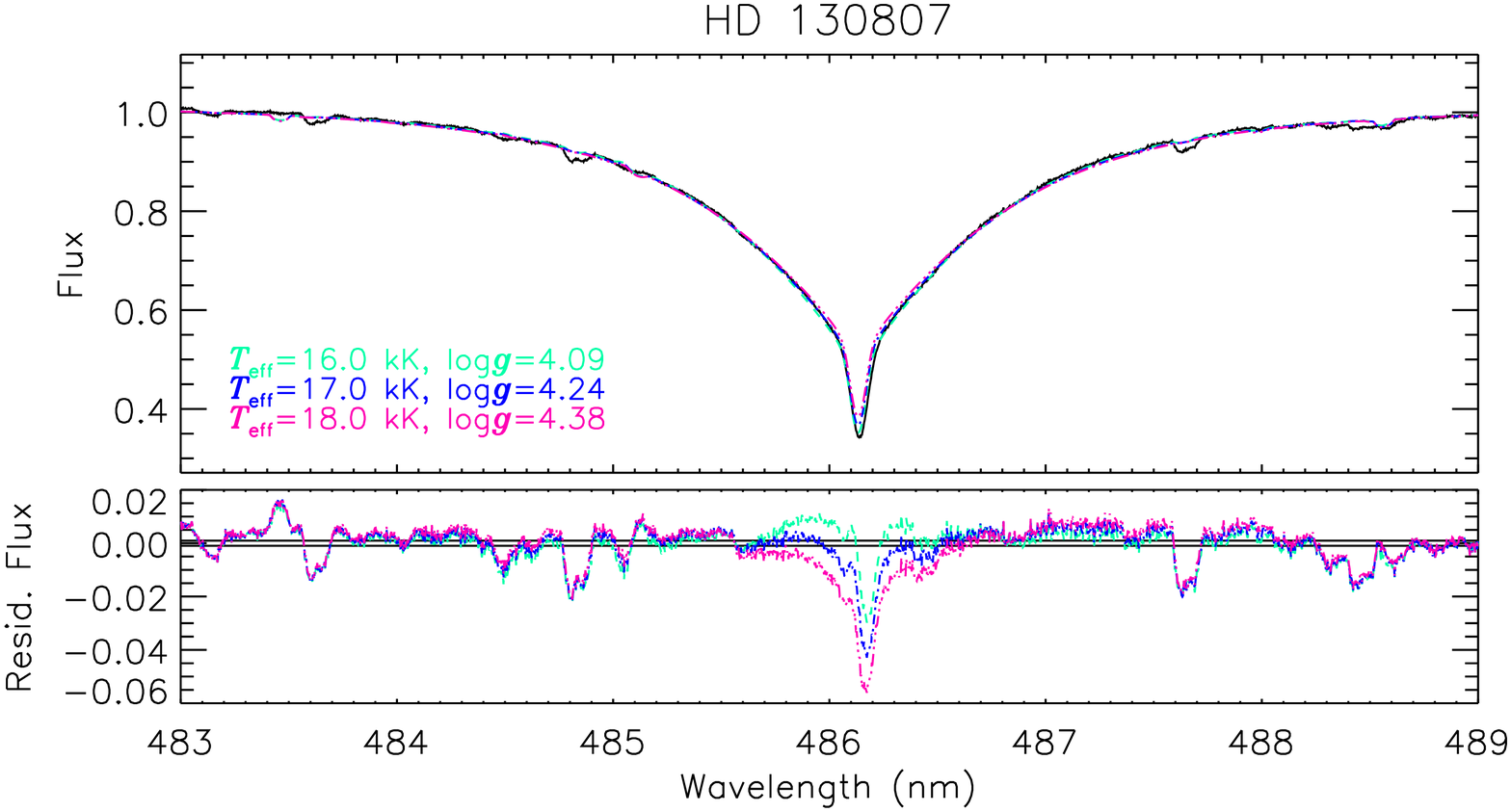} \\
   \includegraphics[trim = 60 0 60 25, width=8.5cm]{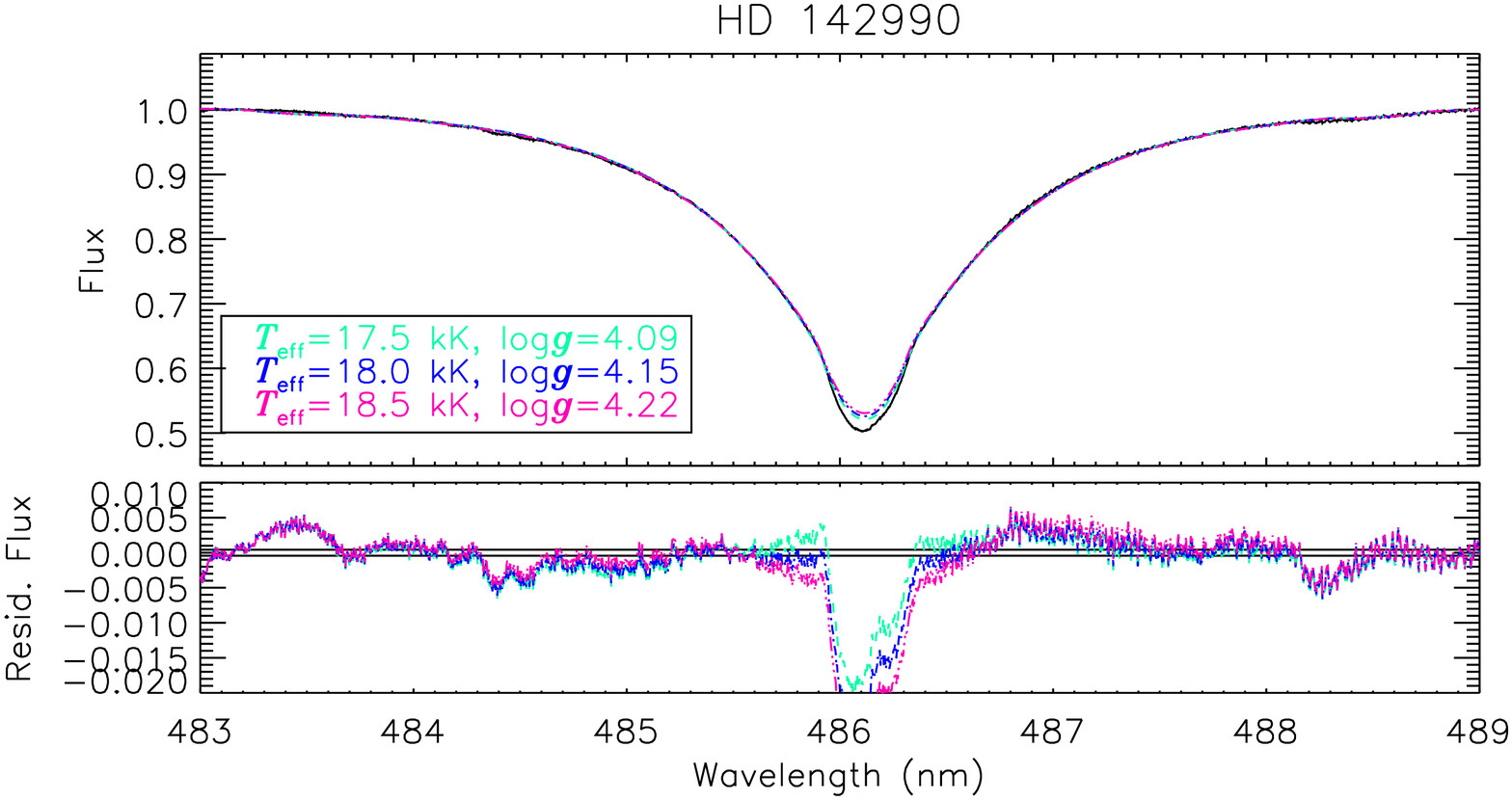} &
   \includegraphics[trim = 60 0 60 25, width=8.5cm]{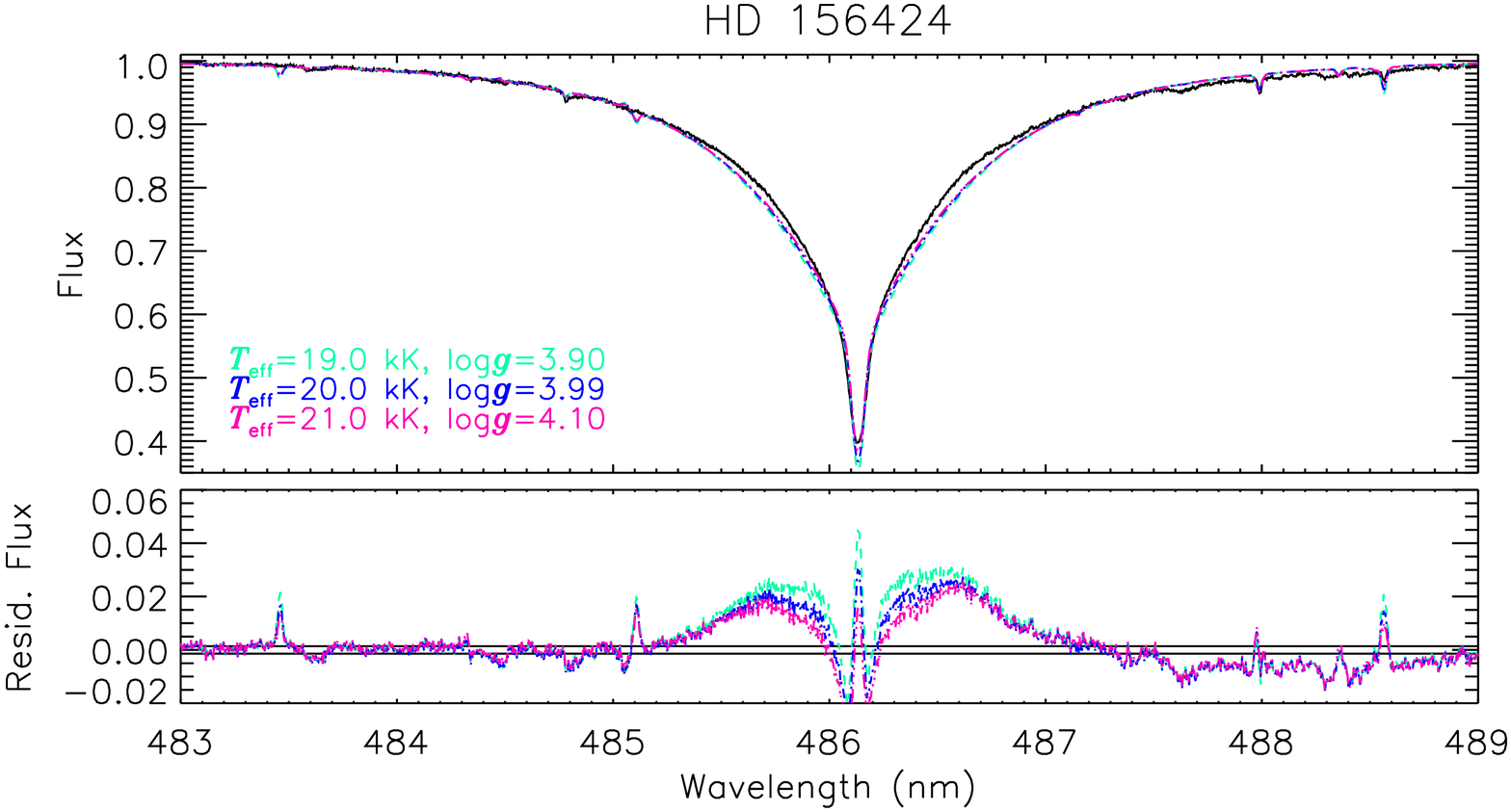} \\
   \includegraphics[trim = 60 0 60 25, width=8.5cm]{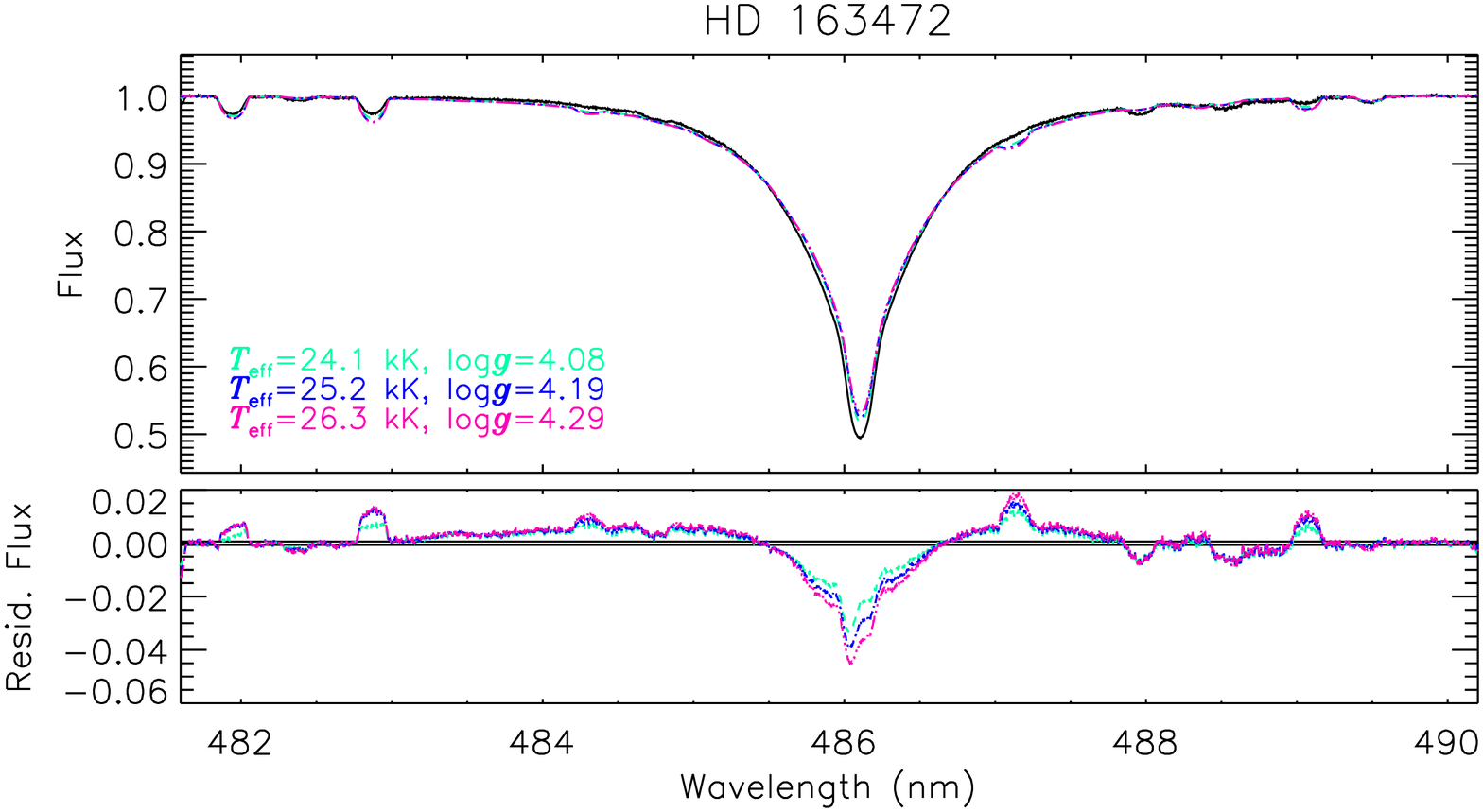} &
   \includegraphics[trim = 60 0 60 25, width=8.5cm]{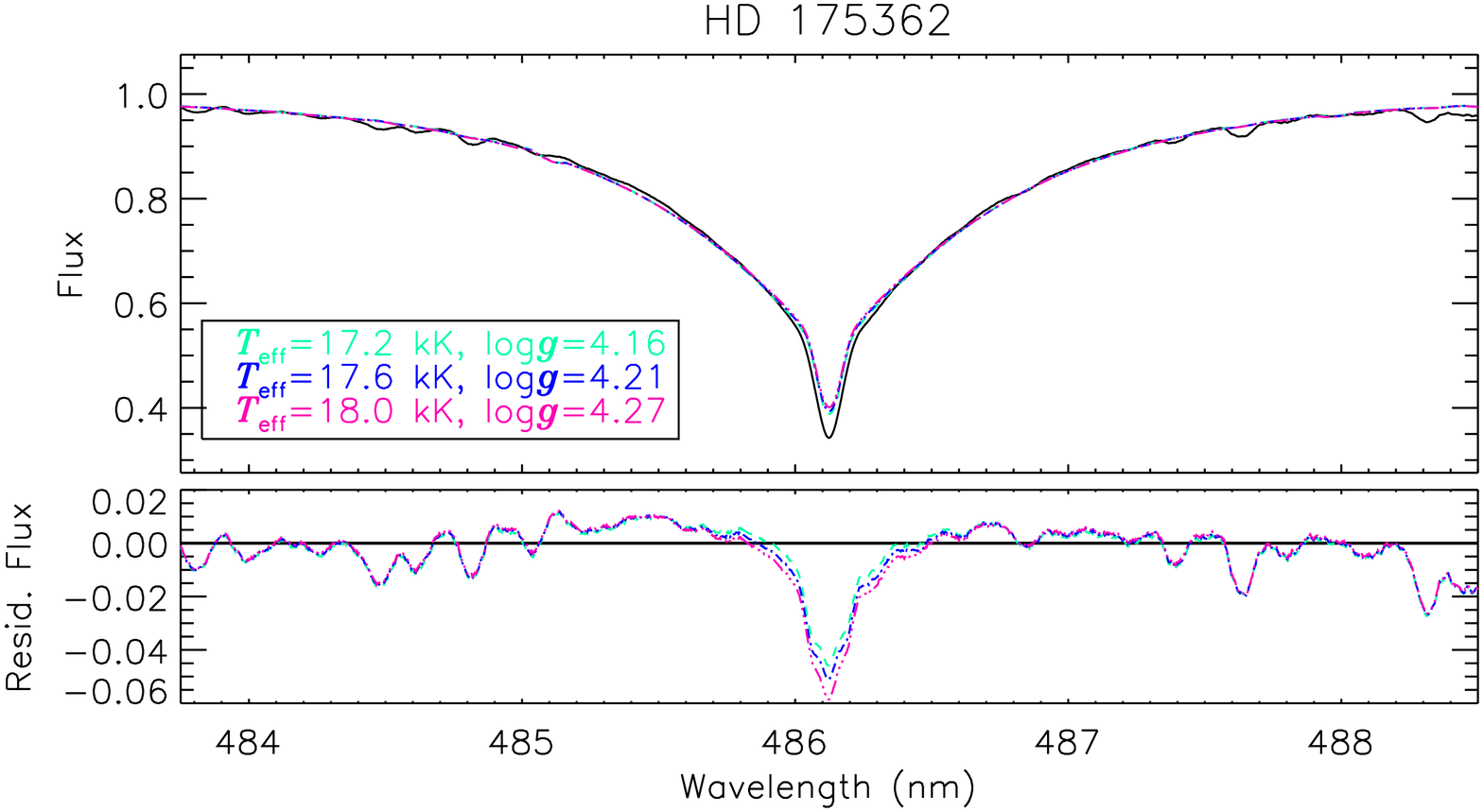} \\
   \includegraphics[trim = 60 0 60 25, width=8.5cm]{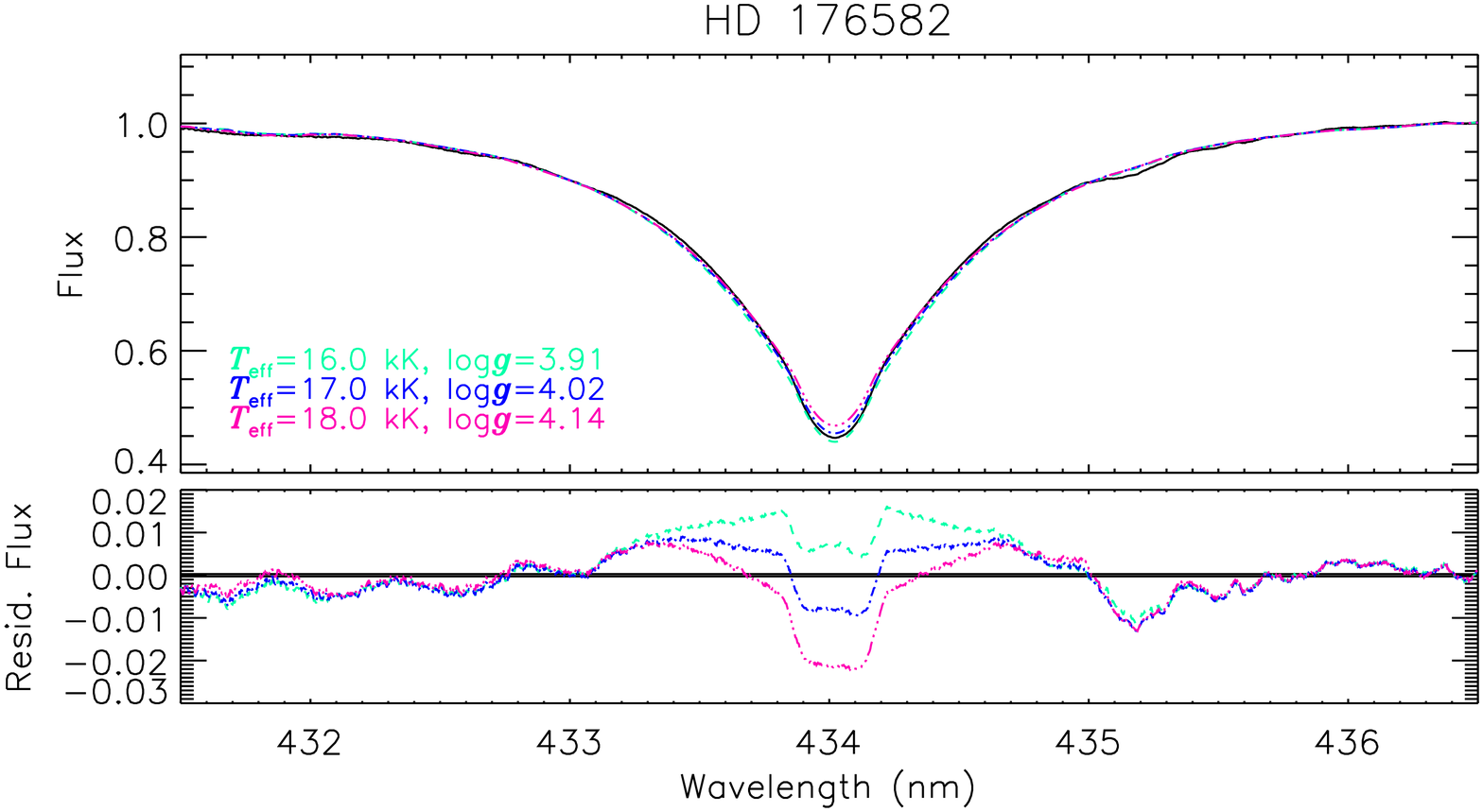} &
   \includegraphics[trim = 60 0 60 25, width=8.5cm]{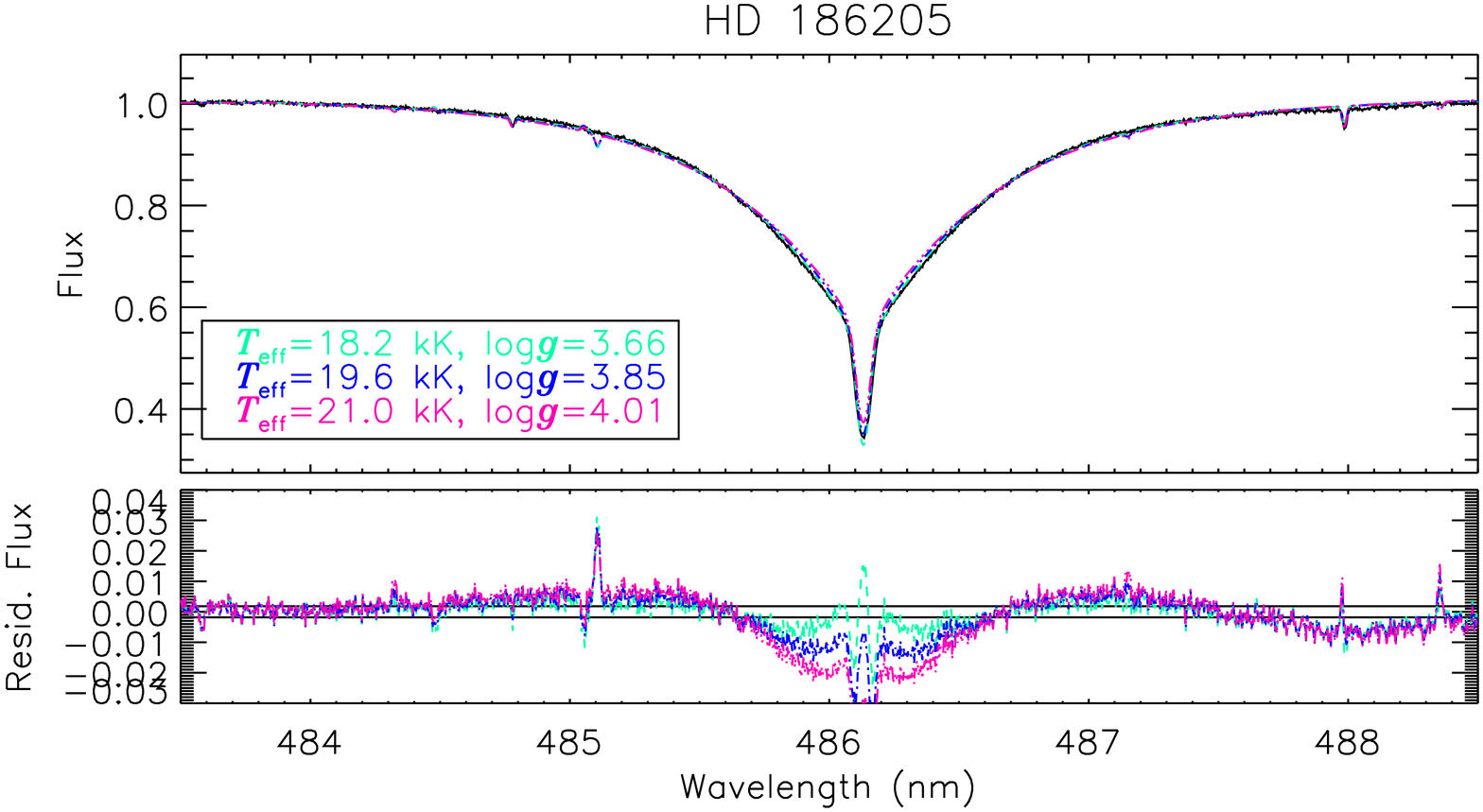} 
\end{tabular}
      \caption[]{As Fig.\ \ref{balmer1}}
         \label{balmer3}
   \end{figure*}

\begin{figure*}
   \centering
\begin{tabular}{cc}
   \includegraphics[trim = 60 0 60 25, width=8.5cm]{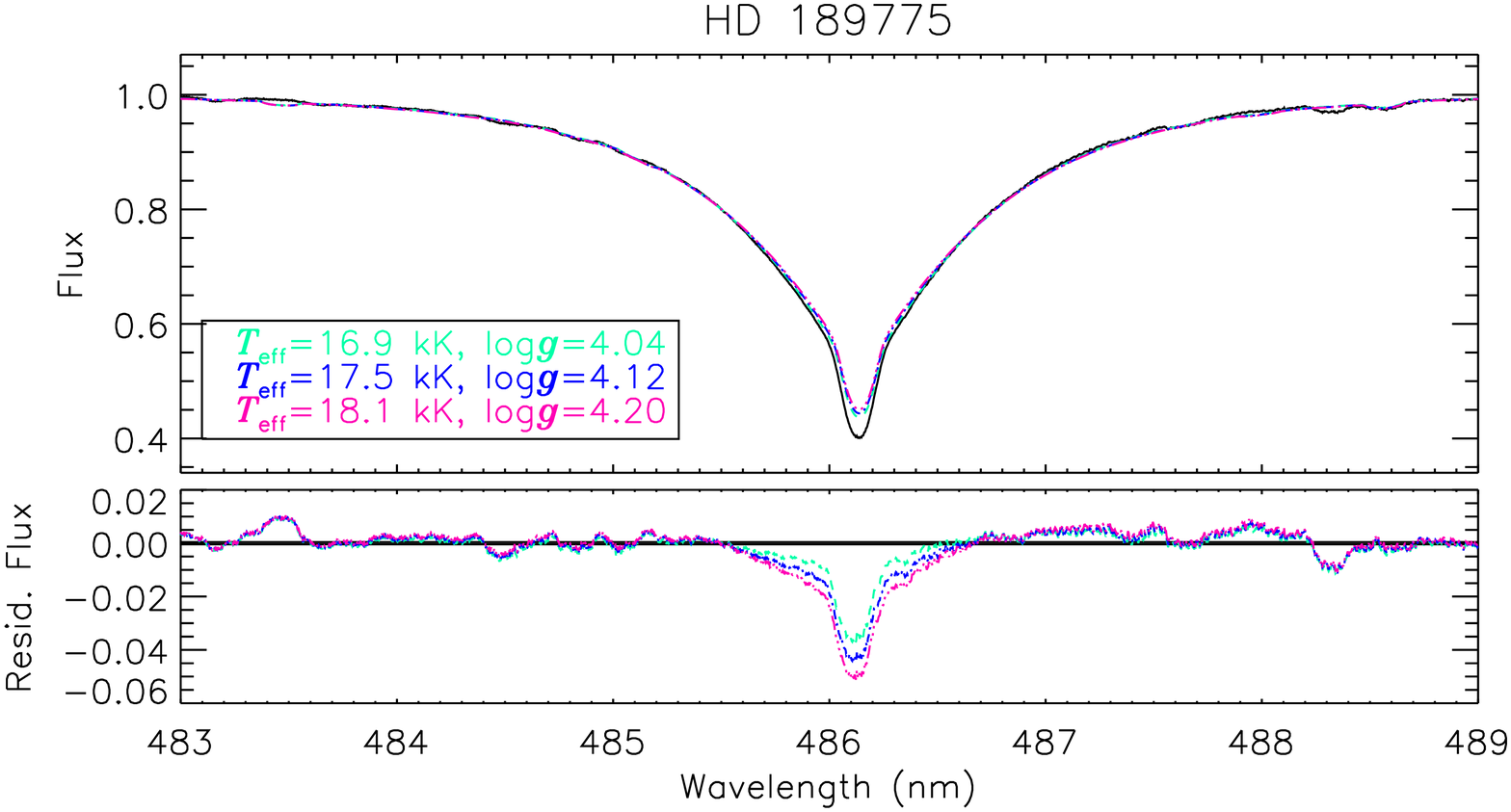} &
   \includegraphics[trim = 60 0 60 25, width=8.5cm]{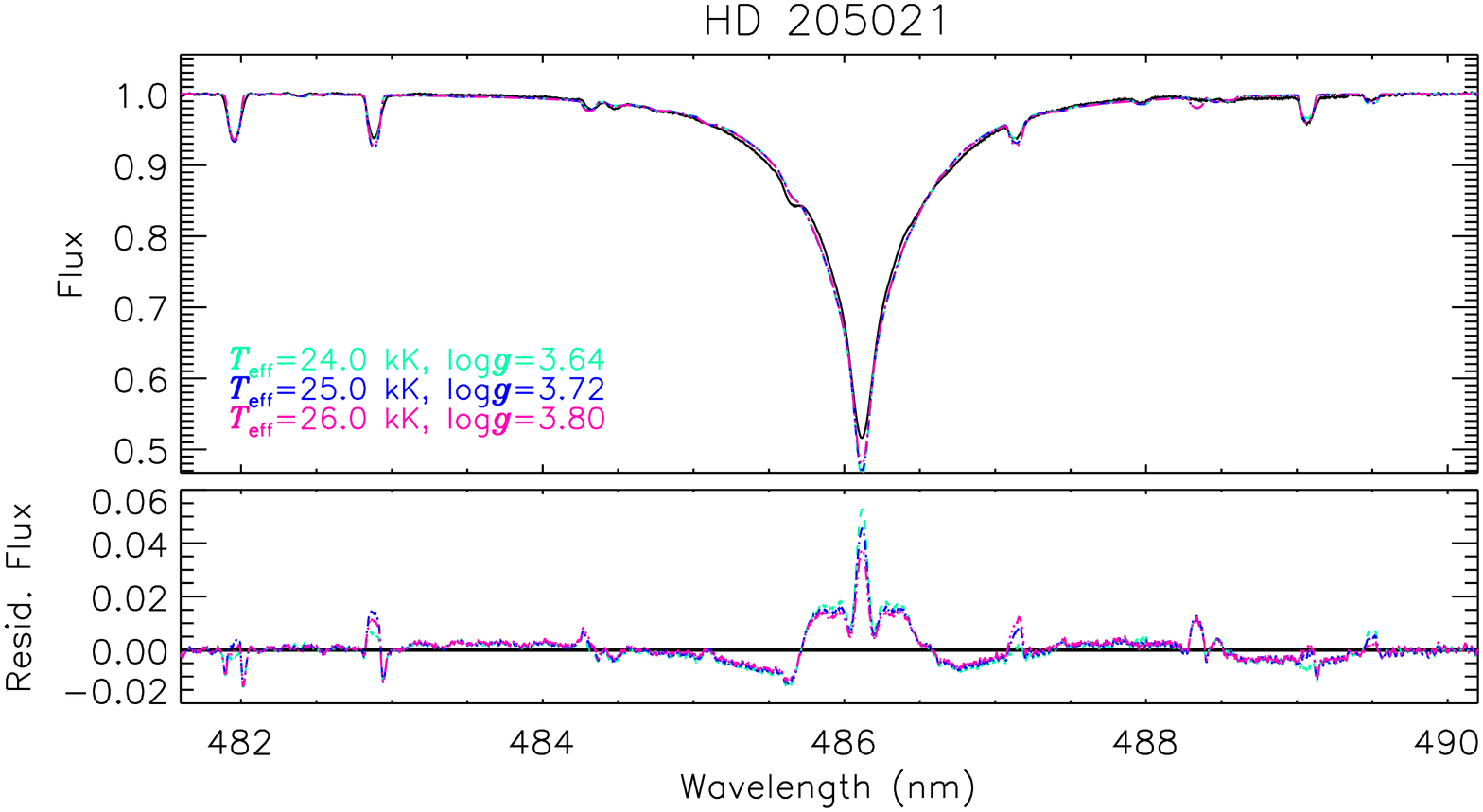} \\
   \includegraphics[trim = 60 0 60 25, width=8.5cm]{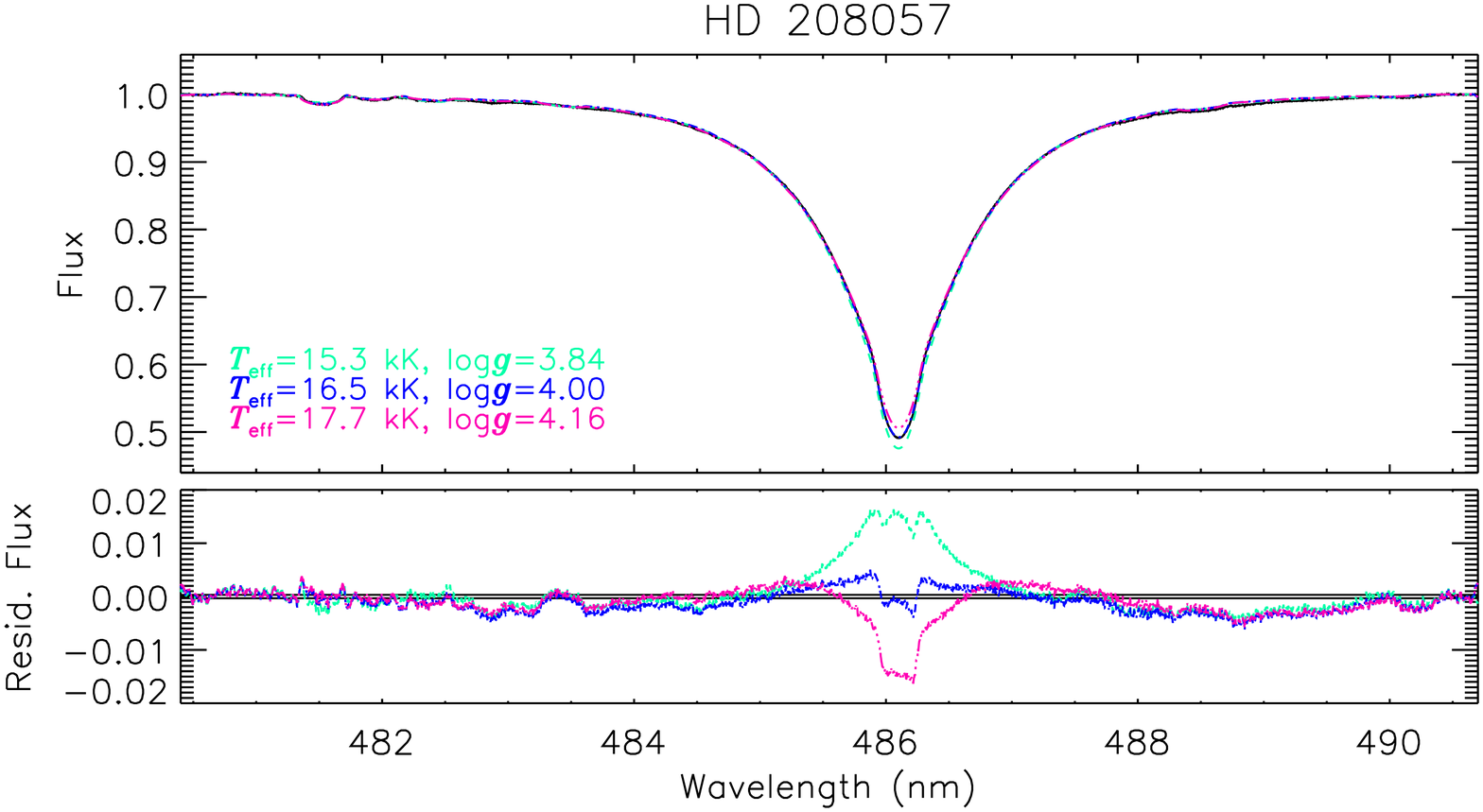} &
   \includegraphics[trim = 60 0 60 25, width=8.5cm]{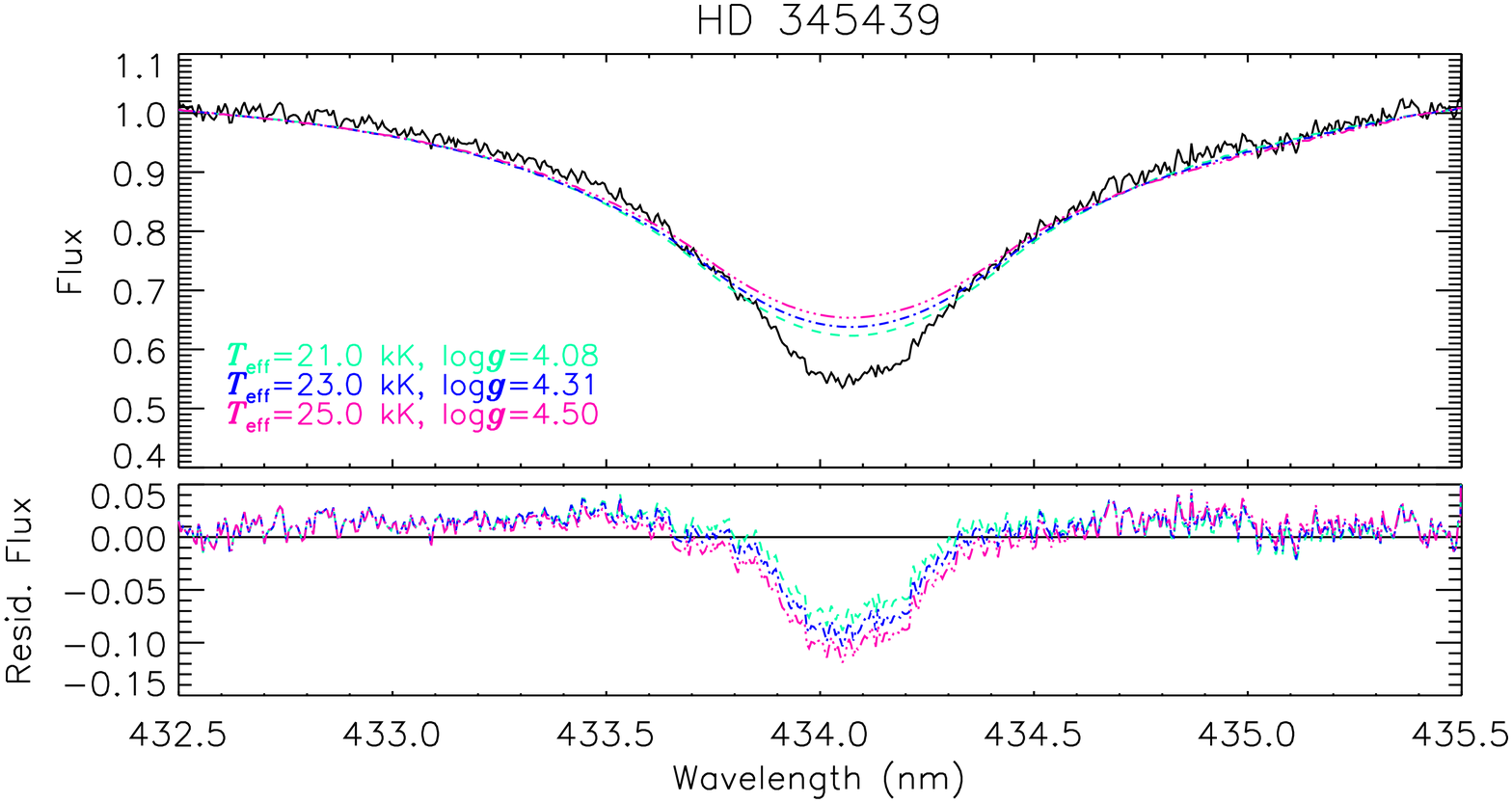} \\
   \includegraphics[trim = 60 0 60 25, width=8.5cm]{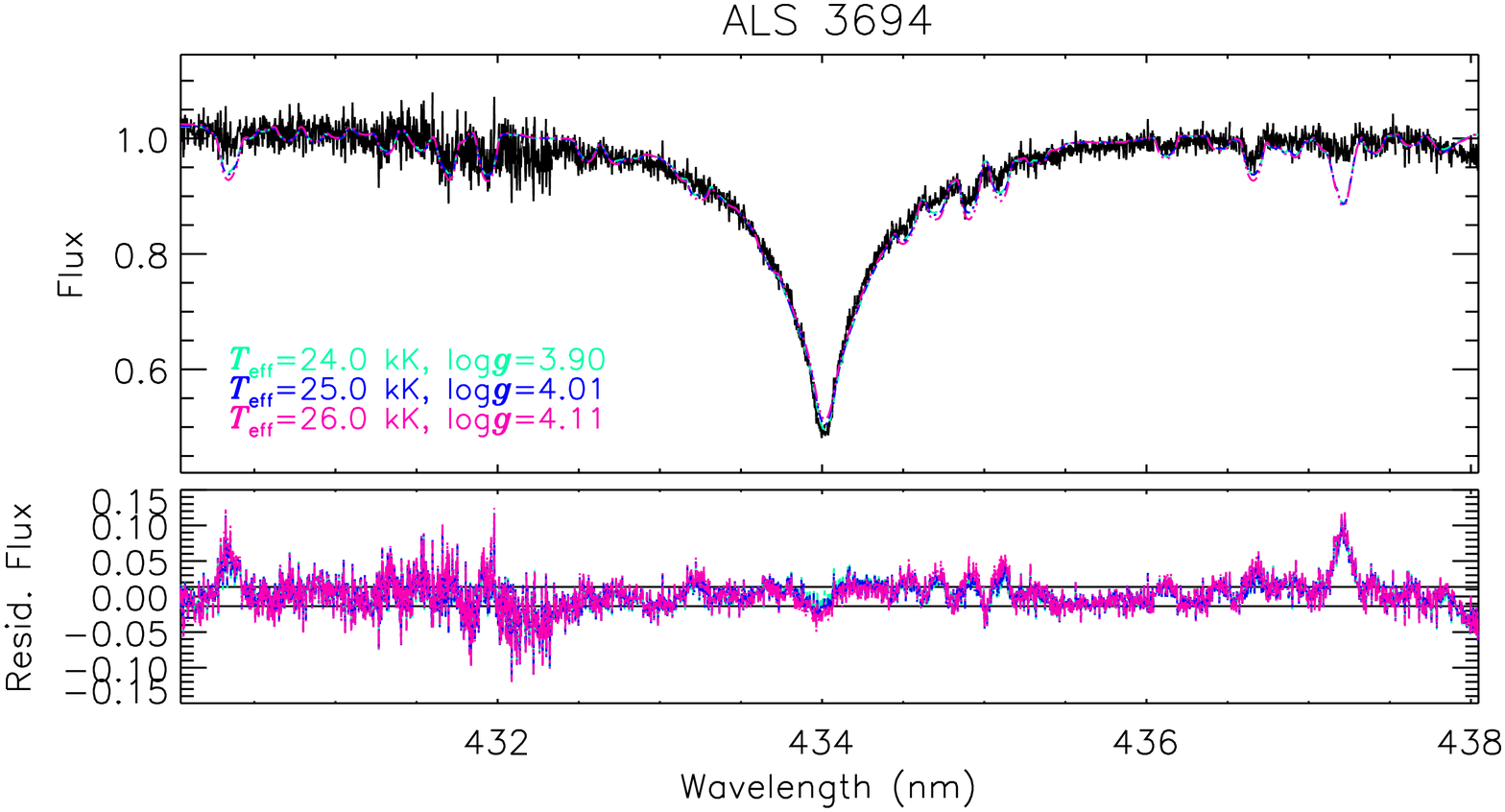} \\
\end{tabular}
      \caption[]{As Fig.\ \ref{balmer1}}
         \label{balmer4}
   \end{figure*}

Figs. \ref{balmer1}-\ref{balmer4} show fits for stars for which only photometric determinations were previously available, or for which we have revisited a previous analysis in light of new data. The best-fit models for each \teff~are shown in the top panel, and the residual flux in the bottom, with the mean flux error indicated by horizontal lines. In almost all cases the residual flux contains features much larger than the mean flux uncertainty. In most cases this can be attributed to chemical peculiarities (i.e., lines not included in the synthetic spectra). Residual flux outside of spectral lines is typically below 1\% of the continuum. 

\noindent {\bf HD 3360}: \cite{2014AA...566A...7N} found $\log{g}=3.80\pm 0.05$ via NLTE quantitative spectroscopy. Adopting the same \teff~as \citeauthor{2014AA...566A...7N}, $20.8\pm0.2$~kK, we find $3.76\pm0.02$, compatible with their result (Fig.\ \ref{balmer1}). There is a systematic bowing in the line wings. This cannot be reproduce with the {\sc tlusty} synthetic spectra, and may reflect a modification of the pressure broadening due to He abundance anomalies and/or magnetic pressure, although the former is unlikely in this case as HD\,3360 is not known to be He-peculiar.





\noindent {\bf HD 35298}: the best fit model (Fig.\ \ref{balmer1}) is not a particularly good fit to the details of the spectrum, likely due to strong chemical peculiarities in the atmosphere of this star. The best-fit $\log{g}$ at the spectroscopic \teff$=15\pm$1 kK is 4.26$\pm$0.13, much higher than the $\log{g}=3.78\pm0.2$ inferred from its temperature and CHORIZOS luminosity. If the slightly higher photometric \teff~(16$\pm$2 kK, \citealt{land2007}) is used instead, $\log{g}$ would need to be even higher to match the spectrum. 



\noindent {\bf HD 36526}: P13 gave $\log{g}=4.0\pm0.3$, consistent with, albeit less precise than the value found here, $\log{g}=4.1\pm0.15$. Model fits are shown in Fig.\ \ref{balmer1}. Similarly to the case of HD\,35298, the residuals are dominated by the strong metallic lines of this highly chemically peculiar star.



\noindent {\bf HD 36982}: \cite{petit2012a} found a reasonable agreement with ESPaDOnS data and {\sc tlusty} synthetic spectra using $\log{g}=4.0\pm0.2$, however determining surface parameters was not the focus of their work and this value was only approximate. Ignoring the innner $\pm0.5$ nm, which is affected by circumstellar and nebular emission, we analyzed H$\beta$ and found $\log{g}=4.4\pm0.2$ (Fig.\ \ref{balmer1}), which seems more plausible given the extremely young age inferred from its association with the Orion Nebula Cluster \citep{1996A&AS..118..503T}.



\noindent {\bf HD 37058}: the EW ratio \teff, 18.5$\pm$0.5 kK, is between that found by spectral fitting, 17 kK \citep{2007AstBu..62..319G}, and the photometric determination of 20 kK \citep{land2007}, however we find a much higher $\log{g}=4.17\pm0.07$ than the value given by P13, 3.8$\pm$0.2 (Fig.\ \ref{balmer1}). 


\noindent {\bf HD 37479}: \cite{1989AA...224...57H} found $\log{g}=3.95 \pm 0.15$ via analysis of high-resolution spectra. Due to the star's extremely strong, broad emission, we used H$\gamma$, ignoring the inner $\pm1.5$ nm (Fig. \ref{balmer1}), and found $\log{g}=4.2\pm0.2$. Outside of the region affected by emission (432.5 nm to 435.5 nm) the residuals are close to the noise level.


\noindent {\bf HD 37776}: \cite{koch2010} adopted $\log{g}=4.0\pm0.1$ as a reasonable fit to high-resolution spectra. Results in the literature from various methods range from 4.1 to 4.5 \citep{2007AA...468..263C}. We measured both H$\beta$ and H$\gamma$ (Fig.\ \ref{balmer1}), excluding the inner $\pm1$ nm in order to avoid the star's circumstellar emission, obtaining $3.97\pm0.07$ and $4.02\pm0.07$, respectively. This is significantly lower than inferred from the star's luminosity, from which $\log{g}=4.30\pm0.07$ would be expected. There are numerous metallic lines that are not fit by the model, and the star is very He-strong; both of these factors decrease the accuracy of the spectroscopic $\log{g}$. We therefore adopt the spectrophotometric value $4.25 \pm0.2$ determined by \cite{2007AA...468..263C}.


\noindent {\bf HD 43317}: \cite{2012A&A...542A..55P} obtained $\log{g}=4.00\pm0.25$ from their analysis of a HARPS spectrum. We excluded the inner $\pm0.5$ nm in order to avoid pulsational variability, and found $\log{g}=4.07\pm0.10$ (Fig.\ \ref{balmer1}). 



\noindent {\bf HD 55522}: \cite{2004AA...413..273B} found \teff~$=17.4\pm0.4$~kK and $\log{g}=4.15\pm0.15$ using Geneva photometry. We obtain the same \teff, but find $\log{g}=3.95\pm0.06$ (Fig.\ \ref{balmer2}). 


\noindent {\bf HD 58260}: the value of $\log{g}$ found here, 3.43$\pm$0.15 (Fig.\ \ref{balmer2}), is in good agreement with some values from the literature \citep[e.g.][]{1989ApJ...346..459B,1999A&A...351..554H,1997A&A...325.1125L}, but is in disagreement with others, which find $\log{g} \ge 4.0$ \citep[e.g.][]{1997AA...324..949Z,2007AA...468..263C}. The spectroscopic value is amongst the lowest surface gravity in the sample, which would imply that HD\,58260 is also one of the most evolved magnetic B-type stars. On the other hand, comparing our sample to the He-strong stars examined by \cite{2007AA...468..263C} using spectrophotometry, all but HD\,58260 are in agreement with our results. The anomalously low value of $\log{g}$ suggests that the Balmer wings may be affected by He overabundance; and indeed, it is one of the most He-strong stars in the current sample \citep{2007AA...468..263C}. Furthermore, applying the same test as in Fig.\ \ref{in_out}, but using the spectrocopic $\log{g}$, yields a disagreement in measured vs.\ inferred surface parameters above the $3\sigma$ level. We therefore suspect the spectroscopic value to be inaccurate, and adopt instead the \cite{2007AA...468..263C} value, $\log{g} = 4.2 \pm 0.2$.



\noindent {\bf HD 63425}: \cite{2011MNRAS.412L..45P} measured $\log{g}=4.0\pm0.1$ from a simultaneous fit of ESPaDOnS and high-resolution UV data using {\sc cmfgen}. We find $4.13\pm0.07$ (Fig.\ \ref{balmer2}), but adopt the \citeauthor{2011MNRAS.412L..45P} result due to their more sophisticated modelling, which accounted for the influence of the stellar wind. 


\noindent {\bf HD 64740}: \cite{1990ApJ...358..274B} found $\log{g}=4.00 \pm 0.14$ based on spectroscopic modelling. We excluded the inner $\pm1$ nm, in order to avoid the weak circumstellar emission. We confirm the \citeauthor{1990ApJ...358..274B} result, obtaining $4.00\pm0.09$ using the mean HARPSpol spectrum  (Fig.\ \ref{balmer2}) and $4.01\pm0.09$ using the mean ESPaDOnS spectrum. 



\noindent {\bf HD 66522}: literature values for $\log{g}$ range widely, from 3.5 to 4.5 \citep{1997AA...324..949Z,1997AA...325.1125L}. We find $3.82 \pm 0.22$ using HARPSpol and $3.88\pm0.23$ with ESPaDOnS (Fig.\ \ref{balmer2}). We adopted the ESPaDOnS value. 


\noindent {\bf HD 66665}: \cite{2011MNRAS.412L..45P} found $\log{g}=3.9 \pm 0.1$, based on simultaneous NLTE {\sc cmfgen} modelling of UV and optical spectra. We find $4.02 \pm 0.07$ (Fig.\ \ref{balmer2}), but adopt the \citeauthor{2011MNRAS.412L..45P} result due to their more sophisticated modelling, which accounted for the influence of the stellar wind. 


\noindent {\bf HD 66765}: \cite{2007AA...468..263C} found $\log{g} = 4.11\pm0.2$ for this star, and \cite{alecian2014} adopted $\log{g}=4.0$ based on their analysis of HARPSpol data. Using the HARPSpol dataset, we find $\log{g}=3.97\pm0.23$; however, the ESPaDOnS and FEROS (Fig.\ \ref{balmer2}) spectra yield $\log{g}=4.13 \pm 0.20$, which we adopted.


\noindent {\bf HD 67621}: \cite{alecian2014} adopted $\log{g}=4.0$ for this star, based on their analysis of HARPSpol data. We confirm their HARPSpol result (finding $4.02\pm0.07$); however, the ESPaDOnS spectrum (Fig.\ \ref{balmer2}) yields $4.18\pm0.07$. We adopted the latter value, due to the systematic offset between HARPS and ESPaDOnS results noted earlier (Fig.\ \ref{esp_harps_logg}).


\noindent {\bf HD 96446}: No detailed measurement of $\log{g}$ appears to have been performed for this star. \cite{neiner2012a} assumed $\log{g}=4.0$, which provided a resonable fit to the HARPSpol spectrum. Our analysis of the mean HARPSpol $H\beta$ profile, in which the inner $\pm0.5$ nm was excluded due to pulsational variability, yielded $\log{g}=3.63\pm0.10$ (Fig.\ \ref{balmer2}). This is despite our adoption of a higher \teff~($23\pm1$ kK) than that found by \cite{neiner2012a} ($21.6\pm0.8$~kK); using the lower \teff~yields an even lower surface gravity. This is lower than is compatible with the star's position on the HRD, which would imply $\log{g}=3.96 \pm 0.06$. Applying the 0.1 dex correction for HARPSpol data (Fig.\ \ref{esp_harps_logg}) narrows the gap. As HD 96446 is He-strong, it is likely that the reduced partial pressure of He leads to a lower apparent $\log{g}$ in H Balmer lines, which might narrow the gap by another 0.1 dex, enough to bring the surface gravity to within 1$\sigma$ of the value inferred from the HRD.


\noindent {\bf HD 105382}: \cite{2001AA...366..121B} used photometric calibrations to obtain \teff$=17.4\pm0.4$ kK and $\log{g}=4.18\pm0.15$. The EW ratio \teff$=18.0\pm0.5$ kK, slightly higher than but compatible with the photometric \teff. This yields $\log{g}=4.03\pm0.07$ (Fig.\ \ref{balmer2}), again compatible with the photometric value albeit somewhat lower. However, only HARPSpol data is available for this star. Applying the 0.1 dex correction brings the photometric and spectroscopic values into close agreement. 








\noindent {\bf HD 121743}: \cite{alecian2014} adopted $\log{g}=4.0$ utilizing HARPSpol data and \teff~$=21$~kK. Using the same HARPSpol dataset we find $3.92 \pm 0.15$; with the mean spectrum from the ESPaDOnS data, we obtain $4.02 \pm 0.13$ (Fig.\ \ref{balmer3}).



\noindent {\bf HD 122451}: While this star is a spectroscopic binary, the two components have very similar luminosities and effective temperatures~\citep{2005MNRAS.356.1362D,2006AA...455..259A,alecian2011,2016A&A...588A..55P}, making it difficult to disentangle the components' Balmer line contributions, but also meaning that they should have similar surface gravities. We therefore treated the system as single for the purposes of determining $\log{g}$, utilizing a \vsini~of 140 \kms~(the approximate mean value of the two components; \citealt{alecian2011}), and ignoring the inner $\pm0.5$ nm in order to avoid the region affected by $\beta$ Cep pulsations \citep[e.g.][]{2016A&A...588A..55P}. The mean spectrum was calculated using only those spectra obtained when the component's radial velocities were close to 0~(in 2013). The result, $\log{g}=3.45 \pm 0.11$ (Fig.\ \ref{balmer3}), is compatible with the results of \cite{alecian2011} (3.5) and \cite{2006AA...455..259A} ($3.5\pm0.4$). We adopted $3.55\pm0.11$, after correcting for the systematic HARPSpol offset.



\noindent {\bf HD 125823}: \cite{2010AA...520A..44B} utilized $\log{g}=4.00 \pm 0.20$, although they did not perform detailed spectroscopic modelling. We find $4.14\pm0.12$ (Fig.\ \ref{balmer3}), where we excluded the inner $\pm0.5$ nm in order to avoid variability in the wings likely arising from the star's considerable horizontal He surface abundance inhomogenities \citep{2010AA...520A..44B}. 






\noindent {\bf HD 130807}: Analyzing HARPSpol and ESPaDOnS data using {\sc atlas9} LTE spectra, \cite{2018arXiv180805503B} found $\log{g}=3.9$ and $\log{g}=3.8$ for the primary and secondary of this spectroscopic binary, significantly lower than the 4.25 found by \cite{alecian2011} in their NLTE analysis of a smaller HARPSpol dataset. While this star is technically a binary, the very similar spectral types and lack of radial velocity variation mean that the components cannot be distinguished in Balmer lines \citep{2018arXiv180805503B}, therefore we treated the star as single for the purpose of measuring $\log{g}$. We find $4.25 \pm 0.15$ with ESPaDOnS (Fig.\ \ref{balmer3}) and $4.14 \pm 0.15$ with HARPSpol, compatible with the results of \cite{alecian2011}. We adopted the ESPaDOnS value, which is closest to the value inferred from the HRD ($4.32\pm0.12$). 



\noindent {\bf HD 142990}: the value found from Balmer line fitting, $4.15\pm0.11$ (Fig.\ \ref{balmer3}), is in agreement with the value determined from the Balmer discontinuity, $4.27\pm0.2$ \citep{2007AA...468..263C}. 



\noindent {\bf HD 156424}: \cite{alecian2014} adopted $\log{g}=4.0$ based on their analysis of HARPSpol spectra. Excluding the inner $\pm0.5$ nm in order to avoid circumstellar emission, we analyzed mean HARPSpol, ESPaDOnS, and FEROS spectra independently, respectively obtaining $\log{g} = 3.85\pm0.15$, $3.96\pm0.08$, and $3.97\pm0.09$. As ESPaDOnS and FEROS yielded compatible results, we utilized the mean ESPaDOnS/FEROS spectrum to obtain $\log{g}=3.99\pm0.10$ (Fig.\ \ref{balmer3}).  




\noindent {\bf HD 163472}: \cite{neiner2003b} found $\log{g}=4.20 \pm 0.11$ via simultaneous fitting of NLTE synthetic spectra to ultraviolet and optical high-resolution spectroscopy. Using H$\beta$ (Fig. \ref{balmer3}), we obtain $\log{g}=4.19 \pm 0.10$, almost identical to the \cite{neiner2003b} result. 



\noindent {\bf HD 175362}: using the EW ratio \teff$=17.6\pm0.4$ kK yields $\log{g}=4.21\pm0.06$ (see Fig.\ \ref{balmer3}). This is a substantially higher surface gravity than that adopted by P13, $\log{g}=3.67\pm0.16$, which was obtained from \cite{1997AA...320..257L} from a simultaneous abundance, \teff, and $\log{g}$ analysis of H$\beta$ and nearby spectral lines, where \teff~was determined from H$\beta$. The discrepancy is due to the much lower \teff~adopted by \citeauthor{1997AA...320..257L}, 14.6 kK. The EW ratio \teff~is consistent with that determined by \cite{2007AA...468..263C} using the Balmer discontinuity, 17.5 kK, and with the mean of photometric \teff~determinations compiled from the literature by \cite{2008AA...491..545N}, $16.8\pm0.6$ kK. 


\noindent {\bf HD 176582}: \cite{bohl2011} found $\log{g}=4.0\pm0.1$ via a careful omission of the central emission-line region of the H$\beta$ line, using a model accounting for non-solar surface abundances. Using the H$\beta$ line, and likewise excluding the central $\pm0.5$ nm from consideration, we find $\log{g}=3.85\pm0.17$; however, using H$\gamma$ (Fig.\ \ref{balmer3}) we find $\log{g}=4.02\pm0.18$, compatible with the results of \cite{bohl2011}. We adopt the \citeauthor{bohl2011} results, as their modelling was more sophisticated.


\noindent {\bf HD 186205}: we find $\log{g}=3.84\pm0.17$, consistent with the photometric determination of $4.0\pm0.2$. While the residuals are within 1\% of the continuum, there is a systematic bowing in the inner wings of the line (see Fig.\ \ref{balmer3}) similar to that of HD\,3360. Since HD\,186205 is He-strong, the bowing might be ascribed to either surface He anomalies, or to magnetic pressure. 



\noindent {\bf HD 189775}: the EW \teff$=17.5\pm0.6$ kK is in reasonable agreement with the photometric \teff$=16.2\pm0.6$ kK \citep{2002MNRAS.333....9L}. Using the EW \teff~yields $\log{g} = 4.12 \pm 0.08$ (Fig.\ \ref{balmer4}), which overlaps within uncertainty with the photometric $\log{g} = 3.97 \pm 0.15$. 



\noindent {\bf HD 205021}: \cite{2010AA...515A..74L} found $\log{g}=3.80 \pm 0.15$ utilizing a {\sc fastwind} NLTE analysis of high-resolution spectra. We excluded the inner $\pm1.0$ nm region in order to avoid the wind (which should not be significant at the star's \teff~$=25\pm1$~kK), the effects of the star's $\beta$ Cep pulsations, and (more importantly) emission from the companion Be star \citep{2006AA...459L..21S} (which is otherwise undetectable in the spectrum). We obtain $\log{g}=3.72\pm0.08$ (Fig.\ \ref{balmer4}).



\noindent {\bf HD 208057}: \cite{2001AA...378..861C} found \teff$=19 \pm 3$~kK and $\log{g} = 4.00 \pm 0.2$ based upon spectral modelling accounting for non-uniform line formation due to gravity darkening. The star's rotation is far from critical ($P_{\rm rot} = 1.4$~d), so gravity darkening should be negligible for this star. Our \teff~is somewhat lower at $16.5 \pm 1.2$~kK. Using this \teff~in conjunction with {\sc tlusty} synthetic spectra, we found $\log{g} = 4.00 \pm 0.16$ (Fig.\ \ref{balmer4}), identical to the value determined by \cite{2001AA...378..861C}. 



\noindent {\bf HD 345439}: No determinations of the star's surface gravity exist in the literature. The star has very strong H$\alpha$ emission extending to $\pm800$~\kms, making the H$\beta$ line unsuitable for determining $\log{g}$. The best available line is H$\gamma$, for which the formal result is $\log{g}=4.3\pm0.2$, as can be seen from Fig.\ \ref{balmer4}. However, only a 2 nm window could be fit due to difficulties in merging the orders of the available ARCES spectra, thus it is likely that the line is over-normalized. Furthermore, due to the very strong CM emission this line is also almost certainly partially filled in. The surface gravity determined here should therefore by re-examined at a later date using spectra with wider spectral orders in conjunction with a model that accounts for circumstellar emission.



\noindent {\bf ALS 3694}: the spectra of this dim ($V = 10.35$ mag) star are quite noisy, and only 4 of the 16 ESPaDOnS spectra are suitable for analysis. Since the star has magnetospheric emission \citep{2016ASPC..506..305S} the inner $\pm0.5$ nm were masked out. We used both H$\beta$ and H$\gamma$, testing the ESPaDOnS and FEROS datasets independently. H$\beta$ yielded $\log{g}=3.7\pm0.1$, while H$\gamma$ (shown in Fig.\ \ref{balmer4}) yielded $4.0\pm0.1$. This is likely because H$\beta$ is still partly filled with weak emission. We adopted the higher value, which is more consistent with the luminosity. 


\section{Surface gravity measurements of binary stars}\label{sec:logg_bin}

   \begin{figure*}
   \centering
\begin{tabular}{cc}
   \includegraphics[trim = 40 0 50 0, width=8.5cm]{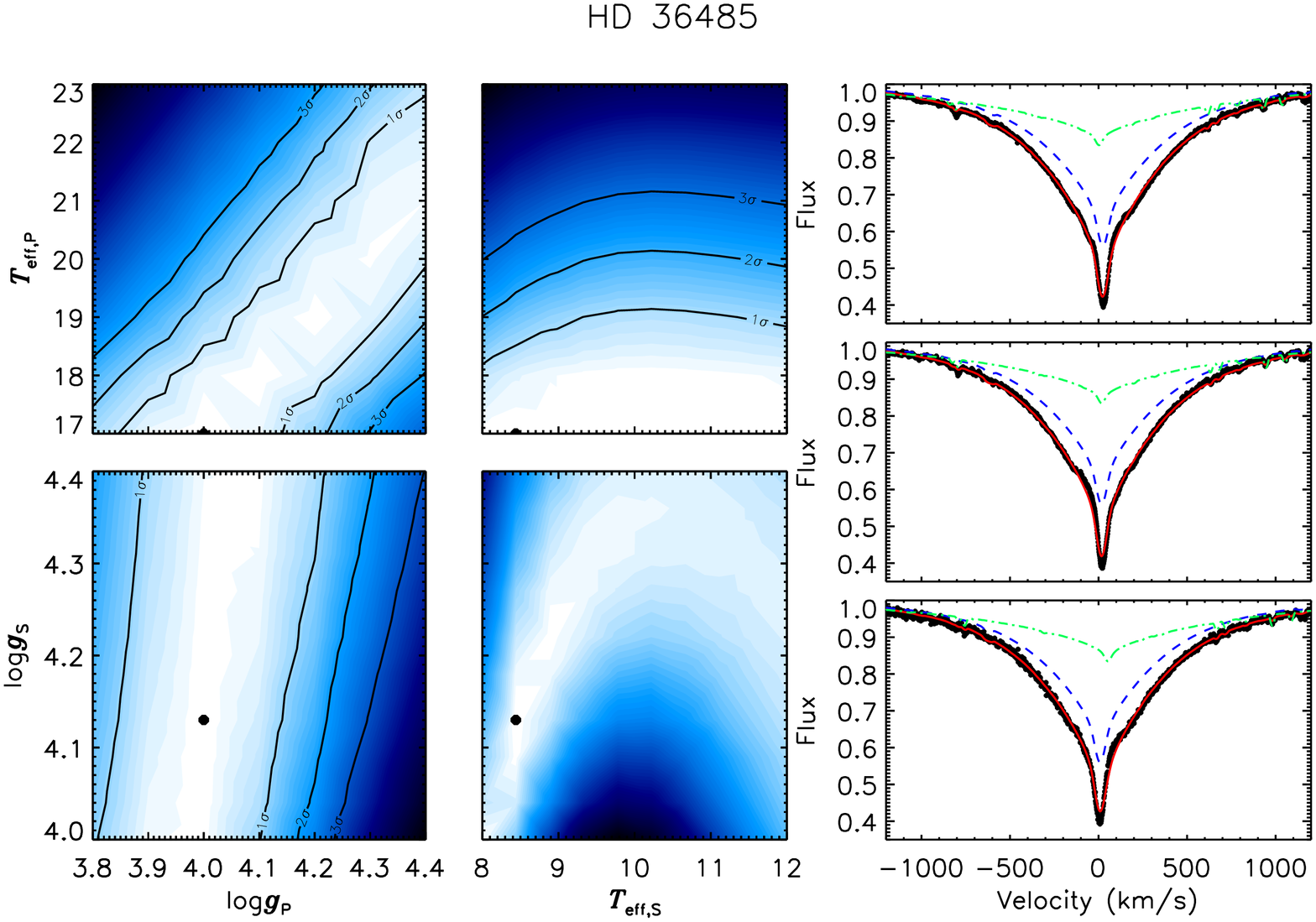} &
   \includegraphics[trim = 40 0 50 0, width=8.5cm]{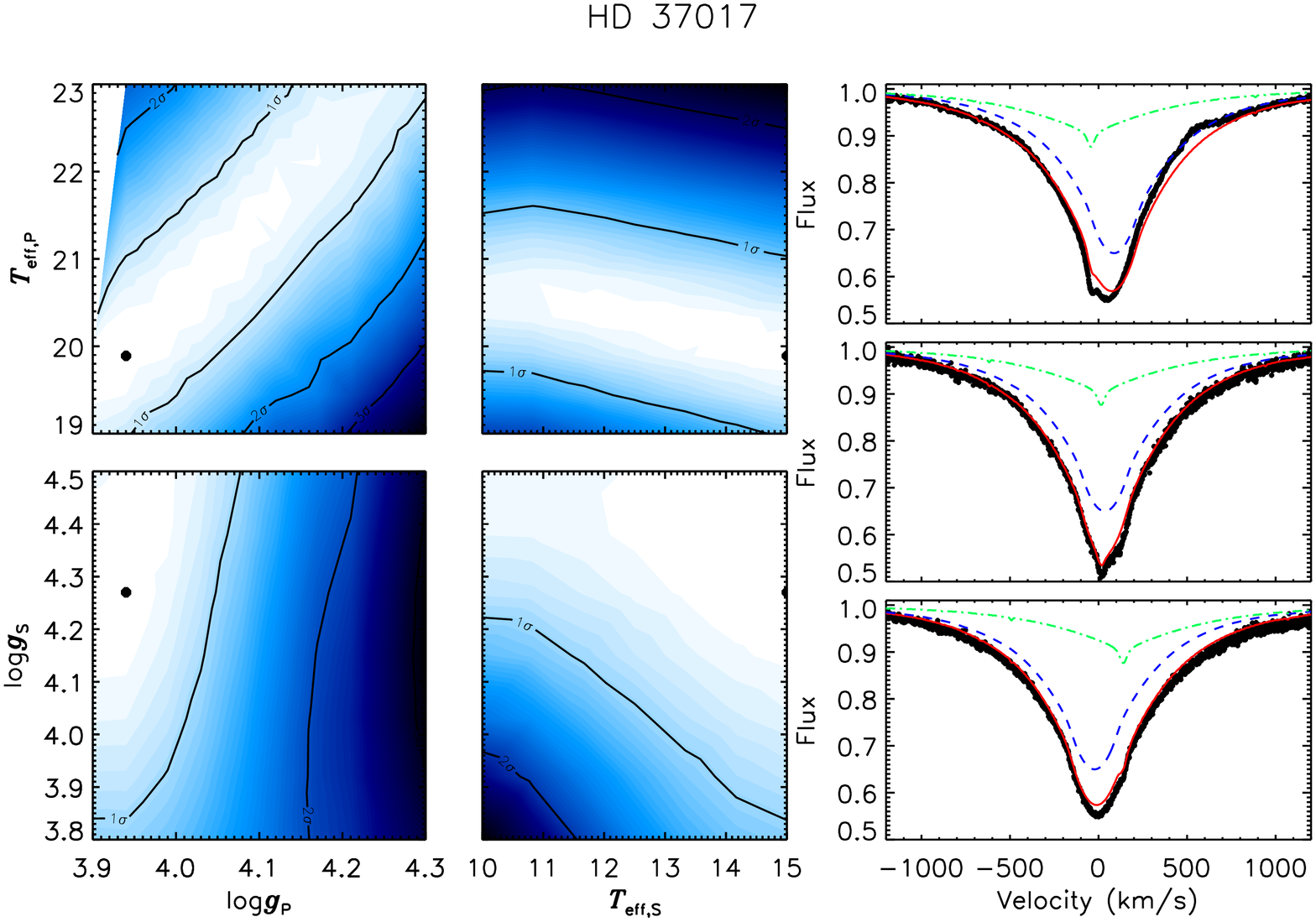} \\
   \includegraphics[trim = 40 0 50 0, width=8.5cm]{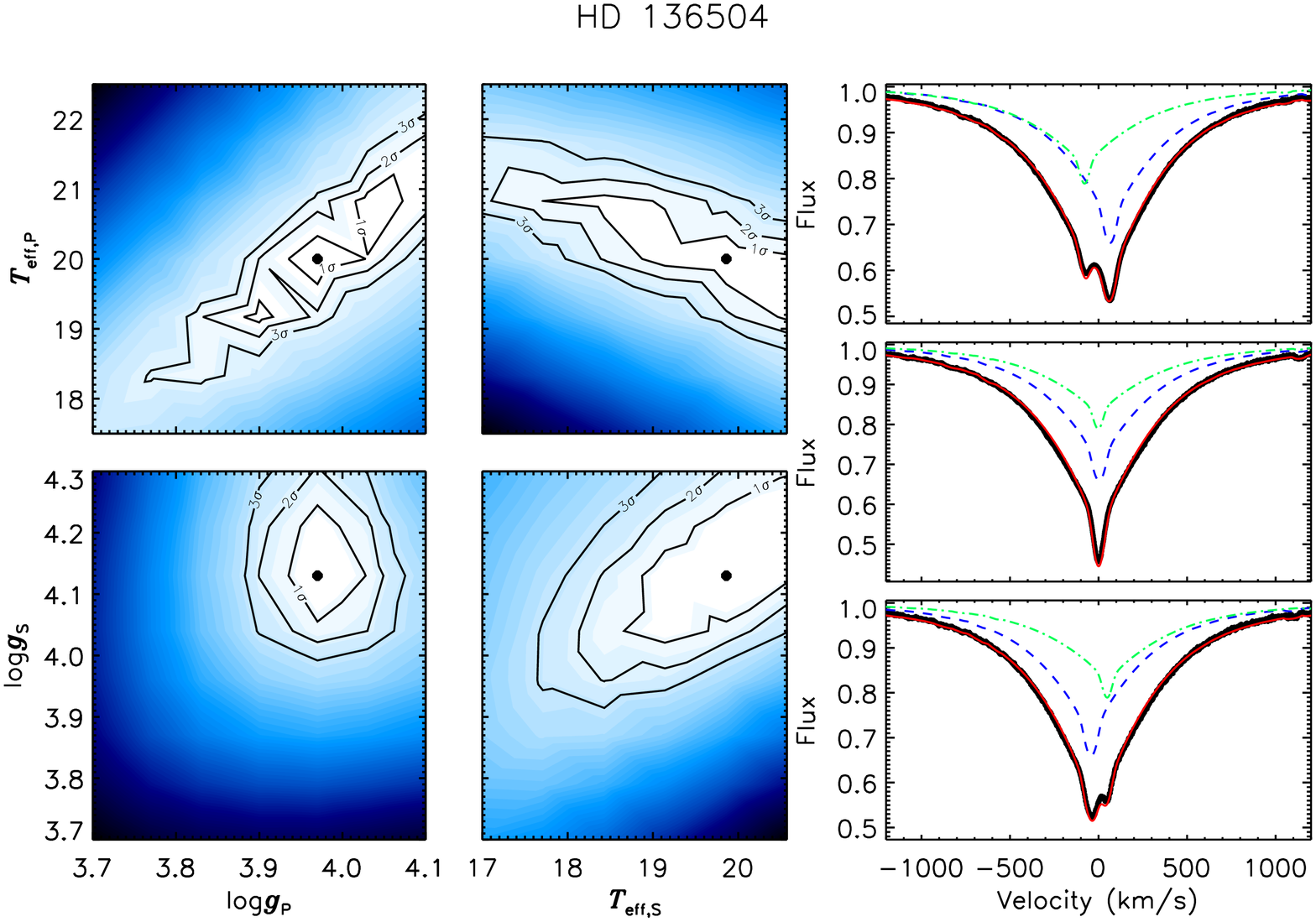} &
   \includegraphics[trim = 40 0 50 0, width=8.5cm]{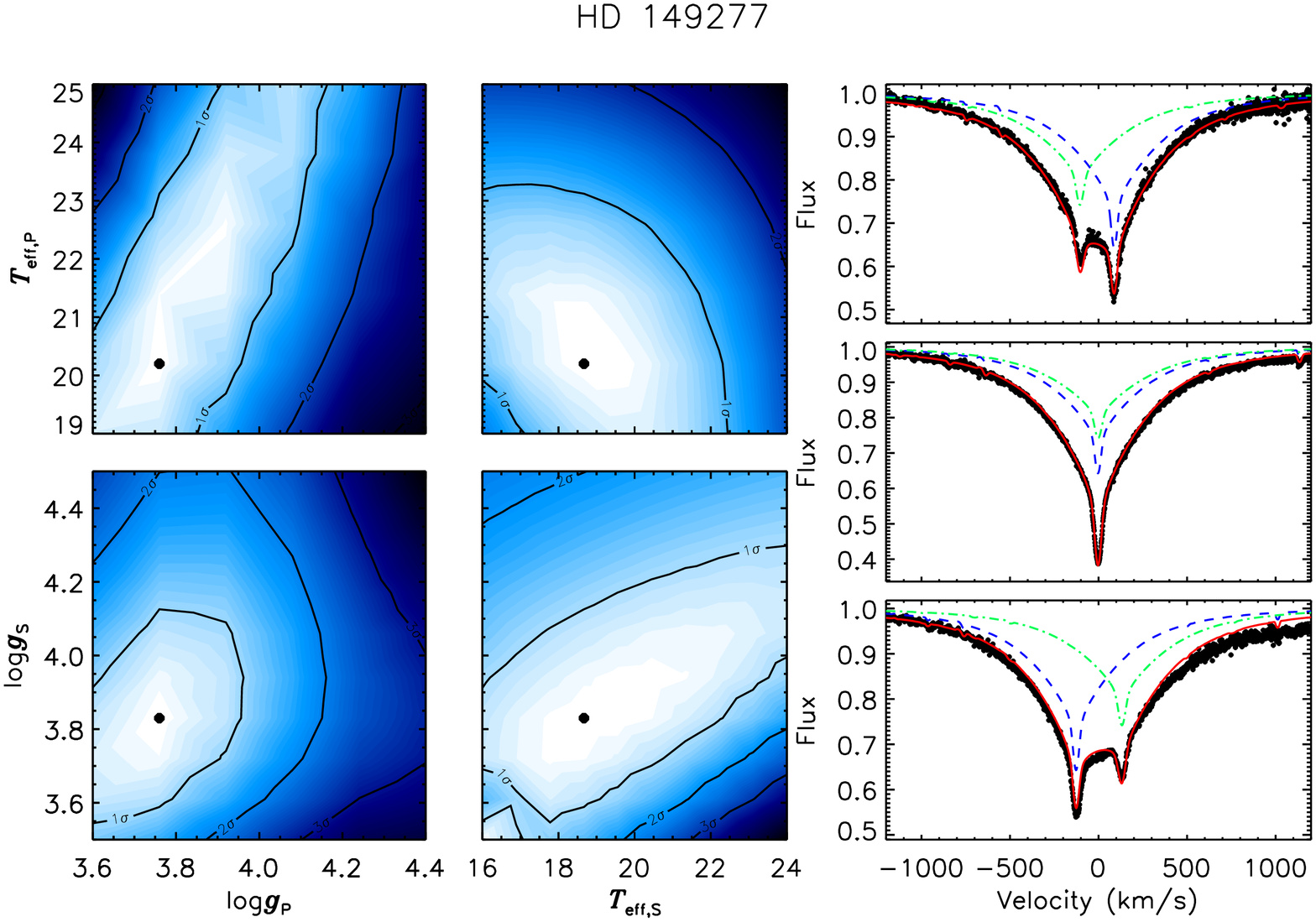} \\
\end{tabular}
      \caption[]{Surface gravity determinations for binary systems. The left sub-panels show reduced $\chi^2$ landscapes for $\log{g_{\rm P}}$, $\log{g_{\rm S}}$, $T_{\rm eff,P}$, and , $T_{\rm eff,S}$, where darker shades correspond to higher $\chi^2$. The location of the best-fit model is indicated by a circle. The right-hand sub-panels show, from top to bottom, the corresponding fits (combined flux in solid red lines, primary in dashed blue, secondary in dot-dashed green) to observations (black dots) at quadrature (top and bottom) and conjunction (middle).}
         \label{bin_logg_fit}
   \end{figure*}

Determining the surface gravities of binary stars requires special care, since all stellar components will contribute to the Balmer line wings. This leads to a degeneracy in their parameters, therefore the uncertainties are typically greater. This can be partly overcome if enough observations are available, and the variability of the components is sufficiently large, that the contributions of the individual stars can be discerned. Spectral modelling accounting for all components has already been performed for HD 25558 \citep{2014MNRAS.438.3535S}, HD 35502 \citep{sikorainprep}, HD 37061 \citep{2011AA...530A..57S,2019MNRAS.482.3950S}, HD 122451 \citep{2006AA...455..259A}, and HD 156324 \citep{alecian2014}. 

No analysis of $\log{g}$ carefully accounting for binarity is available for HD 36485, HD 149277, HD 136504, or HD 37017. For these systems a grid of at least 25 synthetic spectra was prepared for each stellar component, covering the approximate range in \teff~and $\log{g}$ expected for the components, using either {\sc tlusty} BSTAR2006 models (when both components are likely above \teff$=$15 kK) or ATLAS models (when one of the stars is below 15 kK).  The radius and mass of the primary $R_{\rm P}$ and $M_{\rm P}$ were determined by interpolating through evolutionary tracks \citep{ekstrom2012} according to the \teff~and $\log{g}$ of the model. The mass $M_{\rm S}$ of the secondary was then determined from the mass ratio obtained from the radial velocity semi-amplitudes, from which the radius was obtained directly as $R_{\rm S} = \sqrt{(GM_{\rm S}/g)}$. Synthetic spectra were then moved to the measured radial velocities of each component, added together with the contributions of each component scaled by the ratio of the relative stellar radii, normalized to the synthetic continua calculated in the same fashion, and the reduced $\chi^2$ calculated. The overall fit for each pair of models was taken as the weighted mean $\chi^2$ across all observations, with the weights taken from the mean flux error bars of each spectrum in order to keep noisier spectra from biasing the results towards overall worse fits.

\noindent {\bf HD 36485}: \cite{leone2010} found a mass ratio of 2.6. Using high-resolution spectra, \cite{1999A&A...345..244Z} found $\log{g}=4.41$; however, this was without accounting for either the presence of a companion, or the star's circumstellar emission \citep{leone2010}. We excluded the region within $\pm200$~\kms~of the primary's RV in order to avoid bias due to emission. As can be seen in the $\chi^2$ landscapes in Fig.\ \ref{bin_logg_fit} (top left), the best-fit solution is for \teff$_{\rm ,P}=17$~kK, $\log{g}_{\rm P}=4.0$, \teff$_{\rm ,S}=8.5$~kK, and $\log{g}_{\rm S}=4.15$. However, the EW analysis indicated that the primary's effective temperature is $20\pm2$~kK \citep{leone2010}; if we follow the $\chi^2$ contours to this \teff, the best-fit model is $\log{g}_{\rm P}=4.2\pm0.2$.



\noindent {\bf HD 37017}: the primary is a rapidly rotating star, $P_{\rm rot} = 0.901186(1)$ d. Given this, it is potentially important to take rotational distortion of the star into account when constraining the surface gravity, as $\log{g}$ may vary by up to 0.5 dex from pole to equator if the star is rotating near critical. For HD 37017, this is the case for most of the models with $\log{g} \le 3.95$. Models for which the critical rotation fraction $\Omega = \omega / \omega_{\rm crit} < 0.6$ (where $\omega$ is the rotational frequency and $\omega_{\rm crit}$ is the critical rotation frequency) were convolved with \vsini~according to the usual method, as in this regime the effects of rotational distortion are mild enough that they can be ignored. Gravity darkening was accounted for when $0.6 < \omega < 1$, which is the case for about 2/3 of the parameter space under consideration for this star. To account for the rotational distortion, the radius and surface gravity at each point were determined using the formalism supplied by \cite{1996PhDT........92C}, with the gravity darkened flux obtained via the von Zeipel theorem \citep{1924MNRAS..84..665V}. Fig.\ \ref{bin_logg_fit} (top right) shows the resulting best-fit model. Note that the star displays weak emission in its Balmer wings, originating in its magnetosphere, and therefore the inner $\pm500$~\kms~were ignored. For the primary we find the best fit at $\log{g}=3.95$, however this is at the lower end of the star's \teff~measured using EW ratios; following the $\chi^2$ contours to \teff$=21\pm2$~kK as determined spectroscopically suggests instead $\log{g}=4.1\pm0.2$. 



\noindent {\bf HD 136504}: the RV curve implies a mass ratio of 1.19 \citep{uytterhoeven2005,pablo_epslup}. When \teff~and $\log{g}$ are allowed to vary freely, the minimum $\chi^2$ is found with a higher temperature for the secondary. Since ionization balances unambiguously imply a lower temperature for the secondary, the grid was restricted to only those models for which $T_{\rm eff,P} \ge T_{\rm eff,S}$ (see Fig.\ \ref{bin_logg_fit}, bottom left). The result is a somewhat higher surface gravity for the secondary, $\log{g_{\rm S}} = 4.13 \pm 0.1$ as compared to $\log{g_{\rm P}} = 3.97 \pm 0.1$. 



\noindent {\bf HD 149277}: \cite{2016PhDT.......390S} found a mass ratio of 1.1, a result later confirmed by \cite{2018MNRAS.481L..30G}. The best results from this method were obtained for this star, given the large dataset, large radial velocity amplitude, and the sharp spectral lines of both components. The reduced $\chi^2$ landscape (Fig.\ \ref{bin_logg_fit}, bottom right) shows a sharply defined valley around $\log{g_{\rm P}} = 3.75 \pm 0.15$ and $\log{g_{\rm S}} = 3.85 \pm 0.3$, suggesting the system to be somewhat evolved. The effective temperatures preferred by this method are additionally in line with those inferred from ionization balances, albeit less precise. 

\end{document}